\documentclass[lettersize,journal]{IEEEtran}
\usepackage{amsmath,amsfonts}
\usepackage{algorithmic}
\usepackage{algorithm}
\usepackage{array}
\usepackage[caption=false,font=normalsize,labelfont=sf,textfont=sf]{subfig}
\usepackage{textcomp}
\usepackage{stfloats}
\usepackage{url}
\usepackage{verbatim}
\usepackage{graphicx}
\usepackage{booktabs}
\usepackage{cite}
\usepackage{tabularx,booktabs}
\usepackage{diagbox}
\usepackage{array,multirow}
\usepackage{wrapfig}
\usepackage{lipsum}
\usepackage{mathtools}
\usepackage{float}
\usepackage{xcolor}
\usepackage{enumitem}
\usepackage{csquotes}
\usepackage{makecell}
\usepackage{amssymb}

\usepackage{moresize}

\usepackage{booktabs}
\usepackage{adjustbox}

\usepackage{colortbl}

\usepackage{longtable}
\usepackage{lscape}
\usepackage{rotating}

\usepackage{longtable}

\usepackage{booktabs,tabularx,array}

\usepackage[table]{xcolor}
\definecolor{brand}{HTML}{004C99}   
\definecolor{brand2}{HTML}{4DA3FF}  
\definecolor{stripeA}{HTML}{F2F6FC} 
\definecolor{stripeB}{HTML}{E6F0FA} 

\usepackage{tabularx}
\usepackage{colortbl,booktabs}
\usepackage[%
	acronym=true,%
	nonumberlist,%
	nopostdot,%
	seeautonumberlist,%
	shortcuts,%
	section=chapter,%
	toc=false,%
	nogroupskip=true,%
    prefix%
]{glossaries-extra}
\usepackage{textcase,mfirstuc,xspace}
\usepackage{orcidlink}
\usepackage{comment}
\usepackage{svg}
\usepackage{soul}
\usepackage{pifont}
\usepackage[most]{tcolorbox}
\usepackage{eso-pic}
\newcolumntype{C}{>{\centering\arraybackslash}X}
\usepackage[spaces,hyphens]{xurl} 
\usepackage{booktabs,xcolor,siunitx}
\definecolor{headergray}{RGB}{162, 182, 215}    
\definecolor{maincolor}{RGB}{55, 90, 140}       
\definecolor{subcolor}{RGB}{90, 110, 140}      
\definecolor{lightgray}{RGB}{255, 204, 102}       
\usepackage{hyperref}

\definecolor{LightCyan}{rgb}{0.88,1,1}
\definecolor{Gray}{gray}{0.85}

\hypersetup{%
  colorlinks=true,%
  linkcolor={blue!65!black},
  citecolor={blue},
  urlcolor={blue!80!black},
  bookmarksnumbered=true,%
  bookmarksopen=true}
  
\begin{document}


\title{A Survey of OTFS-Based Index Modulation Techniques: Challenges, Benefits, and Future Directions for 6G and Beyond }

\author{Burak Ahmet Ozden,~ Erdogan Aydin,~Emir Aslandogan,~Haci Ilhan,~\IEEEmembership{Senior Member,~IEEE},~Ertugrul Basar, ~\IEEEmembership{Fellow,~IEEE},~Miaowen Wen, ~\IEEEmembership{Senior Member,~IEEE},~Marco Di Renzo,~\IEEEmembership{Fellow,~IEEE},~H. Vincent Poor,~\IEEEmembership{Life Fellow,~IEEE}

\thanks{B. A. Ozden is with the Department of Electrical and Electronics
Engineering, Istanbul Medeniyet University, 34857 Istanbul, Türkiye, and also
with the Department of Computer Engineering, Yıldız Technical University,
34220 Istanbul, Türkiye (e-mail: bozden@yildiz.edu.tr).}

\thanks{E. Aydin is with the Department of Electrical and Electronics En-
gineering, Istanbul Medeniyet University, 34857 Istanbul, Türkiye (e-mail:
erdogan.aydin@medeniyet.edu.tr)}

\thanks{E. Aslandogan and H. Ilhan are with Y{\i}ld{\i}z Technical University, Department of Electronics and Communications Engineering, 34220, Davutpasa, Istanbul, Turkey (e-mail: emira@yildiz.edu.tr and ilhanh@yildiz.edu.tr).}

\thanks{E. Basar is with the Department of Electrical Engineering,
Tampere University, 33720 Tampere, Finland, on leave from the Department
of Electrical and Electronics Engineering, Koc University, 34450 Sariyer,
Istanbul, Turkey (e-mail: ertugrul.basar@tuni.fi).}

\thanks{M. Wen is with School of Information Science and Technology, Nantong
University, Nantong 226019, China, and also with School of Electronic and
Information Engineering, South China University of Technology, Guangzhou
510640, China (e-mail: eemwwen@scut.edu.cn). }

\thanks{M. Di Renzo is with Universit\'e Paris-Saclay, CNRS, CentraleSup\'elec, Laboratoire des Signaux et Syst\`emes, 3 Rue Joliot-Curie, 91192 Gif-sur-Yvette, France. (marco.di-renzo@universite-paris-saclay.fr), and with King's College London, Centre for Telecommunications Research -- Department of Engineering, WC2R 2LS London, United Kingdom (marco.di\_renzo@kcl.ac.uk).}

\thanks{H. Vincent Poor is with the Department of Electrical and Computer Engineering, Princeton University, Princeton, NJ 08544 USA (e-mail: poor@princeton.edu).}
}



\maketitle

\begin{abstract}
Orthogonal time frequency space (OTFS) is a two-dimensional modulation technique that uses the delay-Doppler (DD) domain and is a candidate for providing robust, high-capacity wireless communications for envisioned sixth-generation (6G) and beyond networks. The OTFS technique maps data to the DD domain instead of the traditional time-frequency domain, enabling it to fully utilize channel diversity and transform fast time-varying channels into nearly static channels. Index modulation (IM) is a communication paradigm that conveys information not only through conventional modulation symbols but also by encoding data bits in the indices of the selected communication resources to improve error performance, spectral efficiency, and energy efficiency. In this survey article, a comprehensive review of work on OTFS-based wireless communication systems is presented. In particular, the existing OTFS-based IM (OTFS-IM) schemes in the literature are reviewed and systematically categorized according to their system architectures, detection methods, and performance aspects such as capacity, peak-to-average power ratio (PAPR), diversity, complexity, imperfect channel state information (CSI), spectral efficiency, and outage probability. Furthermore, the operating principles and system models of OTFS-IM variants—including OTFS-based space shift keying (OTFS-SSK), OTFS-based spatial modulation (OTFS-SM), OTFS-based quadrature spatial modulation (OTFS-QSM), OTFS-based media-based modulation (OTFS-MBM), and OTFS-based code index modulation (OTFS-CIM)—are described, followed by a comparative performance analysis in terms of computational complexity, error performance, capacity, energy saving, spectral efficiency, and throughput. Finally, the challenges, benefits, and future directions for OTFS-IM systems are discussed, covering key aspects such as complexity, efficiency, latency, channel estimation, hardware constraints, synchronization, security, and potential integration with other advanced wireless communication techniques. 
\end{abstract}


\section{Introduction}

\begin{figure}
    \centering
    \includegraphics[width=0.9\linewidth]{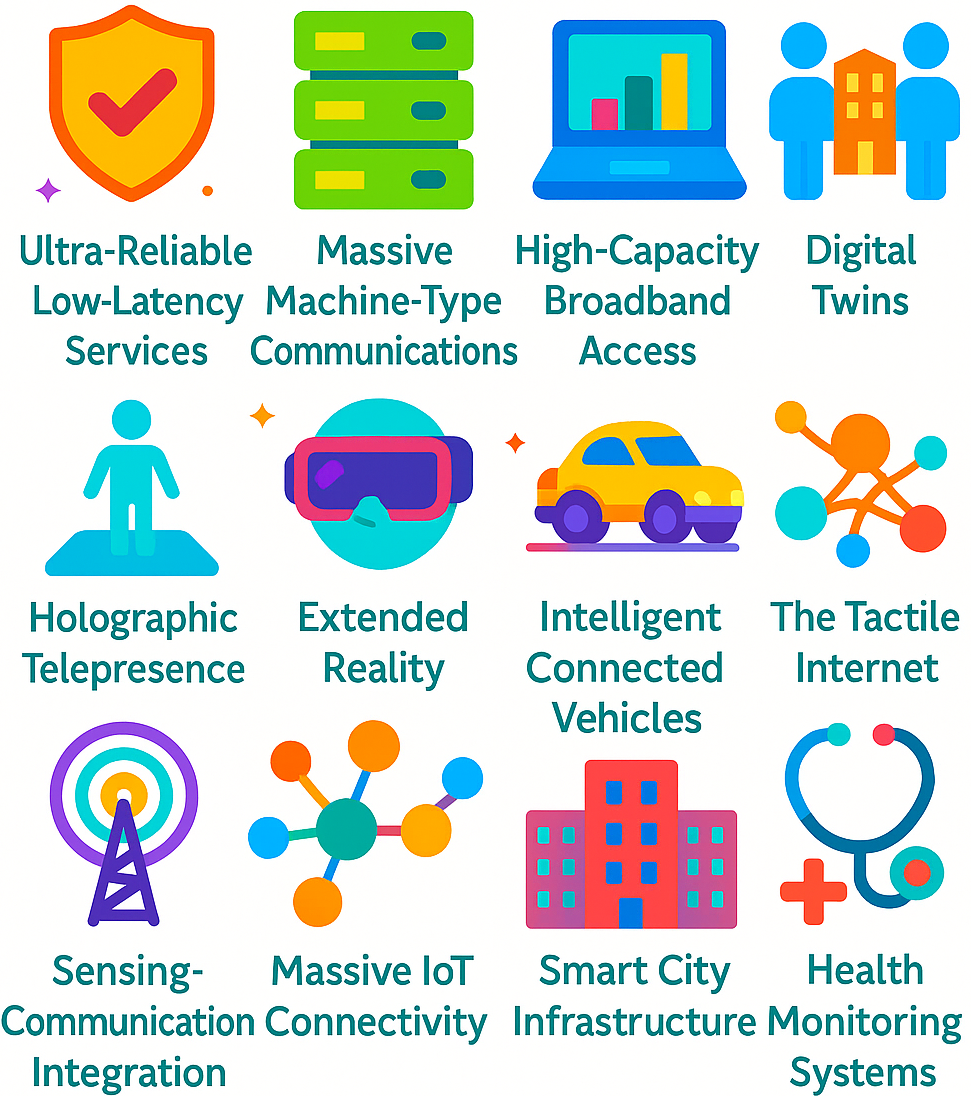}
    \caption{Potential applications of 6G networks.}
    \label{6Gp}
\end{figure}

\IEEEPARstart{B}{y} $2030$, sixth-generation (6G) networks are expected to enable ultra-reliable low-latency services, massive machine-type communications, high-capacity broadband access, holographic telepresence, extended reality, digital twins, sensing–communication integration, the Tactile Internet, intelligent connected vehicles, smart city infrastructure, massive Internet of Things (IoT) connectivity, remote industrial automation, and health monitoring systems \cite{6G3,6G4,6G5,6G6,6G7,6G8,6G9,6G10,6G11,6G12,6G13,6G14,6G15,6G16,6G17,6G18,6G19}, as illustrated in Fig. \ref{6Gp}. In fourth-generation (4G) systems, peak data rates were limited to approximately $300$ Mbps, increasing to $20$ Gbps with fifth-generation (5G) deployments. For 6G, the peak data rate is anticipated to surpass $1$ Tbps to accommodate extremely high data demands. Likewise, the user experience rate is expected to improve substantially from $10$ Mbps in 4G and $100$ Mbps in 5G to around $1$ Gbps in 6G. Furthermore, user plane latency, which was about $50$ ms in 4G and reduced to $1$ ms in 5G, is projected to decrease to the range of $25$ microseconds to $1$ ms in 6G \cite{6G1, 6G2, 6Gek1, 6Gek2}. In 6G networks, the growing demand from mobile users will create a huge volume of data traffic. With billions of devices connected simultaneously—ranging from smartphones and augmented reality/virtual reality (AR/VR) headsets/glasses to autonomous vehicles and IoT sensors—the communication infrastructure must support very high data rates, ultra-low latency, and seamless connectivity under rapidly changing conditions \cite{6G20,6G21,6G22,6G23,6G24}.

\begin{figure}[t]
    \centering
    \includegraphics[width=0.75\linewidth]{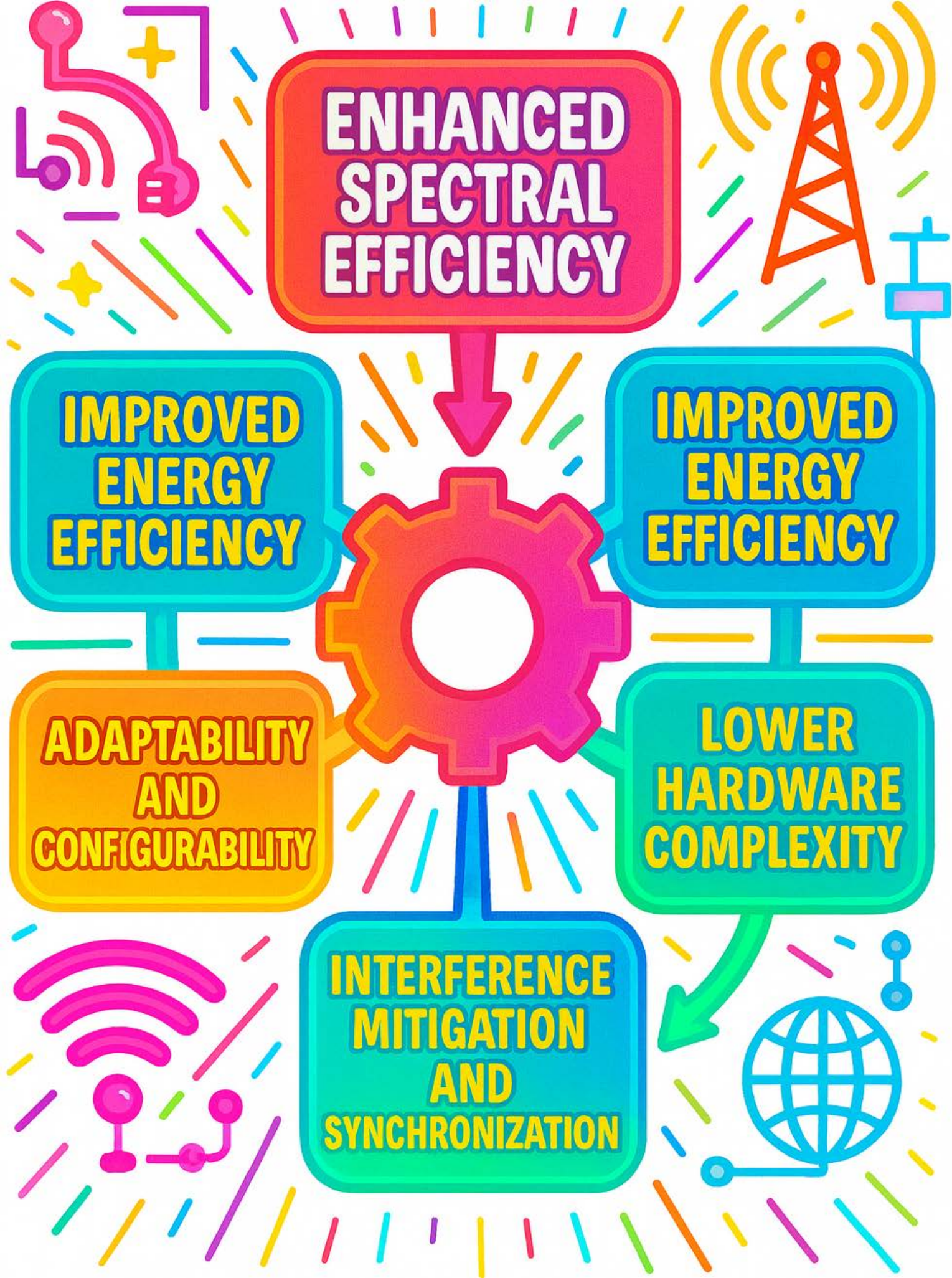}
    \caption{Main advantages of IM-based systems.}
    \label{IMadvantage}
\end{figure}

Index modulation (IM) is a promising approach to enhance spectral and energy efficiency in wireless communication systems. Unlike conventional modulation techniques, which encode data only in the amplitude and phase of symbols, the IM conveys additional data by activating specific communication resources. These can include transmit antennas, subcarriers, time slots, signal constellation points, spatial paths, spreading codes, reconfigurable intelligent surface (RIS) elements, receive antennas, polarization states, radar waveforms, delay-Doppler (DD) resource elements, channel impulsive response (CIR) taps, and mirror activation patterns (MAPs). In contrast to conventional wireless communication schemes, which utilize all available resources simultaneously, the IM techniques transmit data bits by activating only a selected subset and encoding information in the indices of the active elements \cite{IM1, IM2, IM3, IM4, IM13}. IM splits the information bits into two groups: index bits and symbol constellation bits. The index bits specify which resources are active during a transmission period. The constellation bits select a symbol from a conventional modulation constellation such as phase shift keying (PSK) or quadrature amplitude modulation (QAM), and the selected symbol is transmitted through the active resources (in some IM schemes, such as double SM in \cite{DSM}, multiple symbols are selected instead of a single one). Since only a subset of communication resources is active during each transmission period, IM can maintain the same throughput with significantly lower energy consumption, as many resources remain idle and do not require transmission energy. This also reduces hardware complexity. For instance, spatial modulation (SM) requires only a single radio frequency (RF) chain regardless of the number of antennas, unlike conventional multiple-input multiple-output (MIMO) systems, which require one RF chain per antenna \cite{SMint1, SMint2, SMint3, SMint4}. IM inherently mitigates interference challenges such as inter-channel and inter-carrier interference (ICI) by limiting the number of simultaneously active transmission paths. This reduces signal overlap and enables a more streamlined, energy-efficient receiver architecture \cite{IMRC1, IMRC2, IMRC3}. IM improves error performance without requiring additional power or bandwidth. This enhances symbol separability under noise and multipath fading \cite{IM8, IMham1, IMham2}. Fig. \ref{IMadvantage} shows the main advantages of IM-based systems. By conveying extra information with the same resources, IM significantly enhances communication efficiency, making it a strong candidate for 6G networks. The main advantages of IM systems can be summarized as follows \cite{IM5, IM6, IM7, IM9, IM10, IM11, IM12}: IM enables additional bits to be transmitted through communication resource indices without requiring extra bandwidth, thus increasing the data bits transmitted per channel use in a transmission period. In IM systems, index bits are embedded in the selected resources themselves, consuming no additional RF power for transmission. As a result, fewer symbols are needed to convey the same amount of data compared to traditional communication systems. By activating only a fraction of the system resources at a time, IM reduces the number of active RF chains needed, which simplifies transceiver architecture. In IM schemes, only a subset of resources is active while the remaining ones stay inactive.  The inactive elements inherently do not cause interference, which significantly reduces inter-channel interference and simplifies synchronization. IM schemes make transmitted signals more distinguishable and more resistant to noise, which enhances error performance. IM embeds extra information into the indices of already existing system parameters so that it can be directly integrated into almost any communication system.


Orthogonal time frequency space (OTFS) modulation is a strong candidate for meeting the demanding requirements of 6G networks. The OTFS technique was first proposed by Hadani et al. in \cite{OTFSilk}, where it was introduced as a high-performance modulation scheme for time-varying multipath wireless channels. Unlike traditional modulation schemes such as orthogonal frequency division multiplexing (OFDM) that operate in the time-frequency domain, OTFS maps signals into the DD domain, making it highly robust against channel variations and severe Doppler shifts caused by high mobility and multipath propagation. This significant advantage allows OTFS to enable reliable communication in demanding scenarios, including high-speed trains, autonomous drones, RISs, multiple access networks, integrated sensing and communication (ISAC), underwater acoustic communications, millimeter-wave communications, and satellite/space–air–ground integrated networks \cite{isacpaper, OTFSpaper1, OTFSpaper2, OTFSpaper3, OTFSpaper4, OTFSpaper5}. As a result, OTFS enables full exploitation of delay and Doppler diversity, provides nearly constant channel gain per symbol, effectively suppresses ICI, reduces the peak-to-average power ratio, allows dense pilot placement suitable for MIMO and massive MIMO systems, and remains fully compatible with conventional OFDM infrastructure. This makes it a compelling candidate for future wireless systems. The main advantages of the OTFS modulation can be summarized as follows \cite{OTFSSURVEY1, STBC1, STBCOTFS1, STBCOTFS2, OTFSSURVEY2, OTFSSURVEY3, OTFSSURVEY4, OTFSSURVEY5, OTFSSURVEY6, OTFSSURVEY7, OTFSSURVEY8, OTFSSURVEY9, OTFSSURVEY10, OTFSSURVEY11, OTFSSURVEY12, OTFSSURVEY13, OTFSSURVEY14, OTFSSURVEY15, OTFSSURVEY16, OTFSSURVEY17, OTFSSURVEY18}: the OTFS technique provides robustness to high Doppler and delay spreads by transforming the time-varying channel into a quasi-static channel in the DD domain, ensuring reliable communication in high‑mobility scenarios. Unlike the OFDM modulation, which requires guard intervals between every symbol, the OTFS modulation only needs guard intervals between entire frames, resulting in significantly lower idle time. The OTFS scheme exploits the sparse structure of DD channels to achieve accurate channel estimation with significantly fewer pilot symbols than OFDM, especially in rapidly time-varying channels. The OTFS technique modulates symbols over the entire time-frequency plane by multiplexing in the DD domain. This significantly enhances communication reliability. The OTFS modulation offers a lower peak-to-average power ratio (PAPR) compared to OFDM, owing to its reduced cyclic prefix overhead and two-dimensional modulation structure. This leads to improved power amplifier efficiency and makes OTFS more robust against synchronization errors.

In this article, a comprehensive survey and analysis of OTFS and OTFS-IM systems is presented, covering their applications, fundamental principles, performance evaluations, challenges, benefits, and future directions. The main contributions of the article are presented as follows:
\begin{enumerate}
\item A comprehensive review of OTFS applications is presented, addressing various wireless communication systems, including satellite communications, multiple access techniques, RISs, underwater acoustic communications, massive MIMO, ISAC, and deep learning-assisted systems.

\item The fundamental principles and system model of conventional OTFS modulation are detailed, including transmitter and receiver operations. Additionally, the error performance of OTFS systems in single-input single-output (SISO) and multiple-input single-output (MISO) architectures is presented for various system parameters.

\item An extensive literature review covering all OTFS-IM schemes available in the current literature is presented. Furthermore, these schemes are categorized based on their system architectures (SISO,  single-input multiple-output (SIMO), MISO, and MIMO), the detectors employed, and performance analysis aspects including capacity, PAPR, diversity, complexity, imperfect channel state information (CSI), and outage probability. In addition, spectral efficiency expressions for all OTFS-IM schemes in the literature are presented collectively.

\item  A detailed literature review of OTFS-IM schemes, including OTFS-based space shift keying (OTFS-SSK), OTFS-based spatial modulation (OTFS-SM), OTFS-based quadrature spatial modulation (OTFS-QSM), OTFS-based media-based modulation (OTFS-MBM), and OTFS-based code index modulation (OTFS-CIM), is presented, along with the operating principles and system models of each variant. A comparative analysis and benchmarking of the systems is conducted in terms of computational complexity, error performance, capacity, energy saving, spectral efficiency, and throughput, with results compared across all considered systems.

\item A comprehensive investigation of the challenges, benefits, and future directions for OTFS-IM systems is presented, covering aspects such as computational complexity, spectral efficiency, latency, channel estimation, hardware impairments, PAPR, synchronization, physical layer security, and the integration of OTFS with advanced IM schemes, MISO architectures, RIS, MBM, CIM, and deep learning-based detection methods.

\end{enumerate}

\subsection{Organization}
The remainder of this paper is organized as follows. In Section II, OTFS-based wireless communication systems are reviewed, including applications in satellite communications, multiple access communications, RISs, underwater acoustic communications, massive MIMO communications, ISAC, and deep learning. Section III presents the fundamental principles, transceiver model, and error performance results of the conventional OTFS system. Section IV reviews OTFS-IM systems, with a particular focus on OTFS-SM, OTFS-MBM, OTFS-QSM, OTFS-CIM, and OTFS-SSK systems. Section V provides performance analyses of the conventional OTFS and OTFS-IM systems, including computational complexity, error performance, capacity, energy saving, spectral efficiency, and throughput analyses. Section VI presents the challenges and benefits of OTFS-IM systems and outlines open problems and potential future work. Finally, conclusions are presented in Section VII. Also, the list of abbreviations used in the paper is presented in Table \ref{list_abb}.

\subsection{Notation}
The notation used in the paper is as follows. Vectors and matrices are represented by bold lowercase and uppercase letters, respectively. For a complex number $x$, its real and imaginary parts are denoted by $x_{\Re}$ and $x_{\Im}$, respectively. Also, $(\cdot)^T$, $(\cdot)^H$, $|\cdot|$, $||\cdot||$, $\otimes$, $\det(\cdot)$, $\operatorname{tr}(\cdot)$, $\mathbb{E}[\cdot]$, $\log_{10}(\cdot)$, and $\log_{2}(\cdot)$ denote the transpose, Hermitian transpose, absolute value, Euclidean norm, Kronecker product, determinant, trace, expectation, base-10 logarithm, and base-2 logarithm operators, respectively. Further, $\binom{.}{.}$ denotes the binomial coefficient, $\lfloor\cdot\rfloor$ the floor function, and $(\cdot)!$ the factorial.

\definecolor{headerbg}{HTML}{B2EBF2}

\begin{table*}[!t]
\centering
\scriptsize 
\renewcommand{\arraystretch}{1.06}
\setlength{\tabcolsep}{8pt}
\caption{List of Abbreviations}
\label{list_abb}
\rowcolors{3}{stripeA}{stripeB}
\begin{tabularx}{\textwidth}{
  | >{\bfseries}p{2.45cm}  
  >{\raggedright\arraybackslash}p{5.0cm} 
  | >{\bfseries}p{2.45cm}  
  >{\raggedright\arraybackslash}X         
  |
}
\toprule
\rowcolor{headerbg}
\textbf{Abbreviation} & \textbf{Description} & \textbf{Abbreviation} & \textbf{Description}\\
\midrule

4G  & fourth-generation   &  N-AC & norm and antenna correlation \\           
5G  & fifth-generation    &   NOMA & non-orthogonal multiple access     \\   
6G  & sixth-generation             &    ODDM & orthogonal DD division multiplexing    \\         
ABEP & average bit error probability &    ODDM-IM & ODDM with IM      \\  
ABER & average BER                        & OFDM & orthogonal frequency division multiplexing \\     
ADMM & alternating direction method of multipliers  &   OMPFR & orthogonal matching pursuit with fractional refinement              \\  
AEE-JDDIM-OTFS & autoencoder-based joint DD IM & OTSM & orthogonal time sequency modulation  \\  
ASER & average symbol error rate       &     OTFS & orthogonal time frequency space       \\  
AWGN & additive white Gaussian noise      & OTFS-CIM   &  OTFS-based CIM \\
BiLSTM & bidirectional long short-term memory & OTFS-CM-IM & OTFS-based channel modulation IM \\   
BPSK & binary PSK          & OTFS-DFIM & OTFS-based dual frequency IM scheme \\                 
CIM & code index modulation & OTFS-DM-IM & OTFS-aided dual-mode IM  \\  
CIM-QSM & CIM-based QSM        & OTFS-GDM-IM & generalised dual-mode IM   \\                     
CIM-SM & CIM-based SM         &OTFS-$I$/$Q$-IM & OTFS with $I$ and $Q$ IM \\  
CIM-SMBM & CIM-based SMBM            & OTFS-IIM & OTFS modulation scheme with improved IM \\       
CIR & channel impulse response &  OTFS-IM & OTFS-based IM   \\ 
CMP & customized message passing           & OTFS-IM4 & four-dimensional spherical code-based OTFS-IM  \\    
CRB & Cramér–Rao bound &    OTFS-ISAC & OTFS-based ISAC         \\  
CW-OTFS & complex wavelet-based OTFS     & OTFS-MBM & OTFS-based MBM  \\    
DAQSM & diversity-achieving QSM & OTFS-NOMA & OTFS-based NOMA   \\ 
dB & decibel                            & OTFS-QSM & OTFS-based QSM  \\    
DeIM-OTFS & delay-IM with OTFS   &  OTFS-SBIM & OTFS with sub-band IM  \\ 
DD & delay–Doppler                      & OTFS-SM & OTFS-based SM  \\    
DoIM-OTFS & Doppler-IM with OTFS &   OTFS-SSK & OTFS-based SSK      \\  
DS-SS & direct-sequence spread spectrum     &   PAM & pulse amplitude modulation      \\  
E-OTFS-IM & enhanced OTFS with IM &    PSK & phase shift keying       \\  
ELM & extreme learning machines        &  QAM & quadrature amplitude modulation  \\      
EMMSE & enhanced minimum mean square error &    QSM & quadrature SM      \\  
FDD & frequency division duplex           &  RA & reconfigurable antenna   \\   
FSO & free-space optical         &  RDN & residual dense network      \\  
GIM-MIMO-OTFS & generalized IM scheme for MIMO-OTFS &   ResNet & residual network    \\  
GLRT & generalized likelihood ratio test &    RF & radio frequency       \\  
GSDDIM-OTFS & generalized space-DD index modulated OTFS &  RIS & reconfigurable intelligent surface      \\ 
GSIM-OTFS & generalized SIM-OTFS &  RIS-FSO & RIS-assisted FSO    \\  
GSM & generalized SM                       &   RIS-OFDM & RIS-aided OFDM     \\      
GSM-OTFS & GSM-based OTFS &   RIS-OTFS & RIS-based OTFS     \\  
HMIM & hierarchical mode-based IM        &  SFFT & symplectic finite Fourier transform       \\      
ICI & inter-carrier interference &  SIM-OTFS & spatial-IM-based OTFS      \\  
IM & index modulation                        & SIMO & single-input multiple-output  \\  
IM-FSC & IM-aided fragmented spectra centralization &   SISO & single-input single-output       \\  
IoT & Internet of Things                     & SJRs & signal-to-jamming ratios  \\
$IQ$ & in-phase and quadrature &SM-OFDM-IM & SM and IM based OFDM  \\
ISAC & integrated sensing and communication  & SM-STBC & SM-based space time block code   \\
ISFFT & inverse symplectic finite Fourier transform &   SM & spatial modulation        \\  
ISI & inter-symbol interference              & SMBM & spatial MBM  \\ 
JDDIM-OTFS & joint DD IM OTFS &  SS-EP & symbol-by-symbol aided expectation propagation       \\  
LCTAS-TSS & low-complexity TAS based on a tree search scheme & SSK & space shift keying  \\  
LEO & low earth orbit &   STC-OTFS & space-time coded OTFS      \\  
LLR & log-likelihood ratio          &   Str-BP-EP & structured prior-based hybrid belief propagation and expectation propagation   \\           
LMMSE & linear minimum mean square error &  TAS & transmit antenna selection    \\  
MAPs & mirror activation patterns            &  TM-OTFS-IM & tri-mode OTFS-IM          \\  
MBM & media-based modulation & UAS & unmanned aircraft systems          \\  
MCOTFS-IM & modified-constellation-based OTFS with IM &  UAV & unmanned aerial vehicle        \\  
MIMO & multiple-input multiple-output  &   URLLC & ultra-reliable low-latency communication     \\  
MISO & multiple-input single-output          &   V2V & vehicle-to-vehicle      \\   
ML & maximum likelihood &  VBLAST & vertical-Bell Laboratories layered space-time   \\  
MLJSAPD & multi-layer joint symbol and activation pattern detection & VBLAST-OTFS-IM & VBLAST aided OTFS-IM \\  
MLMPD & multi-layer message passing detection &  VV-MP & vector-by-vector-aided message passing     \\
MM-OTFS-IM & multiple-mode OTFS-IM           &   W-3D-OTFS-DM-IM & wavelet-based three-dimensional OTFS with dual-mode IM    \\  
MMIM-OTFS-SM & multi-mode IM aided OTFS-based SM & W-OTFS & wavelet-based OTFS \\ 
MMSE & minimum mean square error        & WH & Walsh–Hadamard  \\ 
MMSE-SIC & minimum-mean-square error with successive interference cancellation & ZF & zero-forcing     \\ 
 MRC & maximum ratio combining            & ZP & zero padding   \\ 

\bottomrule
\end{tabularx}
\end{table*}




\begin{figure}[t]
    \centering
    \includegraphics[width=0.75\linewidth]{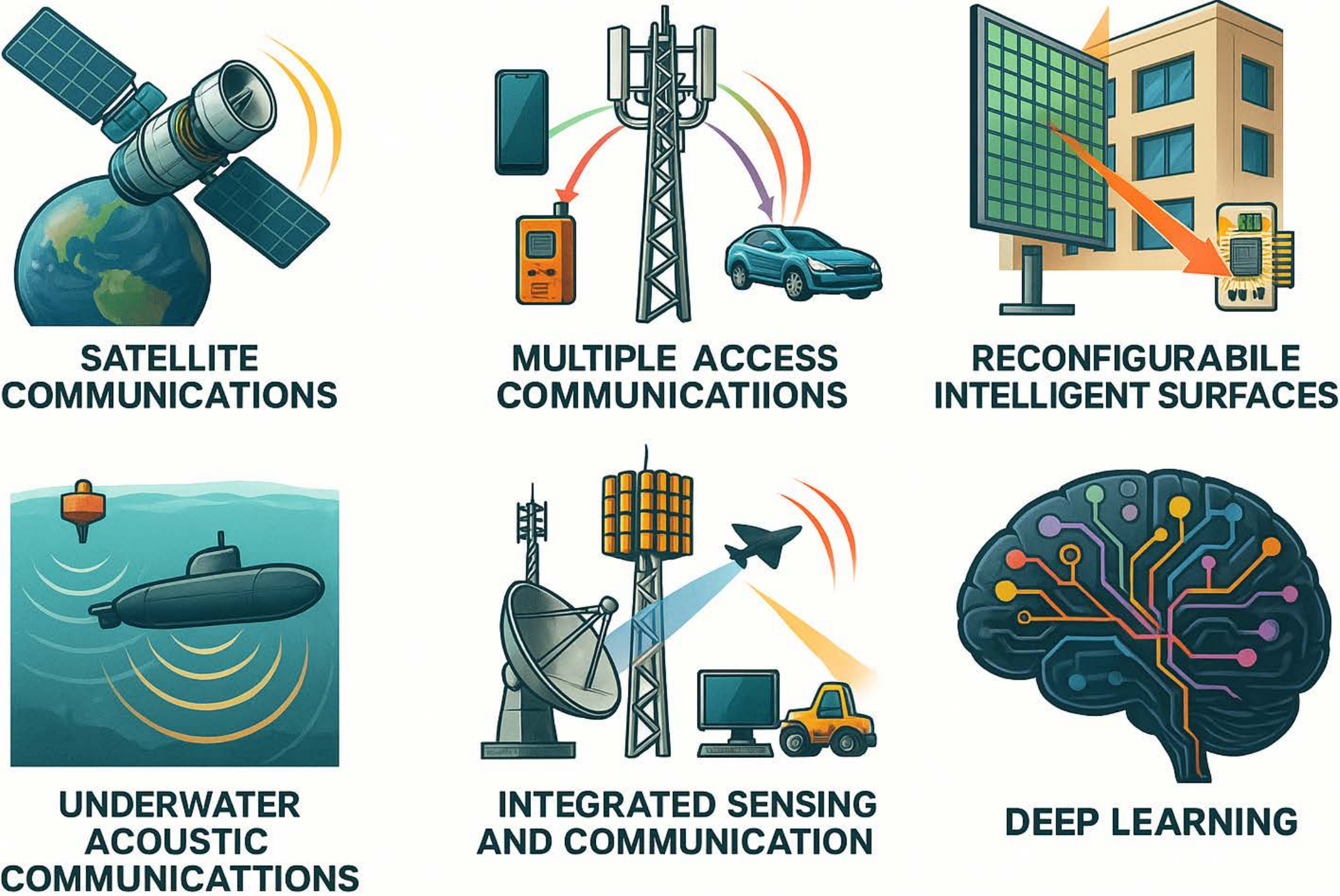}
    \caption{Applications of OTFS modulation in wireless communications.}
    \label{OTFS_systems}
\end{figure}

\section{OTFS-Based Wireless Communication Systems}
This section provides an overview of OTFS-based wireless communication systems, encompassing key areas such as satellite communications, multiple access communications, RISs, underwater acoustic communications, massive MIMO communications, ISAC, and deep learning, as illustrated in Fig. \ref{OTFS_systems}.

\subsection{Satellite Communications}
OTFS-based communication systems have attracted considerable attention in satellite communications and other high-mobility scenarios due to their inherent robustness against severe Doppler effects and rapidly time-varying channels. \cite{OTFSSATE1} applies OTFS modulation for position, navigation, and timing services, showing improved code tracking accuracy and flexibility in pseudorandom noise sequence deployment compared to traditional global navigation satellite system modulation schemes. For free-space optical (FSO)-based LEO satellite communication, \cite{OTFSSATE2} proposes an optical OTFS system that enhances power efficiency and mitigates pointing jitter. \cite{OTFSSATE3} investigates coordinated multi-satellite transmission for OTFS-based low earth orbit (LEO) satellite systems, proposing aggregate channel estimation and fractional Doppler estimation schemes to mitigate inter-satellite interference and Doppler effects. \cite{OTFSSATE4} proposes a complex wavelet-based OTFS (CW-OTFS) scheme that combines discrete wavelet transforms to reduce system complexity and achieve better bit error rate (BER), PAPR, and Cramér-Rao bound (CRB) performance than the traditional wavelet-based OTFS (W-OTFS) system in LEO satellite communications. \cite{OTFSSATE5} analyzes Doppler-induced interferences in OTFS-based LEO systems by deriving closed-form expressions of fractional and squint Doppler interferences, showing that increasing the number of DD bins effectively mitigates these interferences. \cite{OTFSSATE6} develops an iterative channel estimation method for OTFS using comb-type Zadoff-Chu sequences, achieving low PAPR and better estimation accuracy. \cite{OTFSSATE7} proposes a pulsed OTFS signal that cuts acquisition complexity by over $89.4\%$ and improves delay measurement accuracy by about $8$ decibel (dB) compared to conventional  binary PSK (BPSK). \cite{OTFSSATE8} leverages OTFS modulation in a realistic satellite-vehicular network channel model, demonstrating robustness to high Doppler and improved performance compared to the traditional OFDM scheme. \cite{OTFSSATE9} applies OTFS for LEO-based positioning, demonstrating that its DD processing improves estimation accuracy in multipath environments compared to the BPSK scheme. Multi-beam LEO scenarios are analyzed in \cite{OTFSSATE10}, which derives a closed-form outage probability for OTFS-based multi-beam LEO satellites, while \cite{OTFSSATE11} studies unmanned aircraft systems (UAS)-assisted multi-hop OTFS systems and compares zero-forcing (ZF) and linear minimum mean square error (LMMSE) equalizers to evaluate system performance. \cite{OTFSSATE12} proposes a semantic enhanced downlink communication with the OTFS scheme that effectively mitigates Doppler effects, achieving up to $6$ dB SNR gain and $50\%$ overhead reduction for image transmission. \cite{OTFSSATE13} compares OTFS and OFDM systems under multiuser LEO links, demonstrating that the OTFS scheme offers greater robustness to carrier frequency offsets and achieves higher sum-rates with the optimized resource allocation. Channel estimation for multi-user random access in OTFS-based LEO satellite systems is improved in \cite{OTFSSATE15} by designing an OTFS pilot structure and a sparsity-based channel estimation algorithm, leading to better multi-user separation and estimation reliability. \cite{OTFSSATE16} analyzes the secrecy performance of downlink OTFS with cooperative unmanned aerial vehicle (UAV) jamming and ZF and minimum mean square error (MMSE) beamforming, providing improved secure transmission. Beyond these, numerous other works have expanded OTFS applications for satellite communications, exploring research areas such as semantic communication, multiuser resource allocation, carrier frequency offset estimation, random access, joint communication-navigation, and MIMO configurations \cite{OTFSSATE12,OTFSSATE13,OTFSSATE14,OTFSSATE15,OTFSSATE16,OTFSSATE17,OTFSSATE18,OTFSSATE19,OTFSSATE20,OTFSSATE21,OTFSSATE22,OTFSSATE23,OTFSSATE24,OTFSSATE25,OTFSSATE26,OTFSSATE27,OTFSSATE28,OTFSSATE29}.

\subsection{Multiple Access Communications}

OTFS modulation has emerged as a strong candidate for high-mobility communication, and recent studies focus on combining OTFS modulation with different non-orthogonal multiple access (NOMA) techniques to meet the needs of high-mobility users and support many connected devices in future wireless networks. \cite{OTFSNOMA1} combines pattern-domain NOMA with the OTFS scheme to address the demands of 6G networks and proposes a low-complexity power vector expectation propagation detection algorithm for spatially correlated channels, whereas \cite{OTFSNOMA2} introduces a non-orthogonal time frequency space IM multiple access system, an enhanced uplink random access scheme that integrates NOMA with OTFS and IM, enabling grant-free access and demonstrating improved error performance and spectral efficiency. \cite{OTFSNOMA3} investigates a coexistence framework for OTFS, orthogonal time sequency modulation (OTSM), and single carrier waveforms in multi-user uplink scenarios, maintaining error performance while offering increased flexibility. \cite{OTFSNOMA5} investigates the impact of fractional Doppler on OTFS-based NOMA (OTFS-NOMA) systems with high and low mobility users, showing that inter-Doppler interference affects downlink spectral efficiency and outage probability depending on NOMA parameters. \cite{OTFSNOMA6} proposes joint subcarrier and power allocation strategies for the OTFS-NOMA system to maximize sum-rate, significantly outperforming OFDM access schemes. \cite{OTFSNOMA4} considers modulation classification accuracy for OTFS-NOMA systems in the presence of high-power amplifier nonlinearity and impulsive noise using machine learning, highlighting spectrum utilization by combining signals from different mobility profiles, while \cite{OTFSNOMA7} generates a dataset for downlink OTFS-NOMA with heterogeneous user mobility profiles, extracts features, and explores multiple machine learning algorithms for automatic modulation classification. \cite{OTFSNOMA8} proposes spatial division-assisted precoding and user grouping based on angle-domain channel sparsity to improve spectral efficiency in MIMO-OTFS-based NOMA systems. \cite{OTFSNOMA9} combines DD domain mapping, clustering of users with different mobility profiles, and NOMA spectrum sharing to achieve flexible user grouping and enhanced spectral efficiency. \cite{OTFSNOMA10} and \cite{OTFSNOMA21} propose orthogonal time frequency code space modulation-enabled multiple access and OTFS-based tandem spreading multiple access schemes, respectively, to enhance device connectivity and transmission reliability in high-mobility and high-speed railway scenarios. \cite{OTFSNOMA11} proposes a bidirectional long short-term memory (BiLSTM)-based deep learning detection method for OTFS-NOMA systems, achieving better symbol error rate performance than the conventional message passing algorithm. \cite{OTFSNOMA12} proposes cyclic-prefix orthogonal time frequency space-based multiple access schemes using localized and interleaved time-frequency orthogonal multiple access strategies, achieving interference-free multi-user transmission with improved error performance over linear time-varying channels. \cite{OTFSNOMA13} integrates space-time shift keying with OTFS modulation for orthogonal multiple access, significantly enhancing diversity, coding gains, and reducing complexity. \cite{OTFSNOMA14} applies $K$-means clustering for the OTFS-NOMA system to improve sum-rate and interference cancellation performance. \cite{OTFSNOMA15} and \cite{OTFSNOMA17} propose a low-complexity equalization and detection scheme for OTFS-NOMA systems, which combines a least-squares with QR factorization-based equalizer with a new reliability zone detection method, achieving better symbol error rate performance than the conventional minimum-mean-square error with successive interference cancellation (MMSE-SIC) scheme. \cite{OTFSNOMA16} and \cite{OTFSNOMA18} propose an OTFS-based sparse code multiple access (OTFS-SCMA) downlink scheme with a cross-domain receiver that jointly performs OTFS symbol detection and SCMA decoding, achieving higher spectral efficiency and improved error performance in high-mobility channels. \cite{OTFSNOMA19} and \cite{OTFSNOMA27} investigate OTFS-based orthogonal multiple access schemes that allocate non-overlapping DD and/or time-frequency resources to avoid multi-user interference and achieve higher spectral efficiency than guard-band-based methods. \cite{OTFSNOMA20} proposes a two-dimensional OTFS-NOMA scheme that integrates power-domain and code-domain and demonstrates improved spectral efficiency and error performance over the existing power-domain NOMA scheme. In the literature, numerous OTFS-based multiple access schemes have been proposed, such as SCMA-based frameworks, robust beamforming designs, interleaved time-frequency resource allocation methods, and rate-splitting multiple access integrations \cite{OTFSNOMA22, OTFSNOMA23, OTFSNOMA24, OTFSNOMA25, OTFSNOMA26, OTFSNOMA28}.

\subsection{Reconfigurable Intelligent Surfaces}

RISs offer significant potential advantages, including good error performance, energy efficiency, and high throughput, to wireless communication systems in which they are integrated. Signals transmitted through wireless communication channels degrade in quality due to fading effects such as reflection, scattering, and diffraction. RIS technology can help mitigate these disruptive effects by controlling the propagation characteristics of the signals, such as reflection and scattering, leading to improved error performance \cite{RIS1,RIS2,RIS3,RIS4}. The reflective surfaces of RISs are essentially metasurfaces - thin, two-dimensional layers of metamaterials that can manipulate electromagnetic waves. The main advantages of metamaterials include low cost, minimal absorption, and ease of integration \cite{meta1,meta2}. RISs stand out due to its many significant advantages, such as not requiring signal processing, power amplification, or filtering, its low cost, support for full-duplex and full-band transmission, and its flexibility and broad range of applications \cite{risav1,risav2, RISSMBM, risbur1, risbur2}. The integration of RISs with OTFS modulation enables the manipulation of the wireless channel, offering significant performance improvements, particularly under challenging propagation conditions and high-mobility scenarios. Many recent works show that RIS technology can be used to support OTFS systems, providing various benefits such as improved coverage, error performance, reliability, and enhanced spectral and energy efficiency. In \cite{RISOTFS1}, an RIS-assisted OTFS system is shown to outperform OFDM counterparts for space-air-ground integrated networks under high-mobility conditions. A new RIS-aided OTFS-IM scheme is proposed in \cite{OTFS_RIS_IM_1}, where DD domain IM and Frobenius norm-based RIS phase selection are employed to enhance spectral efficiency and error performance, outperforming conventional RIS-aided OFDM (RIS-OFDM)-IM systems in high-mobility scenarios. Sensing and passive beamforming in uplink RIS-based OTFS (RIS-OTFS) scenarios are studied in \cite{RISOTFS2} and \cite{Li_2023_2}, both proposing DD-based schemes that reduce channel training overhead. Machine learning techniques, such as extreme learning machines (ELMs), have been used for channel estimation in \cite{RISOTFS3}, and residual attention networks are used in \cite{RISOTFS20} to improve channel estimation accuracy in dynamic scenarios. While multicast phase shift design for RIS-OTFS systems is investigated in \cite{RISOTFS4}, \cite{RISOTFS11} focuses on phase optimization methods for doubly dispersive channels using approaches such as maximum variance and genetic algorithms. Furthermore, \cite{RISOTFS5} presents an alternative approach by introducing non-diagonal phase shifts to improve error performance. Low-complexity detection for IRS-aided MIMO-OTFS systems is proposed in \cite{RISOTFS6}, which employs maximal ratio combining (MRC) detection to mitigate multipath fading and inter-antenna interference. Additionally, alternating direction method of multipliers (ADMM)-based detection schemes are presented in \cite{RISOTFS7} and \cite{RISOTFS13} as part of a beamforming and precoding design framework, showing improvements in system capacity and performance. \cite{RISOTFS14} and \cite{RISOTFS22} present joint designs of the OTFS system and IRS phase shifts to increase received signal power, and introduce a low-complexity iterative interference cancellation detector to improve symbol detection performance. In \cite{RISOTFS17}, a two-fold physical layer security scheme is introduced by designing a channel-based precoder combined with artificial noise to protect legitimate users from eavesdropping. Also, \cite{RISOTFS23} investigates the fundamental localization limits of RIS-OTFS systems under near-field and far-field conditions, and analyzes how RIS orientation offsets can impact localization accuracy. RIS-assisted sensing employing OTFS modulation is examined in \cite{RISOTFS26}, which demonstrates that RISs can enhance the resolution of OTFS-based radar systems and reduce transmission time by utilizing sparse recovery techniques. In addition, \cite{RISOTFS16} presents a W-OTFS approach combined with RIS technology to achieve improvements in performance, peak-to-average power ratio, and system complexity under high-mobility conditions. Furthermore, \cite{RISOTFS19} proposes a vector differential coding scheme for hybrid RIS-aided OTFS systems, aiming to eliminate the need for channel estimation and simplify symbol detection through a low-complexity maximum likelihood (ML) detector with interference cancellation. Hybrid configurations that integrate RIS with OTFS modulation have been investigated in the literature for both optical and millimeter-wave communication scenarios. For example, \cite{RISOTFS8} introduces RIS-assisted FSO (RIS-FSO) systems that integrate optical OTFS with orbital angular momentum to mitigate atmospheric turbulence and achieve improved energy and spectral efficiency, while \cite{RISOTFS10} proposes a hybrid adaptive biased optical OTFS for RIS-assisted optical wireless links to reduce PAPR and enhance overall efficiency. Also, \cite{RISOTFS28} investigates a hybrid RIS-assisted millimeter-wave OTFS system and presents a joint channel estimation and data detection scheme that combines beamforming design with message passing and expectation-maximization algorithms. Alternative optimization methods such as majorization-minimization, strongest tap maximization, fractional programming, and genetic algorithms are presented for phase shift and beamforming design in RIS-OTFS systems in \cite{RISOTFS4}, \cite{RISOTFS7}, and \cite{RISOTFS11}. Furthermore, \cite{RISOTFS12} and \cite{RISOTFS24} propose phase-shift optimization and passive beamforming designs for RIS-OTFS systems to maximize received SNR, and their results show that the proposed RIS-OTFS systems provide better error performance than RIS-assisted OFDM. \cite{RISOTFS25} and \cite{RISOTFS27} both derive end-to-end input-output relations in high-mobility channels for RIS-OTFS systems, and their results show that RIS-OTFS provides better performance than RIS-OFDM in high-Doppler scenarios. Additionally, \cite{RISOTFS15} proposes a time-domain and DD input-output model for the RIS-OTFS scheme that considers receiver in-phase and quadrature ($IQ$) imbalance and demonstrates that higher gain mismatch degrades error performance. \cite{RISOTFS18} proposes a new energy-efficient phase shift algorithm for RIS-OTFS systems that maximizes received energy and channel gain, demonstrating notable BER improvements over benchmark systems. \cite{RISOTFS9} presents a heuristic phase reconstruction and power optimization algorithm for RIS-OTFS-NOMA systems, which maximizes transmission rates while meeting rate and phase constraints. Numerous other studies in the literature also explore the integration of RISs and OTFS, demonstrating its potential to improve error performance in various system designs, metrics, and scenarios \cite{RISOTFS29, RISOTFS30, RISOTFS31, RISOTFS32, RISOTFS33, RISOTFS34, RISOTFS35, RISOTFS36, RISOTFS37, RISOTFS38}.

\subsection{Underwater Acoustic Communications}
OTFS modulation has also been investigated in various domains beyond terrestrial wireless communications. For example, numerous communication systems have been proposed in the literature that utilize OTFS modulation in underwater acoustic communication systems to address challenges such as Doppler effects, multipath propagation, and channel estimation. Practical system designs and channel estimation methods have been explored in \cite{OTFSUnderWater1, OTFSUnderWater11}, while passive time reversal and adaptive channel estimation techniques are proposed in \cite{OTFSUnderWater2, OTFSUnderWater8, OTFSUnderWater9}. Studies focusing on low-complexity detection and equalization methods include \cite{OTFSUnderWater5, OTFSUnderWater6, OTFSUnderWater7, OTFSUnderWater10, OTFSUnderWater20}. Deep learning-based approaches for channel estimation and signal detection are presented in \cite{OTFSUnderWater13, OTFSUnderWater14, OTFSUnderWater15, OTFSUnderWater16, OTFSUnderWater17}. Furthermore, works addressing mobile underwater communications, pilot optimization, and modulation methods include \cite{OTFSUnderWater3, OTFSUnderWater4, OTFSUnderWater19}, while OFDM-based massive MIMO-OTFS systems are investigated in \cite{OTFSUnderWater18}. Finally, advanced equalization and detection methods, such as turbo equalization and unitary approximate message passing, are studied in \cite{OTFSUnderWater12, OTFSUnderWater21}. Also, in underwater acoustic communication, enhanced modulation techniques such as OTFS combined with IM are being actively studied to improve system error resilience and spectral efficiency \cite{OTFS_IM44, OTFS_IM45}.

\begin{figure*}[t]
    \centering
    \includegraphics[width=1\linewidth]{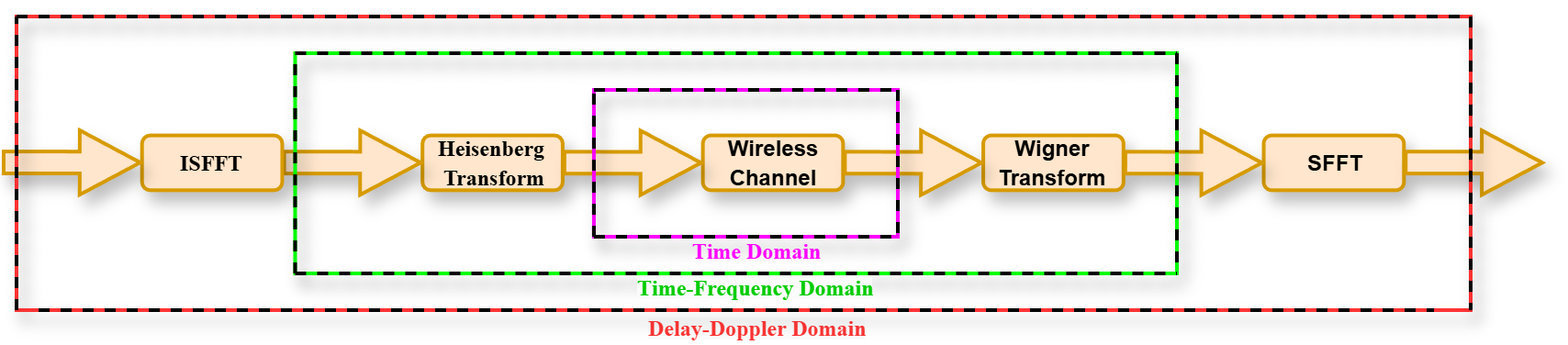}
    \caption{System model of the traditional OTFS scheme.}
    \label{figOTFS}
\end{figure*}

\subsection{Massive MIMO Communications}

The integration of OTFS modulation into massive MIMO systems has emerged as a promising approach in wireless communications, offering the potential for enhanced reliability, spectral efficiency, and robust performance in high-mobility channel conditions. In massive MIMO-OTFS systems, channel estimation and tracking are addressed in \cite{OTFSmass3, OTFSmass8, OTFSmass6}, where \cite{OTFSmass3} proposes a Markov-based model and customized message passing (CMP) algorithm for dynamic channel tracking in the DD-angle domain, \cite{OTFSmass8} introduces a joint sparsity pattern learning approach using spike and slab prior model and orthogonal matching pursuit technique for more accurate channel estimation with reduced pilot overhead, and \cite{OTFSmass6} proposes a new deep learning framework, consisting of three different convolutional neural networks, to improve channel estimation performance and reduce computational complexity in high-mobility scenarios. \cite{OTFSmass9} introduces a novel path division multiple access scheme and a 3D-Newtonized orthogonal matching pursuit algorithm to extract channel parameters and enable interference-free communication. \cite{OTFSmass10} develops new channel models and efficient algorithms for simultaneous localization and channel estimation in massive MIMO-OTFS systems. For a downlink multi-user massive MIMO-OTFS system, \cite{OTFSmass11} proposes a low-complexity precoding and detection scheme that enables practical symbol demodulation at user terminals and improves robustness to Doppler effects. \cite{OTFSmass2} and \cite{OTFSmass7} investigate power control and resource allocation in cell-free massive MIMO-OTFS systems, proposing optimization algorithms that significantly prevent spectral efficiency loss and enhance user SNR fairness in conventional communication and ISAC scenarios. Hybrid OTFS/OFDM modulation and weighted max-min power control for users with different mobility profiles are addressed in \cite{OTFSmass5}, while \cite{OTFSmass4} presents closed-form spectral efficiency analysis for OTFS-aided full-duplex cell-free massive MIMO in high-mobility scenarios. Enhanced user connectivity through angle-DD domain NOMA is proposed in \cite{OTFSmass12}, and comparative evaluation of OTFS and OFDM in cell-free massive MIMO, including new transmit power scaling laws, is addressed in \cite{OTFSmass13}. Finally, \cite{OTFSmass1} addresses joint device identification, channel estimation, and symbol detection for massive MIMO-OTFS-based random access in cooperative LEO satellite constellations, introducing efficient centralized and distributed algorithms that leverage spatial diversity for improved performance. Additionally, numerous recent publications present further applications, derivations, performance analyses, and novel system designs related to OTFS-based massive MIMO systems \cite{OTFSmass14, OTFSmass15, OTFSmass16, OTFSmass17, OTFSmass18, OTFSmass19, OTFSmass20, OTFSmass21}.

\subsection{Integrated Sensing And Communication}
Recent studies of OTFS-based ISAC systems explore various strategies and systems to enhance performance and efficiency in high-mobility environments. \cite{OTFSISAC1} proposes an ISAC-assisted user state refinement method in OTFS systems using a hybrid digital-analog architecture, and a nested array-based technique to perform angle refinement, while \cite{OTFSISAC2} introduces a MIMO-OTFS ISAC system that incorporates signal models accounting for inter-symbol interference (ISI) and ICI effects, generalized likelihood ratio test (GLRT)-based multi-target detection, DD-angle estimation, and adaptive DD multiplexing. \cite{OTFSISAC3} presents a low-complexity minimum BER precoder scheme for OTFS-based ISAC (OTFS-ISAC) systems, and \cite{OTFSISAC4} proposes a NOMA-assisted OTFS-ISAC system with UAVs using 3-D motion prediction topology and robust power allocation. \cite{OTFSISAC5}, \cite{OTFSISAC6}, and \cite{OTFSISAC9} propose deep learning-based, passive sensing, and orthogonal matching pursuit with fractional refinement (OMPFR) algorithms for target parameter estimation and channel estimation in OTFS-ISAC systems. \cite{OTFSISAC7} and \cite{OTFSISAC8} develop OTFS-ISAC schemes for environmental sensing, localization, and achieving highly accurate range-velocity profiles. \cite{OTFSISAC10} overviews DD-ISAC, compares with OFDM-based ISAC, and discusses challenges and opportunities. \cite{OTFSISAC11} proposes a directional transmission scheme for airborne OTFS-ISAC networks with beam prediction and pilot placement schemes, and \cite{OTFSISAC12} addresses the resource allocation and trade-off optimization problem for multi-user MIMO OTFS-ISAC systems. There are also numerous other studies on low-complexity detection, amplitude barycenter calibration, performance analysis, moving vehicle detection, reliability analysis, waveform design, unified variational inference framework, and PAPR reduction algorithms in OTFS-based ISAC systems \cite{OTFSISAC13,OTFSISAC14,OTFSISAC15,OTFSISAC16,OTFSISAC17,OTFSISAC18,OTFSISAC19,OTFSISAC20, OTFSISAC21, OTFSISAC22, OTFSISAC23, OTFSISAC24, OTFSISAC25}.

\subsection{Deep Learning}
In \cite{OTFSDL1} and \cite{OTFSDL11}, a sparse prior-guided and self-adaptive threshold-based deep learning framework is proposed for channel estimation in OTFS systems. A model-driven deep learning-based OAMP detector is proposed to improve detection performance for OTFS systems in \cite{OTFSDL2} and  \cite{OTFSDL3} proposes a deep learning-based scheme with autoencoder codebook mapping and MLP decoder to enhance robustness and reduce latency in downlink OTFS-SCMA systems. Deep learning is applied for radar target detection in \cite{OTFSDL4}, OTFS detection with residual network (ResNet), residual network (ResNet), and residual dense network (RDN) in \cite{OTFSDL5}, channel estimation for OTFS-Assisted ISAC systems in vehicular networks in \cite{OTFSDL6}, automatic modulation recognition for OTFS and OFDM schemes in \cite{OTFSDL7}, target classification with OTFS signaling in \cite{OTFSDL8}, underwater acoustic OTFS channel estimation in \cite{OTFSDL9}, and predictive precoder design for OTFS-based ultra-reliable low-latency communication (URLLC) in \cite{OTFSDL10}. Also, the literature includes numerous studies that integrate deep learning techniques with various OTFS systems to enhance detection, estimation, classification, and system performance \cite{OTFSDL12, OTFSDL13, OTFSDL14, OTFSDL15, OTFSDL16, OTFSDL17, OTFSDL18, OTFSDL19, OTFSDL20}.

\section{The Conventional OTFS System}
As noted previously, OTFS is a recently developed two-dimensional modulation scheme that represents information in the DD domain and offers robust performance in high-mobility wireless channels. The OTFS technique was first proposed as a preprint in \cite{OTFSarxiv} in 2016 and was formally published in 2017 in \cite{OTFSilk}. At the transmitter of the traditional OTFS system, information symbols (typically QAM/QPSK symbols) are mapped onto a two-dimensional grid indexed by DD, and then transformed to time–frequency domain signals via an inverse symplectic finite Fourier transform (ISFFT). Subsequently, the Heisenberg transform is applied to convert these time-frequency samples into a continuous time-domain waveform suitable for transmission. At the receiver, the inverse operations are performed where the Wigner transform brings the received signal to the time-frequency domain, and the symplectic finite Fourier transform (SFFT) converts it back to the DD domain, yielding the recovered information symbols.

Fig. \ref{figOTFS} illustrates the system model of the traditional OTFS scheme, including the signal transformations performed at both the transmitter and receiver. In the OTFS system, each information symbol is associated with a delay $\tau$ and Doppler $\nu$, and is spread over the entire time-frequency domain. An OTFS frame consists of $N$ time slots, each lasting $T$ (symbol duration) seconds, and $M$ subcarriers spaced by $\Delta f$ Hz (subcarrier frequency spacing), resulting in a total frame duration of $N \times T$ seconds and a total bandwidth of $M \times \Delta f$ Hz. A grid of size $N \times M$ is defined in the DD domain, with sampling intervals given by $ \frac{1}{M \Delta f}$ and $ \frac{1}{N T}$. The information symbols $x[k,\ell]$, where $k = 0, \ldots, N-1$ and $\ell = 0, \ldots, M-1$, are placed on the DD grid, with $k$ indexing the Doppler bin and $\ell$ indexing the delay bin. Firstly, the information symbols $x[k,\ell]$ in the DD domain are transformed to the time-frequency domain using a two-dimensional ISFFT. This operation is mathematically expressed as \cite{OTFSilk}
\begin{equation}
X[n,m] = \frac{1}{\sqrt{NM}} \sum_{k=0}^{N-1} \sum_{l=0}^{M-1} x[k,l] e^{j2\pi \left( \frac{nk}{N} - \frac{ml}{M} \right)}, 
\end{equation}
where  $X[n,m]$ are the information symbols in the time-frequency domain. The symplectic Fourier kernel \( e^{j2\pi\left(\frac{nk}{N} - \frac{m\ell}{M}\right)} \) shows that each DD information symbol \( x[k, \ell] \) spreads across all time-frequency samples \( X[n, m] \) with a unique phase pattern. Next, the symbols \( X[n, m] \) in the time-frequency domain are converted to the time domain for transmission using the Heisenberg transform.  The transmitted signal in the time domain can be written as \cite{OTFSilk}
\begin{equation}
s(t) = \sum_{n=0}^{N-1} \sum_{m=0}^{M-1} X[n, m] \, g_{tx}(t - nT) \, e^{j 2\pi m \Delta f (t - nT)},
\end{equation}
where $g_{tx}(t)$ is the transmit pulse shaping function and is expressed as a rectangular pulse of duration $T$. The wireless channel is modeled as a linear time-varying system with an impulse response \( h(\tau, \nu) \) defined in the DD domain. The \( h(\tau, \nu) \) specifies the channel response for a path with delay \( \tau \) and Doppler shift \( \nu \). This response can be expressed as a sum of \( P \) discrete propagation paths, where each path is defined by a delay \( \tau_i \), a Doppler shift \( \nu_i \), and a complex gain \( h_i \), as follows \cite{OTFSilk}:
\begin{equation}
h(\tau, \nu) = \sum_{i=1}^{P} h_i \, \delta(\tau - \tau_i) \, \delta(\nu - \nu_i).
\end{equation}
When the transmitted signal \( s(t) \) passes through the time-varying channel \( h(\tau,\nu) \), the received signal \( r(t) \) corresponds to the 2D convolution of \( s(t) \)'s DD representation with \( h(\tau,\nu) \). The received time domain signal \( r(t) \) is expressed as follows \cite{OTFSilk}:
\begin{equation}
r(t) = \iint h(\tau, \nu) \, s(t - \tau) \, e^{j 2\pi \nu (t - \tau)} \, d\tau \, d\nu + w(t),
\end{equation}
where \( w(t) \) is additive white Gaussian noise (AWGN). The \( s(t - \tau) e^{j2\pi \nu (t - \tau)} \) represents a copy of the transmit signal that is delayed by \( \tau \) and Doppler-shifted by \( \nu \). At the receiver, OTFS demodulation is carried out by reversing the transformations applied at the transmitter. First, the received time domain signal \( r(t) \) is transformed into the time-frequency domain using the Wigner transform. The received signal in the time-frequency domain \( Y[n, m] \) is expressed as follows \cite{OTFSilk}:
\begin{equation}
Y[n, m] = \int_{-\infty}^{\infty} r(t) \, g_{\text{rx}}^*(t - nT) \, e^{-j 2\pi m \Delta f (t - nT)} \, dt,
\end{equation}
where $g_{\text{rx}}^*(t-nT)$ is the receive pulse shaping function. After obtaining the time-frequency domain received signal \( Y[n, m] \), the receiver performs the SFFT to convert the signal back into the DD domain. As a results, the DD domain received signal $y[k,l]$ is obtained by applying the SFFT, a two-dimensional transform with a symplectic kernel, and is given as follows \cite{OTFSilk}:
\begin{equation}
y[k,l] = \frac{1}{\sqrt{NM}} \sum_{n=0}^{N-1} \sum_{m=0}^{M-1} Y[n,m] \, e^{-j 2\pi \left( \frac{n k}{N} - \frac{m l}{M} \right)}.
\end{equation}



\section{OTFS-Based Index Modulation Systems}
This section provides an overview of OTFS-IM systems, exploring their basic principles, advantages, challenges, and recent progress. In particular, the OTFS-IM schemes, including OTFS-SM, OTFS-QSM, OTFS-MBM, OTFS-CIM, and OTFS-SSK, are presented in individual sections, where their system models, operating principles, and overall structures are described in detail. Unlike conventional modulation schemes that operate in the time-frequency domain, OTFS utilizes the DD domain to combat the effects of time-varying multipath channels and provides better error performance than its counterpart OFDM under high mobility conditions \cite{OTFSURVEY1, OTFSURVEY2, OTFSURVEY3}. IM has emerged as a promising technique in wireless communications, providing improved spectral and energy efficiency by transmitting data not only through conventional symbols but also through indices of activated communication sources \cite{IM10, IM11, IM12}. The rational combination of these two innovative modulation systems, OTFS and IM, offers significant potential for high spectral efficiency, low error data transmission, low energy consumption, robustness to channel distortion, and high performance in high-mobility scenarios, making it a critical research area.  

\subsection{Literature Review of OTFS-IM Systems}
In \cite{OTFS_IM1}, a scheme called joint DD IM OTFS (JDDIM-OTFS) is proposed, where the delay or Doppler source elements in a subframe are activated, and different constellations are used for each active source block to achieve higher spectral efficiency. To achieve lower complexity, a greedy detector is used for parameter estimation of the proposed system. Furthermore, a theoretical upper bound for the BER is derived. Verification via simulation results shows that the proposed JDDIM-OTFS scheme outperforms conventional OTFS and other conventional wireless communication schemes in terms of BER performance under perfect and imperfect channel conditions. This paper in \cite{OTFS_IM2} proposes two new block-wise IM schemes for OTFS: delay-IM with OTFS (DeIM-OTFS) and Doppler-IM with OTFS (DoIM-OTFS), which simultaneously enable blocks of delay/Doppler source boxes. Using an ML detector, upper bounds on average bit error rates are analyzed, and performance comparisons with existing OTFS-IM systems are presented. It also presents two low-complexity detector algorithms, the multi-layer joint symbol and activation pattern detection (MLJSAPD) algorithm, and the CMP detection algorithm. We also analyze imperfect CSI and show that the proposed algorithms are robust to imperfect CSI. \cite{OTFS_IM3} introduces DoIM-OTFS, where a block of Doppler resource bins is activated simultaneously based on index bits. A CMP algorithm is developed for implementing DoIM-OTFS. The BER results show that DoIM-OTFS achieves better error performance than conventional OTFS without IM. It is also shown that the CMP algorithm is robust to imperfect CSI. A novel IM approach called IM-aided fragmented spectra centralization (IM-FSC) for OTFS systems is proposed in \cite{OTFS_IM4}. The IM-FSC scheme aims to collect fragmented energy-free subcarriers from subblocks to form a completely unused frequency band. The unused frequency band created by combining fragmented spectra from multiple subblocks is used as zero padding (ZP) and guard interval for interference cancellation and channel estimation in OTFS systems. Closed-form expressions for BER and guidelines for generalizing the IM-FSC scheme are provided. Simulation results demonstrate improved communication capability and channel estimation accuracy compared to conventional OTFS-IM systems. Performance analysis shows the IM-FSC system outperforms OTFS-IM under the same receiver and spectral efficiency conditions. The publication in \cite{OTFS_IM5} proposes a modulation technique called OTFS-IM that provides high spectral efficiency and BER performance at high mobility. The proposed system transmits information via constellation symbols and active point indices in the DD domain. Since the proposed system involves high receiver complexity, a low-complexity MMSE-ML detector is presented. Simulation results show that the proposed system exhibits better error performance than traditional OTFS and OFDM-IM, especially at higher spectral efficiency, and is robust to Doppler effects in high mobility channels. In \cite{OTFS_IM6}, a new OTFS modulation scheme with improved IM (OTFS-IIM) is proposed to improve the BER performance on frequency-selective and rapidly time-varying fading channels. Each index bit is transmitted twice to improve the BER performance of the index bits. Simulation results show that the proposed OTFS-IIM can achieve better BER performance than traditional OTFS and OTFS-IM schemes on rapidly time-varying frequency-selective fading channels. In \cite{OTFS_IM7}, a new wireless communication scheme called OTFS-aided dual-mode IM (OTFS-DM-IM) is proposed to balance reliability and spectral efficiency. It is shown that the proposed OTFS-DM-IM system transmits data with fewer errors than the traditional OTFS and OTFS-IM systems in the literature. Also, a modified log-likelihood ratio (LLR) detector is proposed, which performs better than the traditional LLR detector. It is seen that the proposed modified LLR detector provides approximately the same BER performance as the ML detector. The paper \cite{OTFS_IM8} presents a novel uplink multiple access scheme with OTFS-IM support in the DD domain. The proposed system allows multiple users to share the same DD resources by activating a subset of DD resources without requiring coordination from the base station. Although data collisions are possible among users, simulation results show that the error performance of the proposed system is better than that of traditional schemes. In \cite{OTFS_IM9}, the authors propose OTFS with sub-band IM (OTFS-SBIM), an enhanced OTFS-IM scheme that restricts IM to sub-bands and employs a band-controlling bit to achieve higher spectral efficiency. By utilizing independent $I$/$Q$ modulation, the proposed system provides improved BER performance and enhanced diversity. The proposed scheme achieves superior performance compared to OTFS-IM in terms of both BER and PAPR, while maintaining higher spectral efficiency, as demonstrated through low-complexity ML detection and analytical average bit error probability (ABEP) and PAPR metrics. \cite{OTFS_IM10} introduces a wavelet-based three-dimensional OTFS with dual-mode IM (W-3D-OTFS-DM-IM) for next-generation wireless communications. The proposed W-3D-OTFS-DM-IM system utilizes dual-mode IM with 3D signal constellations in the DD domain to improve spectral efficiency. Simulation results demonstrate that the proposed system outperforms benchmark systems in error performance and spectral efficiency under high-mobility conditions. \cite{OTFS_IM13} proposes a tri-mode OTFS-IM (TM-OTFS-IM) scheme that improves error performance and spectral efficiency by carrying extra bits through both multiple constellations and inactive resource block indices. Simulation results demonstrate that the proposed TM-OTFS-IM system achieves lower BER compared to traditional OTFS-IM and dual-mode OTFS-IM systems. \cite{OTFS_IM17} proposes an OTFS with generalised dual-mode IM (OTFS-GDM-IM) scheme that improves spectral efficiency by flexibly varying subcarrier allocations and using different modulation schemes across grid points within an OTFS subblock. Simulation results show that OTFS-GDM-IM achieves better BER performance than conventional OTFS-IM and dual-mode OTFS-IM, while the proposed LLR detector offers lower computational complexity than the ML detector. Similarly, \cite{OTFS_IM33} proposes a multiple-mode OTFS-IM (MM-OTFS-IM) scheme that activates all grids with different modulation modes and uses a distance-based detection algorithm, demonstrating enhanced performance in time-varying channels. \cite{OTFS_IM14} proposes an OTFS-based channel modulation IM (OTFS-CM-IM) scheme that combines channel modulation and DD indexing with a compact input-output model and improved distance properties, achieving better BER than standard OTFS and OTFS-CM systems. \cite{OTFS_IM24} proposes a modified-constellation-based OTFS with IM (MCOTFS-IM) system that optimizes the positions of QAM constellation points close to the origin to enlarge the Euclidean distances in the DD grid, thereby improving the detection accuracy of active and inactive DD grid points. Simulation results indicate that the proposed system outperforms conventional OTFS-IM in terms of error performance under different system parameters. \cite{OTFS_IM15} proposes a new OTFS-based dual frequency IM scheme (OTFS-DFIM) that improves spectral efficiency by using cascaded index mapping and updating to keep the OTFS frame sparse and achieve improved BER performance, for which tight closed-form BER expressions are derived. \cite{OTFS_IM16} proposes an enhanced OTFS with IM (E-OTFS-IM) scheme that allows variable numbers of active grids and multiple constellations, with an efficient encoding method and an updated message passing detector. Also, simulation results verify improved BER performance and the derived ABEP upper bound. \cite{OTFS_IM18} introduces a new autoencoder-based joint DD IM (AEE-JDDIM-OTFS) scheme that selectively activates delay or Doppler resource bins using multidimensional symbol vectors with increased squared minimum Euclidean distance, providing error performance improvements over conventional OTFS schemes. \cite{OTFS_IM26} and \cite{OTFS_IM35} propose spatial-IM-based OTFS (SIM-OTFS) systems that employ three-dimensional indexing across transmit antennas, delay, and Doppler domains to boost transmission rate and reliability in high-mobility scenarios. \cite{OTFS_IM26} derives the average BER (ABER) for the proposed SIM-OTFS scheme, analyzes diversity, coding gain, and complexity, and illustrates the impact of resolvable multipaths on system performance. Meanwhile, \cite{OTFS_IM35} analyzes the ABER performance of the proposed SIM-OTFS system and shows that it achieves better ABER performance compared to MIMO-OTFS and SM and IM based OFDM (SM-OFDM-IM) systems in high mobility channels. \cite{OTFS_IM27} proposes a spatial multiplexing aided OTFS-IM system by combining vertical-Bell Laboratories layered space-time (VBLAST) with OTFS-IM, and derives an equivalent channel models and ABEP upper bounds to analyze the theoretical performance gains of the proposed system. \cite{OTFS_IM28} proposes a generalized IM scheme for MIMO-OTFS (GIM-MIMO-OTFS), conveying extra information through spatial and DD indices and deriving ABEP expressions validated via simulation results. Similarly, \cite{OTFS_IM29} presents a generalized space-DD index modulated OTFS (GSDDIM-OTFS) scheme that extends multi-domain indexing across space, delay, and Doppler resource units to transmit additional information, with ABEP results confirmed under doubly-selective fading channels. \cite{OTFS_IM30} proposes a generalized SIM-OTFS (GSIM-OTFS) system that combines the benefits of generalized SM, IM, and OTFS to improve reliability in highly mobile scenarios. A framework for ABER analysis is developed and validated through simulations, demonstrating that the proposed GSIM-OTFS system offers superior spectral efficiency and error performance compared to peer IM schemes. \cite{OTFS_IM31} proposes a physical layer security scheme for OTFS-IM operating in frequency division duplex (FDD) mode. The proposed system uses secure mapping to protect transmitted bits, derives a closed-form ergodic secrecy rate, and shows through simulations that it maintains reliable BER and secrecy performance even when there are angle estimation errors. \cite{OTFS_IM25} proposes a new iterative algorithm for channel estimation and data detection in OTFS-IM systems, where superimposed pilots are embedded to improve spectral efficiency and enable reliable performance in high-mobility scenarios. \cite{OTFS_IM20} introduces a block-wise IM-aided OTFS system and develops a new multi-layer message passing detection (MLMPD) algorithm that leverages a priori information to accelerate convergence, achieving improved BER performance while reducing computational complexity. \cite{OTFS_IM32} proposes a symbol-by-symbol aided expectation propagation (SS-EP) detector for the OTFS-IM system, which uses null symbols and a varied zero sequence to improve detection accuracy and reduce complexity. Simulation results confirm that the proposed detector outperforms existing detectors for both uncoded and coded OTFS-IM schemes. An IM-aided $IQ$ imbalance compensator scheme is proposed for OTFS systems in \cite{OTFS_IM36} to reduce the negative impact of hardware impairments, and its effectiveness in maintaining reliable BER performance as well as energy and spectral efficiency is demonstrated through simulation results. In \cite{OTFS_IM19}, the BER performance of the OTFS-IM scheme under barrage jamming is analyzed by employing ML, MMSE, and LLR detectors. It is shown that the OTFS-IM system outperforms conventional OTFS and OFDM-IM schemes at high signal-to-jamming ratios (SJRs). In \cite{OTFS_IM21}, a supply IM-aided OTFS scheme is proposed, where the indices of the active DD grids are used to convey extra information, improving spectral efficiency and robustness in high-mobility environments. A theoretical BER analysis for the proposed system is presented, and a low complexity detector based on the energy is developed, which reduces the complexity of the MMSE-ML detector with less than $0.7$ dB performance loss. \cite{OTFS_IM22} proposes a novel OTFS-IM scheme combined with NOMA to improve spectral efficiency while reducing transmit power and BER. Simulation results show that the proposed system achieves better error performance than conventional OTFS and OTFS-IM systems. \cite{OTFS_IM23} analyzes the performance of an OTFS-IM system with decode-and-forward relaying, deriving a closed-form expression for the end-to-end pairwise error probability and the asymptotic diversity order. Simulation results confirm that indexing improves relaying performance and that the BER upper bound matches the theoretical diversity predictions. In \cite{OTFS_IM24}, a modified-constellation-based OTFS-IM (MCOTFS-IM) system is proposed, where zero symbols are incorporated as part of the constellation set and the positions of symbols close to the origin are optimized to enhance error performance. The results indicate that the proposed MCOTFS-IM system provides improved BER performance compared to the traditional OTFS-IM system. \cite{OTFS_IM25} addresses channel estimation in OTFS-IM systems under high-mobility scenarios, introducing an iterative algorithm based on superimposed pilots, which jointly performs LMMSE-based data detection and data-aided channel estimation, resulting in improved spectral efficiency and error performance compared to state-of-the-art schemes. In \cite{OTFS_IM26}, a new spatial IM-based OTFS system (SIM-OTFS) is introduced for vehicular networks, employing transmit antenna, delay, and Doppler indices to increase spectral efficiency and reliability. The proposed SIM-OTFS system provides improved ABER performance compared to MIMO-OTFS and SM-OFDM-IM under high mobility, as demonstrated by numerical results and performance analysis. In \cite{OTFS_IM27}, a VBLAST aided OTFS-IM (VBLAST-OTFS-IM) system is proposed by combining VBLAST and OTFS-IM schemes. The results show that the proposed VBLAST-OTFS-IM system provides ABEP improvements over OTFS-IM and OTFS-SM. In \cite{OTFS_IM28}, a GIM-MIMO-OTFS is presented, activating multiple indices in spatial and DD domains to improve spectral efficiency. Also, the proposed two-stage index detector achieves lower complexity and better error performance than the ML detector in high mobility environments. In \cite{OTFS_IM29}, a GSDDIM-OTFS system is proposed, which enables the activation of space-DD resource units. Analytical ABEP expressions of the proposed system are derived, and simulation results demonstrate improved BER performance for doubly-selective fading channels. \cite{OTFS_IM30} introduces a generalized spatial-IM OTFS (GSIM-OTFS) system, integrating IM, generalized SM (GSM), and OTFS techniques to achieve improved spectral efficiency and reliability in high-mobility scenarios. Analytical expressions of the proposed system for the ABER are derived, and the results are validated through simulation results. In \cite{OTFS_IM31}, a chaos-based secure mapping scheme is proposed for an FDD IM-OTFS system to enhance physical layer security, utilizing angular reciprocity to generate chaos sequences. The secrecy performance of the proposed scheme is validated analytically and via simulation, demonstrating robustness to angle estimation errors. \cite{OTFS_IM32} introduces a new SS-EP detector for the OTFS-IM system, where zero symbols are used as constellation points and a varied zero sequence scheme is proposed. Both the proposed SS-EP detector and its proposed low-complexity variant provide substantially better error performance than existing detectors. In \cite{OTFS_IM33}, a MM-OTFS-IM scheme is proposed, activating all grids with different modulation modes to transmit modulation bits and employing the designed distance-based signal detection algorithm. The proposed MM-OTFS-IM system improves spectral and energy efficiency, and simulation results demonstrate the error performance advantages of the proposed system under time-varying channels. \cite{OTFS_IM34} presents OTFS with $I$ and $Q$ IM (OTFS-$I$/$Q$-IM) system, where additional data bits are conveyed through the grid indices of the OTFS block in the $I$ and $Q$ dimensions to increase spectral and energy efficiency. Analytical ABEP and PAPR expressions of the OTFS-I/Q-IM system are derived, and simulation results indicate improved error performance in high-mobility scenarios. In \cite{OTFS_IM35}, a SIM-OTFS system is proposed, employing transmit antenna, delay, and Doppler indices for higher spectral efficiency and improved performance in high mobility communication. Theoretical and simulation results of the proposed system demonstrate improved performance over conventional MIMO-OTFS and SM-OFDM-IM. In \cite{OTFS_IM36}, an IM-aided $IQ$ imbalance compensator for OTFS systems is proposed, which mitigates $IQ$ imbalance effects without requiring iterative processing or training/pilot sequences, and achieves reliable error performance for various system parameters. \cite{OTFS_IM37} introduces a new efficient vector-by-vector-aided message passing (VV-MP) detector for OTFS-IM schemes, providing error performance improvements and a complexity-performance trade-off compared to MMSE-based detectors. In \cite{OTFS_IM38}, a structured prior-based hybrid belief propagation and expectation propagation (Str-BP-EP) algorithm is proposed for an iterative receiver scheme in coded OTFS-IM systems, demonstrating superior error performance in both coded and uncoded scenarios. \cite{OTFS_IM39} proposes a four-dimensional spherical code-based OTFS-IM (OTFS-IM4) system for air-to-ground communication, which enhances BER and power consumption efficiency under high-mobility scenarios. In \cite{OTFS_IM40}, an AEE-JDDIM-OTFS system is proposed, which uses deep learning to optimize the mapping and demapping, and employs a low-complexity greedy detector. Simulation results of the proposed system show improved spectral and energy efficiency, as well as robust error performance even for imperfect CSI. \cite{OTFS_IM41} proposes a novel multi-mode co-directional orthogonal index OTFS modulation scheme, where $I$ and $Q$ components of subgrids in the DD domain independently transmit indices of one-dimensional pulse amplitude modulation (PAM) constellations, thereby significantly enhancing spectral/energy efficiency and error performance compared to the traditional OTFS-IM scheme in underwater acoustic communications. In \cite{OTFS_IM42}, a hierarchical mode-based IM (HMIM) scheme is proposed for orthogonal DD division multiplexing (ODDM) systems. Results indicate that, compared to ODDM and ODDM with IM (ODDM-IM), the proposed ODDM-HMIM scheme offers comparable or better error performance, while showing low complexity.


\subsection{OTFS-based SM System}

The SM technique, introduced in 2008, utilizes antenna indices in a MIMO system to transmit additional data, leading to increased data transmission, high energy efficiency, and other benefits \cite{smilk}. It eliminates the need for ICI and antenna synchronization, resulting in reduced hardware complexity and energy consumption \cite{SM1, SM2, SM3, SM4, SM5}. A number of studies have considered combining OTFS and SM techniques in a single communication system to increase spectral, energy efficiency and enhance performance in high-Doppler-shift and mobile communication scenarios. \cite{SMOTFS1} introduces a novel GSM-based OTFS (GSM-OTFS) system and a decision feedback detector based on the MMSE criterion, while \cite{SMOTFS2} also proposes a GSM-OTFS system but shows its superior BER performance and spectral efficiency through theoretical analysis and simulations. \cite{SMOTFS3} and \cite{SMOTFS4} focus on the design and analysis of OTFS-SM systems, with \cite{SMOTFS3} deriving and verifying closed-form expressions for average symbol error rate (ASER) and BER, and \cite{SMOTFS4} demonstrating that OTFS-SM achieves higher spectral efficiency and lower detection complexity compared to space-time coded OTFS (STC-OTFS). \cite{SMOTFS5} improves transmission reliability with the enhanced OTFS-SM system by introducing a new sparse signal estimation detector that uses Variational Bayesian Inference to reduce complexity. Also, a low-complexity distance-based detection algorithm for OTFS-SM in doubly-selective channels is proposed in \cite{SMOTFS7}, demonstrating a trade-off between BER performance and complexity. \cite{SMOTFS8} proposes an SM-based space-time block code (SM-STBC) aided OTFS system with a block message passing detector that enhances performance in high mobility scenarios compared to traditional OTFS systems. The paper in \cite{SMOTFS9} proposes a multi-mode IM aided OTFS-based SM (MMIM-OTFS-SM) system to enhance spectral efficiency and transmission reliability in high-mobility wireless environments. \cite{SMOTFS10} introduces an OTFS-SM system for LEO satellite constellations to overcome Doppler effects and spatial underutilization, enhancing system reliability through multi-satellite cooperation with a channel capacity-driven satellite selection technique, and provides a superior error performance over existing schemes. In \cite{SMOTFS11}, a transmit antenna selection (TAS) based OTFS-SM system is proposed to improve transmit diversity and reliability in mobile communication environments, including Euclidean distance-based TAS, low-complexity TAS based on a tree search scheme (LCTAS-TSS), and norm and antenna correlation (N-AC) based TAS scheme. Additionally, \cite{SMOTFS12} investigates OTFS-SM performance under imperfect CSI, providing practical insights for real-world high-mobility scenarios. \cite{SMOTFS13} proposes a multi-domain channel estimation technique for OTFS-SM to enhance spectral efficiency in beyond 5G systems. For indoor visible light communications, \cite{SMOTFS14} introduces quad-LED OTFS schemes that achieve better BER performance than existing methods. In the context of vehicle-to-vehicle (V2V) communications, \cite{SMOTFS15} analyzes a cooperative multi-relay OTFS-SM system with optimal relay selection to improve outage probability and capacity, while \cite{SMOTFS16} focuses on capacity and outage analysis of OTFS-SM with imperfect CSI, demonstrating its impact under high-mobility conditions.

\begin{table*}[t!]
\centering
\addtolength{\tabcolsep}{-2.5pt}
\caption{Mapping procedure for the OTFS-SM system}
\label{OTFS_SM_MAP}
\begin{tabular}{|c|c|c|c|c|c|} 
\hline
\hline
\textbf{\small System Parameters}       & \textbf{\begin{tabular}[c]{@{}c@{}}\small Data Bits\\ \end{tabular}} & \textbf{\begin{tabular}[c]{@{}c@{}} \small First DD grid \\ \end{tabular}} & \textbf{\begin{tabular}[c]{@{}c@{}} \small Second DD grid \\ \end{tabular}} & \textbf{\begin{tabular}[c]{@{}c@{}} \small Third DD grid \\ \end{tabular}} & \textbf{\begin{tabular}[c]{@{}c@{}} \small Fourth DD grid \\ \end{tabular}} \\ \hline\hline
\begin{tabular}[c]{@{}c@{}} \small $M_Q=2$\\ $n_T=2$\\ $N=2$\\ $M=2$ \end{tabular} & 
\begin{tabular}[c]{@{}c@{}} {[}00101101{]} \\ $n_{OTFS-SM}=8$ \end{tabular}                                                             & \begin{tabular}[c]{@{}c@{}}{[}00{]}\\ $s=-1$, $\ell=1$ \\ $\textbf{x}_1=[-1 \ \ 0]^T$ \end{tabular}                   & \begin{tabular}[c]{@{}c@{}}{[}10{]} \\ $s=1$, $\ell=1$ \\ $\textbf{x}_2=[1 \ \ 0]^T$ \end{tabular}                 & \begin{tabular}[c]{@{}c@{}}{[}11{]}\\ $s=1$, $\ell=2$ \\ $\textbf{x}_3=[0 \ \ 1]^T$  \end{tabular}  & \begin{tabular}[c]{@{}c@{}}{[}01{]} \\ $s=-1$, $\ell=2$ \\ $\textbf{x}_4=[0 \ \ -1]^T$ \end{tabular} \\ \hline \hline
\end{tabular}
\end{table*}

The spectral efficiency of the OTFS-SM system for each DD grid point is the same as that of the conventional SM and can be given as follows:
\begin{equation}
    n_{SM}=\log_2(M_Q n_T),
\end{equation}
where $M_Q$ is the modulation order and $n_T$ is the number of transmit antennas. However, for a DD grid of size $N M$, the spectral efficiency of the OTFS-SM system during the duration of an OTFS frame can be expressed as follows:
\begin{equation}
    n_{OTFS-SM}=NM\log_2(M_Q n_T),
\end{equation}

\begin{figure*}[t]
\centering{\includegraphics[width=0.98\textwidth]{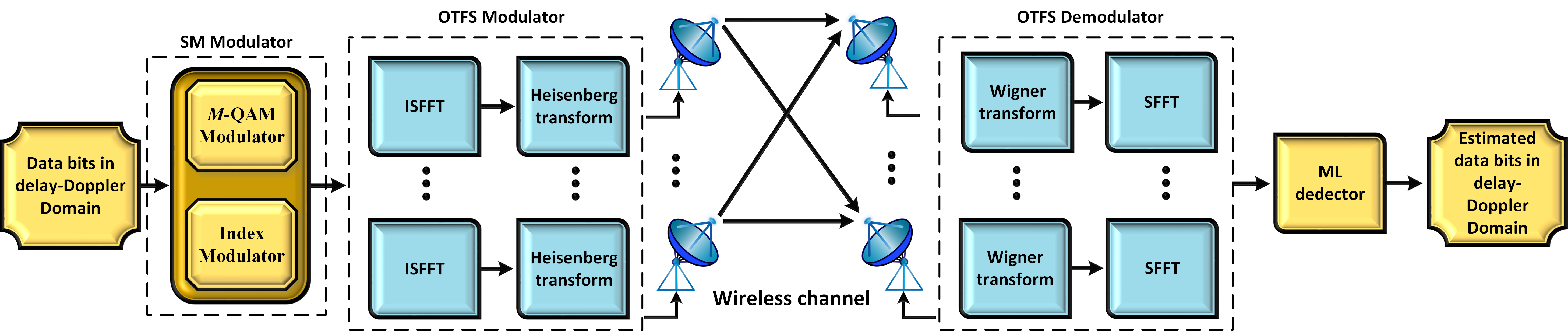}}	
	\caption{System model of the proposed OTFS-SM scheme.}
	\label{system_model_SM_OTFS} 
\end{figure*}
The system model of the OTFS-SM system is presented in Fig. \ref{system_model_SM_OTFS}. At the transmitter terminal of the OTFS-SM system, in each OTFS-SM transmission frame, a bit sequence vector $\textbf{d}$ containing $n_{OTFS-SM}$ data bits is transmitted. First, the $n_{OTFS-SM}=NM\log_2(M_Q n_T)$ expression is split into $NM$ parts. Then $\log_2(M_Q n_T)$ bits of $\log_2(M_Q)$ are mapped to the selected symbol ($s_q$) and $\log_2(n_T)$ bits to the active transmit antenna index ($\ell$). The transmission vector of the OTFS-SM system for the $ith$ DD grid can be defined as follows:    
\begin{eqnarray}\label{transmission_signal}
\textbf{x}_i = {\Bigg[ 0\, \,0\cdots 0\,  \, \cdots \underset{\stackrel{\uparrow}{\ell^{\text{th}} \ \text{position}}}{s_q} \cdots \, \,0\, \,\cdots 0\, \,0
	\Bigg]^T},
\end{eqnarray}
where, $q \in \{1,2,\cdots, M_Q\}$, $\ell \in \{1,2,\cdots, n_T\}$, and $i \in \{1,2,\cdots, NM\}$. The transmission matrix of OTFS-SM containing all DD grid points is expressed as follows:
\begin{equation} \textbf{X}=[\textbf{x}_1, \textbf{x}_2, \cdots, \textbf{x}_{NM} ].
\end{equation}

Table \ref{OTFS_SM_MAP} presents the mapping process of the OTFS-SM system using BPSK modulation for system parameters $n_T$, $N=2$, and $M=2$. $NM=4$, therefore the system contains $4$ DD grid points in total. At each DD grid point, the conventional SM technique can be considered as realized. Therefore, as can be seen from Table \ref{OTFS_SM_MAP}, each DD grid contains one conventional SM transmission vector. Then the OTFS-SM transmission vector for the example in Table \ref{OTFS_SM_MAP} can be expressed as follows:
\begin{equation}
    \mathbf{X} = [\textbf{x}_{1} \textbf{x}_{2} \textbf{x}_{3} \textbf{x}_{4}] = \begin{bmatrix}
        -1 & 1 & 0 & \ 0 \\
        \ 0 & 0 & 1 & -1
    \end{bmatrix}.
\end{equation}

The transmitter performs ISFFT to convert the DD domain signals, represented in each row vector of $\textbf{X}$, into time-frequency domain signals for each transmit antenna. Subsequently, the OFDM modulators, each corresponding to a specific transmit antenna, apply the Heisenberg transform to convert these time-frequency domain input signals into time-domain signals, denoted as $\textbf{Q}(t)$. Also, the time-variant channel matrix \(\mathbf{H}(t) \in \mathbb{C}^{n_R NM \times n_T NM}\), each of length $P$ (number of multipaths), with distinct channel delays and Doppler shifts for each path and expressed as follows:
\begin{equation}
    \mathbf{H}(t) = 
\begin{bmatrix}
    \mathbf{H}_{1,1}(t) & \mathbf{H}_{1,2}(t) & \cdots & \mathbf{H}_{1,n_T}(t) \\
    \mathbf{H}_{2,1}(t) & \mathbf{H}_{2,2}(t) & \cdots & \mathbf{H}_{2,n_T}(t) \\
    \vdots & \vdots & \ddots & \vdots \\
    \mathbf{H}_{n_R,1}(t) & \mathbf{H}_{n_R,2}(t) & \cdots & \mathbf{H}_{n_R, n_T}(t)
\end{bmatrix}
\end{equation}
where $\mathbf{H}_{r,\ell} (t)$ is the $NM \times NM$ dimensional channel matrix between the $\ell\text{th}$ transmit antenna and the $r\text{th}$ receive antenna. Here $r \in \{1,2,\cdots, n_R\}$.

The time domain signal $\mathbf{y}(t)$ obtained at the receiver, can be mathematically given as follows:
\begin{equation}
    \mathbf{y}(t) = \mathbf{H}(t) \otimes \mathbf{Q}(t) + \mathbf{w}(t),
\end{equation}
where $\mathbf{w}(t)$ denotes the AWGN vector with zero mean and variance $N_0$. After OTFS demodulation with the Wigner transform and SFFT, the received time domain signal is converted into the received DD domain signal. Then, the transmitted symbol and the active transmit antenna index are estimated with the ML detector over the received signal as shown in Fig. \ref{system_model_SM_OTFS}. The ML detector estimates the selected symbol and the active transmit antenna index for the $i\text{th}$ DD grid as follows:  
\begin{eqnarray}\label{eq_ML_SM}
 [\hat{q}_i, \hat{\ell}_i] & =& \text{arg}\underset{q, \ell}{\mathrm{min}} \ \Big|\Big| {\mathbf{y}} - \textbf{h}_{i,\ell} s_q \Big|\Big|^2,
\end{eqnarray}
where ${\mathbf{y}} \in \mathbb{C}^{n_RNM \times 1}$ is the received DD domain signal and $\textbf{h}_{i,\ell} \in \mathbb{C}^{n_RNM \times 1}$ is the DD domain channel vector for the $i\text{th}$ DD grid. The ML detector of the OTFS-MBM system operates $NM$ times to estimate the selected symbol and the active transmit antenna index for each DD grid. As a result, the estimated $NM$ symbols and active transmit antenna indices corresponding to all $NM$ DD grid points in an OTFS frame are expressed as follows:
\begin{eqnarray}\label{eq15}
\hat{\textbf{s}} = [\hat{s}_1, \hat{s}_2, \cdots, \hat{s}_{NM}], \ \hat{\boldsymbol{\ell}} = [\hat{\ell}_1, \hat{\ell}_2, \cdots, \hat{\ell}_{NM}].
\end{eqnarray}
Finally, the estimated bit sequence vector $\hat{\textbf{d}}$ is obtained by using the parameters estimated $\hat{\textbf{s}}$ and  $\hat{\boldsymbol{\ell}}$ with the SM demapper.

\subsection{OTFS-based QSM System}
The QSM technique, first proposed by Mesleh et al. in 2014 in \cite{QSMilk}, is an IM technique that extends the conventional SM technique by simultaneously activating two antennas for the $I$ and $Q$ components of the signal. It improves the spectral efficiency of SM by transmitting the real and imaginary parts of each complex symbol through different antennas. Thereby, compared to conventional SM, the QSM technique doubles the transmitted information per channel use and, thanks to $I–Q$ orthogonality, prevents inter-channel interference. It provides low complexity, requires only a single RF chain, eliminates interference, enables improved error performance, and easy integration with MIMO systems \cite{QSM1, QSM2, QSM3, QSM4, QSM5, IM13, QSM7, QSM8, QSM9, QSM10, QSM11, QSM12}.

The OTFS-QSM system was first introduced in \cite{OTFSQSM1}, where the authors proposed a novel transmission scheme combining QSM with OTFS to enhance spectral efficiency, achieve diversity gains, and mitigate the Doppler effect in high-mobility scenarios. This study provided a comprehensive system model, theoretical ABER analysis, and an enhanced minimum mean square error (EMMSE) detection method to reduce detection complexity. In \cite{OTFSQSM2}, a diversity-achieving QSM (DAQSM)-aided MIMO-OTFS scheme was proposed to offer both transmit diversity and a higher data rate. The simulation results demonstrate that the proposed system outperforms peer communication techniques in high-mobility scenarios. Additionally, \cite{OTFSQSM3} proposes two multi-LED OTFS schemes: the OTFS-QSM and a dual-mode IM-based OTFS. The proposed schemes achieve higher data rates without needing Hermitian symmetry and DC bias operations, and they outperform traditional OFDM-based schemes in terms of error performance. 

In the QSM technique, two transmit antennas are active per transmission time: one transmits the real part and the other transmits the imaginary part of the selected symbol.  Thus, the spectral efficiency of the QSM scheme is expressed as follows:
\begin{equation}
  n_{\mathrm{QSM}} = \underbrace{\log_2(M_Q)}_{\text{symbol bits}} + \underbrace{\log_2(n_T)}_{\text{real-antenna index}} + \underbrace{\log_2(n_T)}_{\text{imag-antenna index}} \!\!\!\!\!\!.
\end{equation}
Over the OTFS frame of size $NM$, the total spectral efficiency of the OTFS-QSM system is given as follows:
\begin{equation}
  n_{\mathrm{OTFS\text{-}QSM}} = N M \Bigl[\,\log_2(M_Q) + 2\log_2(n_T)\Bigr].
\end{equation}

In the OTFS-QSM technique, the data bits for each DD grid are mapped to the transmit antenna indices and symbol as follows:
\begin{itemize}
  \item First $\log_2(M_Q)$ bits select QAM symbol $s_q\in\mathcal{S}_{M_Q}$.
  \item Next $\log_2(n_T)$ bits select real-part antenna index $\ell_1\in\{1,2,\dots,n_T\}$.
  \item Last $\log_2(n_T)$ bits select imaginary-part antenna index $\ell_2\in\{1,2,\dots,n_T\}$,
\end{itemize}
this bit mapping procedure is repeated $NM$ times in the OTFS-QSM technique. The transmission vector of the OTFS-QSM system for the $i$-th DD grid can be defined as
\begin{equation} \label{eq:QSM_transmission}
\mathbf{x}_i = \left[
0, \dots, 0, \underset{\ell_1\text{-th position}}{s_{q_I}}, 0, \dots, 0, 
\underset{\ell_2\text{-th position}}{j\,s_{q_Q}}, 0, \dots, 0
\right]^T
\end{equation}
where $s_{q_I}$ and $s_{q_Q}$ denote the real and imaginary components of the complex number $s_q$, respectively. To include all $NM$ DD grid points the transmission matrix can be rewritten as a matrix as follows:
\begin{equation}\label{eq:QSM_matrix_explicit}
\mathbf{X} = [\mathbf{x}_1, \mathbf{x}_2, \cdots, \mathbf{x}_{NM}]. 
\end{equation}

In the transmitter of the OTFS-QSM system, to modulate the information symbols onto the time-frequency plane, the ISFFT is applied row-wise to $\mathbf{X}$, generating the time-frequency domain matrix $\widetilde{\mathbf{X}}$. Each row $\widetilde{\mathbf{x}}_a$ is then passed through the Heisenberg transform, which implements OFDM modulation for each transmit antenna $a = 1, \dots, n_T$, yielding the time-domain signal $\mathbf{q}_a(t) = \mathrm{Heis}(\widetilde{\mathbf{x}}_a)$. These time-domain signals form the transmitted signal matrix $\mathbf{Q}(t) = [\mathbf{q}_1(t), \dots, \mathbf{q}_{n_T}(t)]$.

The transmitted signal propagates through a time-varying multipath MIMO channel modeled as $\mathbf{H}(t) = [h_{r,a}(t)] \in \mathbb{C}^{n_R \times n_T}$, where the channel coefficient between the $a$-th transmit and $r$-th receive antenna is given as follows:
\begin{equation}
    h_{r,a}(t) = \sum_{p=1}^P \alpha_{r,a}^{(p)} \, \delta(t - \tau_{r,a}^{(p)}) \, e^{j 2\pi \nu_{r,a}^{(p)} t}.
\end{equation}
The received signal at the receiver of the OTFS-QSM system is expressed as
\begin{equation}
    \mathbf{y}(t) = \mathbf{H}(t) \otimes \mathbf{Q}(t) + \mathbf{W}(t),
\end{equation}
At the receiver side, time-frequency samples are first extracted by applying the Wigner transform to $\mathbf{y}(t)$. Subsequently, the SFFT is applied to map the samples back to the DD domain, yielding the estimated matrix $\widehat{\mathbf{X}} \in \mathbb{C}^{n_T \times NM}$.

At the receiver side, time-frequency samples are first extracted by applying the Wigner transform to $\mathbf{y}(t)$. Subsequently, the SFFT is applied to map the samples back to the DD domain, resulting in the estimated matrix $\widehat{\mathbf{X}} \in \mathbb{C}^{n_T \times NM}$. Finally, symbol and antenna index detection is carried out over $\widehat{\mathbf{X}}$ to jointly estimate the transmitted QAM symbol and the indices of the antennas that are activated at the transmitter.

At the receiver side, time-frequency samples are first extracted by applying the Wigner transform to $\mathbf{y}(t)$. Subsequently, the SFFT is applied to these samples to obtain the received signal in the DD domain. Based on this DD-domain received signal, it is estimated that the transmitted QAM symbol and the indices of the activated transmit antennas.

\subsection{OTFS-based MBM System}
The MBM scheme, first introduced by Khandani in 2013 as a new approach to wireless transmission in \cite{MBMilk}, conveys information in the indices of the different channel fade realizations (referred to as MAPs, in the literature)  created through switching on/off a set of RF mirrors placed near the transmit antennas. The MBM technique enhances spectral efficiency by transmitting data bits not only through the selected conventional symbol but also through the indices of the selected MAP, allowing for additional information transfer without increasing the number of transmit antennas or RF chains. This technique not only significantly enhances spectral efficiency but also improves error performance by exploiting the inherent diversity of the wireless channel, particularly in rich-scattering environments. Furthermore, MBM improves energy efficiency, reduces hardware complexity by eliminating the need for large antenna arrays \cite{MBM1, MBM2, MBM3, MBM4, MBM5, MBM6, MBM7, MBM8, MBM10, MBM9}. 

\cite{OTFS_IM11} proposes a novel OTFS-MBM scheme for high-mobility wireless communication scenarios. In the proposed system, the MBM scheme is integrated with OTFS modulation in a SIMO configuration to obtain better error performance and higher data rates compared to the conventional OTFS system. By combining the robustness of the OTFS technique in the DD domain with the high spectral efficiency advantages of the MBM scheme using RF mirrors, the proposed system achieves improved error performance compared to conventional OTFS and OTFS-SM systems. The results demonstrate the superiority of the proposed OTFS-MBM in terms of throughput, spectral efficiency, and energy efficiency. 

The proposed system comprises one reconfigurable antenna (RA) at the transmitter and $n_R$ receive antennas at the receiver. Near the RA, $n_{RF}$ RF mirrors generate $2^{n_{RF}}$ distinct MAPs. For a DD grid of size $N \times M$, the spectral efficiency of the proposed OTFS-MBM system is expressed as follows:
\begin{equation}
\eta_{OTFS-MBM}=NM\left(\log_2(M_q) + n_{RF}\right).
\end{equation}
where $M_q$ is the modulation order. The vector of data bits $\mathbf{b}$ containing $\eta_{OTFS-MBM}$ bits is divided into $NM$ segments, then $\log_2(M_q)$ bits are mapped to the selected symbol ($x_q$) and $n_{RF}$ bits to the active MAP index ($\varpi$). The transmission vector $\mathbf{s}_i \in \mathbb{C}^{2^{n{RF}}\times1}$ of the proposed system for the $i^{\text{th}}$ DD grid is expressed as
\begin{equation}
\mathbf{s}_{i} = \left[0, \dots, 0, x_q, 0, \dots, 0\right]^T,
\end{equation}
where the selected symbol $x_q$ is located at position $\varpi^{\text{th}}$. Then, the transmission vectors $\mathbf{s}_i \in \mathbb{C}^{2^{n{RF}} \times 1}$ at each of the $NM$ DD grid points are combined to form the transmission matrix $\mathbf{S} \in \mathbb{C}^{2^{n_{RF}} \times NM}$. OTFS modulation converts the DD-domain signals to the time-frequency domain via an ISFFT, followed by the Heisenberg transform to produce time-domain signals $\mathbf{C}(t)$. In the proposed system, signals are transmitted from the transmitter to the receiver over a multipath Rayleigh fading channel, represented by the time-varying channel matrix $\mathbf{H}_{\text{M}}(t) \in \mathbb{C}^{n_R NM \times 2^{n{RF}} NM}$. At the receiver, the received signal $\mathbf{y}(t)$ can be written as follows:
\begin{equation}
\mathbf{y}(t) = \mathbf{H}_{\text{M}}(t) \otimes \mathbf{C}(t) + \mathbf{w}(t),
\end{equation}
By applying OTFS demodulation (i.e., the Wigner transform and SFFT), the receiver converts $\mathbf{y}(t)$ into the DD-domain received signal $\bar{\mathbf{y}}$, which can be written as follows: 
\begin{equation}
\bar{\mathbf{y}} = \mathbf{H}_{\text{M}}^{\text{eff}} \mathbf{s^{*}} + \mathbf{w}^{\text{eff}},
\end{equation}
Here, $\mathbf{s^{*}} \in \mathbb{C}^{2^{n_{RF}} NM \times1}$ is the vector formed by vertically concatenating all $s_i$ transmission vectors for each DD grid point into a single column vector, $\mathbf{H}_{\text{M}}^{\text{eff}}$ denotes the equivalent DD-domain channel matrix, and $\mathbf{w}^{\text{eff}}$ represents the equivalent DD-domain noise vector. Assuming that the receiver has perfect CSI, the ML detector of the proposed OTFS-MBM system, which jointly estimates the selected symbol and the active MAP index, can be expressed as follows:
\begin{equation}
[\hat{q}_i, \hat{\varpi}i] = \text{arg}\underset{q, \varpi}{\mathrm{min}} \left|\left|\bar{\mathbf{y}} - \mathbf{h}_{i,\varpi} x_q\right|\right|^2.
\end{equation}
Finally, the estimated symbols and MAP indices reconstruct the estimated data bit vector for each DD grid point.

\subsection{OTFS-based CIM System} 
CIM is a notable IM technique introduced in 2014 by Kaddoum et al. to enhance the efficiency and performance of direct-sequence spread spectrum (DS-SS) systems \cite{CIMilk}. CIM transmits additional information by encoding data in the indices of selected spreading codes, thereby increasing spectral efficiency without increasing the modulation order. Moreover, energy efficiency improves because fewer actively modulated symbols are required to convey the same information compared to conventional wireless communication techniques, which reduces power consumption. In the CIM system, using spreading codes for extra data transmission enables the use of simpler receivers, such as correlators and ML detectors, without requiring complex demodulation or additional RF chains. Assigning a unique spreading code to each transmitted signal improves detection reliability under multipath fading and interference, as the receiver can distinguish signals based on the mutual orthogonality of the codes. It also enhances physical-layer security, as varying the code indices for each transmission complicates unauthorized interception and decoding by potential eavesdroppers \cite{CIM1, CIM2, CIM3, CIM4, CIM5, CIM6, CIM7, CIM8, CIM9, CIM10}.

\cite{OTFS_IM12} proposes a new OTFS-IM scheme, termed OTFS-CIM, which integrates OTFS modulation for robust error performance in high-mobility Rayleigh channels with the CIM technique to enhance spectral and energy efficiency in a SIMO architecture. Comparative analyses show that the proposed OTFS-CIM system outperforms traditional OTFS and OTFS-SM systems in terms of error performance, throughput, spectral efficiency, and energy efficiency.  

In the proposed OTFS-CIM system, the transmitter uses a single transmit antenna, while the receiver employs $n_R$ receive antennas. Each OTFS frame transmits $NM$ information symbols mapped onto a DD grid with $N$ time slots and $M$ subcarriers. For each grid point, $\log_2(M_q)$ bits select an $M_q$-QAM symbol $s = s_\Re + js_\Im$, and $2\log_2(N_C)$ bits determine the indices of two orthogonal Walsh-Hadamard (WH) spreading codes for the $I$ and $Q$ components. Each spreading code has length $L$ chips with elements $\pm 1/\sqrt{L}$.
Thus, the spectral efficiency of the proposed system for an OTFS frame is:
\begin{equation}
\eta_{\text{OTFS-CIM}} = NM \Big( \log_2(M_q) + 2\log_2(N_C) \Big).
\end{equation}
In the proposed OTFS-CIM system, the real and imaginary symbol vectors in the DD domain, denoted by $\textbf{s}_\Re$ and $\textbf{s}_\Im$, each contain $NM$ complex $M_q$-QAM symbols. To obtain the corresponding time-frequency domain symbols, the ISFFT is applied. In this process, each symbol is multiplied by the selected WH spreading code before transformation. As a result, the real and imaginary components of the time-frequency domain information symbols, $S^\ell_\Re[n,m]$ and $S^\ell_\Im[n,m]$, are obtained as follows:
\begin{equation}
S^\ell_\Re[n,m] = \frac{1}{\sqrt{NM}} \sum_{k=0}^{N-1} \sum_{l=0}^{M-1} z_{c_\Re,\ell}\, s_\Re[k,l]\, e^{j2\pi(\frac{nk}{N} - \frac{ml}{M})},
\end{equation}
\begin{equation}
S^\ell_\Im[n,m] = \frac{1}{\sqrt{NM}} \sum_{k=0}^{N-1} \sum_{l=0}^{M-1} z_{c_\Im,\ell}\, s_\Im[k,l]\, e^{j2\pi(\frac{nk}{N} - \frac{ml}{M})}.
\end{equation}
Then, the time-frequency domain signals $S^\ell_\Re[n,m]$, $S^\ell_\Im[n,m]$ are transformed to the time domain signals $x^{\ell}_{\Re}(t)$, $x^{\ell}_{\Im}(t)$ by the Heisenberg transform. The subcarrier spacing is defined as $\Delta f = 1/T_c$, where $T_c$ denotes the chip duration, and the symbol duration is given by $T_s = L T_c$. At the receiver, the received signal for each chip of the $I$ and $Q$ components for the $v^{th}$ DD grid can be expressed as
\begin{equation}
\textbf{y}^I_{\ell,v} = s_\Re z_{c_\Re,\ell} \textbf{h}_v + \mathbf{w}^{\text{eff}}_I,
\end{equation}
\begin{equation}
\textbf{y}^Q_{\ell,v} = s_\Im z_{c_\Im,\ell} \textbf{h}_v + \mathbf{w}^{\text{eff}}_Q,
\end{equation}
where $\textbf{h}_v$ represents the Rayleigh channel coefficient for the $v^{th}$ grid point and $\mathbf{w}^{\text{eff}}_I$, $\mathbf{w}^{\text{eff}}_Q$ denote the effective noise vectors for $I$ and $Q$ components.
In the proposed OTFS-CIM system, a despreading-based ML detector is employed to estimate the transmitted QAM symbol and the active spreading code indices. At the receiver, the received $I$ and $Q$ signals are correlated with the entire WH code set. Thanks to the orthogonality, only the selected codes in the transmitter maximize the received signal norm. As a result, the estimated spreading code indices that yield the maximum norms are given as follows:
\begin{equation}
\hat{c}_\Re = \arg\max_{c} \| \tilde{\textbf{y}}^I_{v,c} \|^2,
\end{equation}
\begin{equation}
\hat{c}_\Im = \arg\max_{c} \| \tilde{\textbf{y}}^Q_{v,c} \|^2.
\end{equation}
Using the received signals corresponding to the estimated spreading code indices $\hat{c}_\Re$, $\hat{c}_\Im$, the ML detector estimates the transmitted QAM symbols for the $v$-th DD grid point as follows:
\begin{equation}
[\hat{s}_\Re, \hat{s}_\Im] = \arg\min_{s_\Re,s_\Im}
\Big\|
\big( \tilde{\textbf{y}}^I_{v,\hat{c}_\Re} + j\,\tilde{\textbf{y}}^Q_{v,\hat{c}_\Im} \big)
- E_z\, (s_\Re + j\,s_\Im)\,\textbf{h}_v
\Big\|^2,
\end{equation}
where $E_z$ represents the average energy of the WH spreading codes. Finally, the estimated QAM symbols and code indices are used to obtain the data bits in the all $NM$ DD domain.

\subsection{OTFS-based SKK System}
SSK, a member of the IM family, differs from SM in that only antenna indices are used for data transmission. This reduces the complexity of the SSK compared to the SM technique, but as a trade-off, the spectral efficiency is also reduced. In \cite{otfssk1}, SSK is presented as a modulation scheme that conveys information through antenna indices, achieving better performance than conventional amplitude- and phase-modulation techniques.

Combining SSK and OTFS techniques provides significant benefits such as high Doppler robustness and low energy consumption. Unlike the SM system, the SSK system transmits data only in the antenna indices, not in the symbol. Therefore, the spectral efficiency of the SSK technique is expressed as follows:
\begin{equation}
    n_{SSK}=\log_2(n_T)
\end{equation}
Similar to OTFS-SM, in the OTFS-based SSK technique, for a DD grid of size $N\times M$, the spectral efficiency of the OTFS-SSK system during an OTFS frame can be expressed as follows:
\begin{equation}
    n_{OTFS-SSK}=NM\log_2(n_T).
\end{equation}
Also, the transmission vector of the OTFS-SSK system is defined as follows:
\begin{eqnarray}\label{transmission_signal_ssk}
\textbf{x}_i = {\Bigg[ 0\, \,0\cdots 0\,  \, \cdots \underset{\stackrel{\uparrow}{\ell^{\text{th}} \ \text{position}}}{1} \cdots \, \,0\, \,\cdots 0\, \,0
	\Bigg]^T},
\end{eqnarray}
The transmission matrix of the OTFS-SSK system containing $NM$ DD grid points is expressed as follows:
\begin{equation} 
\mathbf{X} = [\mathbf{x}_1, \mathbf{x}_2, \cdots, \mathbf{x}_{NM} ],
\end{equation}
where each column has a $1$ at the $\ell^{\text{th}}$ position corresponding to the active transmit antenna, and $0$ in all other positions, for the $i$-th DD grid point. The transmission matrix $\mathbf{X}$ in the DD domain is first processed at the transmitter by the OTFS modulator. Specifically, each row of $\mathbf{X}$, corresponding to one active transmit antenna, is transformed to the time–frequency domain using the ISFFT. The resulting time–frequency domain signals are then converted to the continuous time-domain waveforms through the Heisenberg transform. These time-domain signals are transmitted from the active antennas. The transmitted waveforms propagate through the time-varying Rayleigh wireless channel, where each link between a transmit and receive antenna is characterized by distinct delay and Doppler parameters. At the receiver, AWGN is added to the received signals. The OTFS demodulator then processes the noisy received waveforms by first applying the Wigner transform to obtain the time–frequency domain samples, followed by the SFFT to map the signals back to the DD domain. Subsequently, based on the obtained DD domain signal at the receiver, the indices of the active transmit antennas are estimated. 

\begin{table}[t!]
\centering
\addtolength{\tabcolsep}{-2.5pt}
\caption{System Architectures of OTFS-IM Schemes.}
\label{system_arc_IM}
\begin{tabular}{|c|c|c|c|} 
\hline
\hline
\rowcolor{LightCyan}
 \textbf{\begin{tabular}[c]{@{}c@{}}\small SISO \\ \end{tabular}} & 
 \textbf{\begin{tabular}[c]{@{}c@{}} \small SIMO \\ \end{tabular}} & 
 \textbf{\begin{tabular}[c]{@{}c@{}} \small MISO \\ \end{tabular}} & 
 \textbf{\begin{tabular}[c]{@{}c@{}} \small MIMO \\ \end{tabular}}  \\ \hline\hline
 \rowcolor{Gray}
 \begin{tabular}[c]{@{}c@{}}  \cite{OTFS_IM1}, \cite{OTFS_IM2}, \cite{OTFS_IM3}, \cite{OTFS_IM4} \\ \cite{OTFS_IM5}, \cite{OTFS_IM6}, \cite{OTFS_IM7}, \cite{OTFS_IM8} \\ 
\cite{OTFS_IM9}, \cite{OTFS_IM10}, \cite{OTFS_IM13}, \cite{OTFS_IM15}  \\ 
\cite{OTFS_IM16},  \cite{OTFS_IM17}, \cite{OTFS_IM18},  \cite{OTFS_IM19} \\ 
\cite{OTFS_IM20}, \cite{OTFS_IM21}, \cite{OTFS_IM22}, \cite{OTFS_IM23} \\
\cite{OTFS_IM24}, \cite{OTFS_IM25}, \cite{OTFS_IM31},  \cite{OTFS_IM32} \\ 
\cite{OTFS_IM33}, \cite{OTFS_IM34}, \cite{OTFS_IM36},  \cite{OTFS_IM37} \\ 
\cite{OTFS_IM38}, \cite{OTFS_IM39}, \cite{OTFS_IM40}, \cite{OTFS_IM41} \\ 
\cite{OTFS_IM42}, \cite{OTFS_IM43}, \cite{OTFS_IM44}, \cite{OTFS_IM45} \\
\cite{OTFS_IM46},  \cite{OTFS_RIS_IM_1} \end{tabular} & 
\begin{tabular}[c]{@{}c@{}} \cite{OTFS_IM11} \\\cite{OTFS_IM12} \\ \cite{OTFS_IM14}
 \end{tabular} & 
\begin{tabular}[c]{@{}c@{}} -- \end{tabular} &   \begin{tabular}[c]{@{}c@{}}          \cite{OTFS_IM26}, \cite{OTFS_IM27}, \cite{OTFS_IM28} \\ \cite{OTFS_IM29},  \cite{OTFS_IM30},  \cite{OTFS_IM35} \\   \cite{OTFSQSM1}, \cite{OTFSQSM2},  \cite{OTFSQSM3} \\ \cite{SMOTFS1}, \cite{SMOTFS2}, \cite{SMOTFS3} \\ \cite{SMOTFS4}, \cite{SMOTFS5}, \cite{SMOTFS7} \\ \cite{SMOTFS8}, \cite{SMOTFS9},  \cite{SMOTFS10} \\ \cite{SMOTFS11},  \cite{SMOTFS12}, \cite{SMOTFS13} \\ \cite{SMOTFS14},\cite{SMOTFS15}, \cite{SMOTFS16} \end{tabular}   \\ \hline \hline
\end{tabular}
\end{table}

\begin{table*}[t!]
\centering
\caption{Detectors employed in OTFS-IM schemes.}
\label{table_detectors}
\setlength{\tabcolsep}{6pt}
\renewcommand{\arraystretch}{1.1}
\begin{tabularx}{\textwidth}{>{\columncolor{gray!15}\raggedright\arraybackslash}p{7.2cm}|>{\raggedright\arraybackslash}X}
\toprule
\rowcolor{red!80}
\textcolor{white}{\textbf{Detector}} & \textcolor{white}{\textbf{References}} \\
\midrule

\textbf{ML} & \scriptsize \makecell{\cite{OTFSQSM2}, \cite{OTFSQSM3}, \cite{OTFS_IM2}, \cite{OTFS_IM4}, \cite{OTFS_IM10}, \cite{OTFS_IM11},  \cite{OTFS_IM12}, \cite{OTFS_IM13},  \cite{OTFS_IM14}, \cite{OTFS_IM15},  \cite{OTFS_IM16}, \\ \cite{OTFS_IM17}, \cite{OTFS_IM19},  \cite{OTFS_IM22}, \cite{OTFS_IM23}, \cite{OTFS_IM26},   \cite{OTFS_IM29},  \cite{OTFS_IM30}, \cite{OTFS_IM31}, \cite{OTFS_IM34},  \cite{OTFS_IM35}, \cite{OTFS_IM36}, \\ \cite{OTFS_IM39}, \cite{OTFS_IM41}, \cite{OTFS_IM42}, \cite{SMOTFS2}, \cite{SMOTFS3}, \cite{SMOTFS4}, \cite{SMOTFS9}, \cite{SMOTFS11}, \cite{SMOTFS13}, \cite{SMOTFS14}, \cite{SMOTFS15}, \cite{SMOTFS16}}  \\  \hline

\textbf{MLJSAPD / CMP} & \scriptsize \makecell{\cite{OTFS_IM2}, \cite{OTFS_IM3}, \cite{OTFS_IM20}} \\ \hline

\textbf{MMSE} & \scriptsize \makecell{\cite{OTFSQSM2}, \cite{OTFS_IM14}, \cite{OTFS_IM6}, \cite{OTFS_IM19}, \cite{OTFS_IM24}, \cite{OTFS_IM29}, \cite{OTFS_IM33}, \cite{OTFS_IM36}, \cite{OTFS_IM37}, \cite{SMOTFS1}, \cite{SMOTFS5}, \cite{SMOTFS10}} \\ \hline

\textbf{LMMSE} & \scriptsize \makecell{\cite{OTFS_IM25}} \\ \hline

\textbf{Updated message passing} & \scriptsize \makecell{\cite{OTFS_IM16}} \\ \hline 

\textbf{Low-complexity message passing} & \scriptsize \makecell{\cite{OTFS_IM27}} \\ \hline 

\textbf{Block message passing} & \scriptsize \makecell{\cite{SMOTFS8}} \\ \hline 

\textbf{SS-EP / low-complexity SS-EP} & \scriptsize \makecell{\cite{OTFS_IM32}} \\ \hline

\textbf{LLR} & \scriptsize \makecell{\cite{OTFS_IM17}, \cite{OTFS_IM19}, \cite{OTFS_IM38}} \\ \hline

\textbf{Modified LLR} & \scriptsize \makecell{\cite{OTFS_IM7}} \\ \hline

\textbf{Jacobi preconditioning conjugate gradient} & \scriptsize \makecell{\cite{OTFS_IM24}} \\ \hline

\textbf{Low complexity detection strategy based on energy} & \scriptsize \makecell{\cite{OTFS_IM6}} \\ \hline

\textbf{Low complexity ML} & \scriptsize \makecell{\cite{OTFS_IM9}} \\ \hline

\textbf{MMSE-ML} & \scriptsize \makecell{\cite{OTFS_IM5}, \cite{OTFS_IM21}} \\ \hline

\textbf{MLMPD} & \scriptsize \makecell{\cite{OTFS_IM20}} \\ \hline

\textbf{EMMSE} & \scriptsize \makecell{\cite{OTFSQSM1}} \\ \hline

\textbf{\makecell[l]{Matched-filtered Gauss-Seidel\\with ML/low-complex greedy}} & \scriptsize \makecell{\cite{OTFS_IM1}, \cite{OTFS_IM18}, \cite{OTFS_IM40}} \\ \hline

\textbf{Two stage detection} & \scriptsize \makecell{\cite{OTFS_IM28}} \\ \hline

\textbf{Distance-based signal detection} & \scriptsize \makecell{\cite{OTFS_IM33}, \cite{SMOTFS7}} \\ \hline

\textbf{VV-MP} & \scriptsize \makecell{\cite{OTFS_IM37}} \\ \hline

\textbf{Gradient descent} & \scriptsize \makecell{\cite{OTFS_IM41}} \\ \hline

\textbf{Successive interference cancellation-based MMSE} & \scriptsize \makecell{\cite{OTFS_IM42}} \\ \hline

\textbf{Decision feedback based on MMSE} & \scriptsize \makecell{\cite{SMOTFS1}} \\ \hline 

\textbf{MMSE-ML based energy greedy detector} & \scriptsize \makecell{\cite{OTFS_IM21}} \\ \hline 

\textbf{Sparse signal estimation-based detector} & \scriptsize \makecell{\cite{SMOTFS5}} \\ 

\bottomrule
\end{tabularx}
\end{table*}

Finally, for each DD grid point, the active transmit antenna index is estimated using the ML detector, expressed as
\begin{equation}
\hat{\ell}_i = \arg\min_{\ell \in \{1,\dots,n_T\}} \left\| \mathbf{y}_i - \mathbf{h}_{i,\ell} \right\|^2,
\end{equation}
where $\mathbf{y}_i$ is the received DD-domain vector for the $i$-th grid point, and $\mathbf{h}_{i,\ell}$ denotes the $\ell$-th column of the corresponding DD-domain channel matrix $\mathbf{H}_i$. The ML detector of the OTFS-SSK system operates $NM$ times to estimate the active transmit antenna indices.  
As a result, the estimated active transmit antenna indices corresponding to all $NM$ DD grid points in an OTFS frame are expressed as follows:
\begin{equation}\label{eq_est_ell}
\hat{\boldsymbol{\ell}} = [\hat{\ell}_1, \hat{\ell}_2, \cdots, \hat{\ell}_{NM}].
\end{equation}
Finally, the estimated bit sequence vector is obtained by mapping the estimated antenna indices $\hat{\boldsymbol{\ell}}$ back to the corresponding bit combinations using the SSK demapper.

\definecolor{lightblue}{RGB}{204, 229, 255}
\definecolor{lightgreen}{RGB}{204, 255, 204}
\definecolor{lightred}{RGB}{255, 204, 204}
\definecolor{headerblue}{RGB}{0, 102, 204}
\captionsetup[table]{justification=centering, labelsep=colon}

\settowidth\rotheadsize{Complexity}
\begin{table}[t!]
\centering
\addtolength{\tabcolsep}{-2.5pt}
\caption{The performance analyses in OTFS-IM systems.}
\label{performace_OTFS_IMs}
\begin{tabular}{|c|c|c|c|c|c|} 
\hline
\hline
\rowcolor{LightCyan}
 \textbf{\begin{tabular}[c]{@{}c@{}}\small \rothead{Capacity}  \\ \end{tabular}} & 
 \textbf{\begin{tabular}[c]{@{}c@{}} \small \rothead{PAPR}  \\ \end{tabular}} & 
 \textbf{\begin{tabular}[c]{@{}c@{}} \small \rothead{Diversity}  \\ \end{tabular}} & 
 \textbf{\begin{tabular}[c]{@{}c@{}} \small \rothead{Complexity}  \\ \end{tabular}} & 
 \textbf{\begin{tabular}[c]{@{}c@{}} \small \rothead{Imperfect \\ CSI}  \\ \end{tabular}} & 
 \textbf{\begin{tabular}[c]{@{}c@{}} \small \rothead{Outage \\  Probability}  \\ \end{tabular}}\\ \hline\hline
 \rowcolor{Gray}
 \begin{tabular}[c]{@{}c@{}}  \cite{SMOTFS7} \\ \cite{SMOTFS9} \\ \cite{SMOTFS15} \\ \cite{SMOTFS16} \\ \cite{OTFS_IM31} \end{tabular} & 
\begin{tabular}[c]{@{}c@{}} \cite{OTFS_IM9} \\ \cite{OTFS_IM34}\\ \cite{OTFS_IM39}
 \end{tabular} & 
\begin{tabular}[c]{@{}c@{}} \cite{OTFS_IM16} \\ \cite{OTFS_IM23} \\ \cite{OTFS_IM26}\\
  \cite{SMOTFS11} \\ \cite{OTFSQSM2} \end{tabular} &   \begin{tabular}[c]{@{}c@{}}          \cite{OTFS_IM1}, \cite{OTFS_IM2}, \cite{OTFS_IM5}\\
\cite{OTFS_IM6}, \cite{OTFS_IM7}, \cite{OTFS_IM9} \\ \cite{OTFS_IM10},
\cite{OTFS_IM15}, \cite{OTFS_IM16} \\  \cite{OTFS_IM17}, \cite{SMOTFS5},
\cite{OTFS_IM20} \\ \cite{OTFS_IM21},  \cite{OTFS_IM24}, \cite{OTFS_IM26} \\ \cite{OTFS_IM27},  \cite{SMOTFS7}  
\cite{OTFS_IM28}  \\ \cite{OTFS_IM32}, \cite{OTFSQSM1},  \cite{SMOTFS11} \\
\cite{OTFS_IM33}, \cite{OTFS_IM37},  \cite{OTFS_IM38} \\ \cite{OTFS_IM42}, \cite{SMOTFS5} \end{tabular}   &   \begin{tabular}[c]{@{}c@{}}          \cite{OTFS_IM2} \\ \cite{OTFS_IM3}\\
\cite{OTFS_IM36} \\ \cite{OTFS_IM40} \\ \cite{SMOTFS12} \\ \cite{SMOTFS16} \end{tabular} 
&   \begin{tabular}[c]{@{}c@{}}          \cite{SMOTFS15} \\ \cite{SMOTFS16} \end{tabular} \\ \hline \hline
\end{tabular}
\vspace{-1em}
\end{table}

In Table \ref{system_arc_IM}, the system architectures of OTFS-IM schemes are classified into SISO, SIMO, MISO, and MIMO configurations, with the relevant literature references provided for each category. As observed in Table \ref{system_arc_IM}, fewer works have investigated SIMO systems, while no study has been reported for MISO configurations. In contrast, a significant number of contributions have addressed SISO and MIMO architectures.

Table \ref{table_detectors} lists the detectors used in OTFS-IM schemes, ranging from classical methods to advanced and hybrid approaches. As presented in Table \ref{table_detectors}, ML and MMSE are the most frequently utilized detectors in OTFS-IM schemes.

Table \ref{performace_OTFS_IMs} presents the key performance analyses in OTFS-IM systems in the literature. The studies are categorized according to the performance metrics they address, including capacity, PAPR, diversity, computational complexity, imperfect CSI, and outage probability.

\section{Performance analyses of the OTFS-IM Systems}
This section presents a comprehensive performance evaluation of conventional OTFS and various OTFS-IM schemes, including analyses of computational complexity, error performance, capacity, energy saving, spectral efficiency, and throughput under different system configurations. The system parameters used for all systems in the comparisons are as follows: carrier frequency of $4\text{GHz}$, subcarrier spacing of $15\text{kHz}$, maximum speed of $506.2\text{km/h}$, and number of channel paths equal to $6$.

\vspace{-1em}

\subsection{Computational Complexity Derivations Of OTFS-IM systems}
In this section, the computational complexities of OTFS, OTFS-SSK, OTFS-SM, OTFS-QSM, OTFS-MBM, and OTFS-CIM systems are derived. Also, the considered systems are compared for various scenarios with the same spectral efficiency.

The ML detector for the SM system jointly searches across all possible transmitted symbol values and transmit antenna indices.  Specifically, for each point on the DD grid, the ML detector jointly evaluates all possible combinations of $M_Q$ constellation symbols and $n_T$ transmit antennas. This means that for each DD grid, there are $M_Q n_T$ possible cases to evaluate, since the detector tries every combination of symbol and transmit antenna to estimate a symbol and transmit antenna index. Therefore, the ML metric in (\ref{eq_ML_SM}) is jointly calculated for the $M_Q n_T$ case. As seen in (\ref{eq_ML_SM}), each metric calculation involves the multiplication of two complex expressions, $s_q \in \mathbb{C}^{1 \times 1}$ and $\textbf{h}_{i,\ell} \in \mathbb{C}^{n_RNM \times 1}$. Since $\mathbf{h}_{i,\ell}$ contains $n_RNM$ complex numbers, a total of $n_RNM$ complex number multiplications are performed for each metric calculation. The multiplication of two complex numbers results in $4$ real multiplications (RMs). Also, the squared norm of a complex number $z = a + jb$ is calculated as $|z|^2 = a^2 + b^2$, which requires $2$ RMs. Therefore, the squared norm requires $2n_R NM$ RMs. As a result, for the $i\text{th}$ DD grid, the complexity of the ML detector of the SM system in RMs is obtained as follows: 
\begin{equation} \label{comp_sm_dd1}
\mathcal{O}_{\mathrm{OTFS-SM-ML}}^{(i)} = M_Q n_T \times 4 \times  2n_R NM = 8 n_R NM M_Q n_T.
\end{equation}
To estimate the $NM$ symbols and transmit antenna indices corresponding to all DD grids, the ML metric in (\ref{eq_ML_SM}) is evaluated $N \times M$ times. Therefore, the computational complexity of the ML detector of the OTFS-SM system for an OTFS frame, in terms of RMs, is obtained as follows:
\begin{equation} \label{comp_sm_dd2}
\mathcal{O}_{\mathrm{OTFS-SM-ML}} = 8 n_R (NM)^2 M_Q n_T.
\end{equation}
As shown in \cite{QSMilk}, the ML detectors of SM and QSM systems have the same computational complexity. Therefore, the complexity of the ML detector in the QSM system is expressed as follows:
\begin{equation} \label{comp_qsm_dd}
\mathcal{O}_{\mathrm{OTFS-QSM-ML}} = 8 n_R (NM)^2 M_Q n_T.
\end{equation}

In the SSK technique, unlike the SM technique, information is carried only in the indices of the transmit antennas, and not in the symbols. Therefore, the computational complexity of the ML detector of the SSK system for an OTFS frame can be expressed in terms of RMs as follows:
\begin{equation} \label{comp_ssk_dd}
\mathcal{O}_{\mathrm{OTFS-SSK-ML}} = 8 n_R (NM)^2 n_T.
\end{equation}

In the MBM technique, unlike in SM, additional information is conveyed through the indices of MAP states, which are formed by the on/off configurations of RF mirrors, rather than using transmit antenna indices. Therefore, since the ML detector jointly considers the existing $2^{n_{RF}}$ MAP states and $M_Q$ symbols, the computational complexity of the ML detector of the MBM system is given as follows:
\begin{equation} \label{comp_mbm_dd3}
\mathcal{O}_{\mathrm{OTFS-MBM-ML}} = 8 n_R (NM)^2 M_Q 2^{n_{RF}}.
\end{equation}

In the proposed OTFS-CIM system in \cite{OTFS_IM12}, a despreading-based ML detector is used to estimate the spreading codes and the symbol. The detector first estimates the spreading code indices, then uses them to estimate the symbols. In (15) of the \cite{OTFS_IM12}, to obtain the expression $\tilde{\textbf{y}}^I_{v} \in \mathbb{C}^{n_RNM \times 1}$ for the $I$ component, $\textbf{Y}^I_{v} \in \mathbb{C}^{n_RNM \times L}$ is multiplied by $\textbf{z}_{c} \in \mathbb{R}^{L \times 1}$. Since there are $N_C$ spreading codes, a total of $2 N_C L n_RNM$ RMs arise. Subsequently, in (19) of \cite{OTFS_IM12}, for spreading code estimation, it is necessary to calculate the squared norm of the expression $\tilde{\textbf{y}}^I_{v} \in \mathbb{C}^{n_RNM \times 1}$. As explained above, the norm square operator yields $2n_R NM$ RMs. As a result, for spreading code estimation of the $I$ component, a total of $4 N_C L (n_RNM)^2$ RMs are required. Since the same operations are performed for the $Q$ component, the computational complexity arising from the estimation of the spreading code indices in the OTFS-CIM system is presented in terms of RMs as follows:
\begin{equation} \label{comp_mbm_dd1}
\mathcal{O}_{c} = 8 N_C L  (n_RNM)^2.
\end{equation}
In the OTFS-CIM system, symbol estimation is performed based on the estimated spreading code indices. The conventional ML detector complexity for symbol estimation is $8n_RNM M_Q$. Consequently, the computational complexity of the despreading-based ML detector used by the OTFS-CIM system for an OTFS frame in terms of RMs is given as follows:
\begin{equation} \label{comp_mbm_dd2}
\mathcal{O}_{\mathrm{OTFS-CIM-ML}} = NM \Big( 8 N_C L  (n_RNM)^2 + 8n_RNM M_Q \Big).
\end{equation}
Using a similar approach, the ML complexity of the conventional OTFS technique for SIMO systems can be given in terms of RMs as follows:
\begin{equation} \label{comp_otfs_dd}
\mathcal{O}_{\mathrm{OTFS-ML}} = 8 n_R (NM)^2 M_Q.
\end{equation}

\begin{figure}[t!]
\centering{\includegraphics[width=0.47\textwidth]{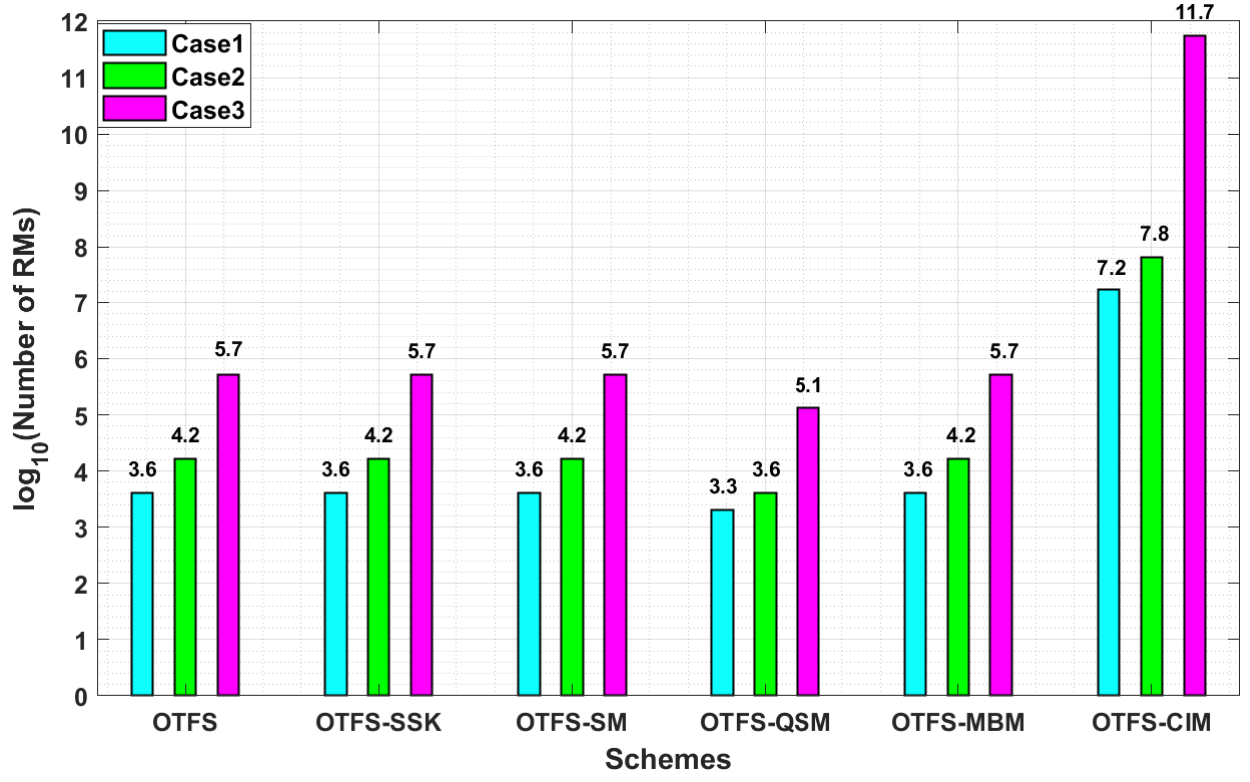}}	
	\caption{Computational complexity comparisons for the same spectral efficiency.}
	\label{OTFS_IM_comps} 
    \vspace{-1em}
\end{figure}

Fig. \ref{OTFS_IM_comps} shows the computational complexity order comparison of conventional OTFS and various OTFS-IM schemes for the same spectral efficiency under three different cases. In all cases, the Doppler and delay bins are set as $N = 2$, $M = 2$ for Case 1 and Case 2, and $N = 4$, $M = 4$ for Case 3. The number of receive antennas is $n_R = 2$ in Cases 1 and 2, and $n_R = 4$ in Case 3. The OTFS employs $n_T = 1$ and $M_Q = 16$ (Case 1), $M_Q = 64$ (Cases 2 and 3); OTFS-SSK uses $n_T = 16$ (Case 1), $n_T = 64$ (Cases 2 and 3); OTFS-SM is configured with $n_T = 4$, $M_Q = 4$ (Case 1), $n_T = 16$, $M_Q = 4$ (Cases 2 and 3); OTFS-QSM has $n_T = 2$, $M_Q = 4$ (Case 1), $n_T = 4$, $M_Q = 4$ (Cases 2 and 3); OTFS-MBM employs $n_{RF} = 2$, $M_Q = 4$ (Case 1), $n_{RF} = 4$, $M_Q = 4$ (Cases 2 and 3); and OTFS-CIM uses $n_C = 2$, $L = 4$, $M_Q = 4$ (Case 1), $n_C = 4$, $L = 8$, $M_Q = 4$ (Cases 2 and 3). In Fig. \ref{OTFS_IM_comps}, all schemes have the same spectral efficiency for each case: $16$ bits per channel use (bpcu) in Case 1, $24$ bpcu in Case 2, and $96$ bpcu in Case 3. Fig. \ref{OTFS_IM_comps} shows that the OTFS-CIM system exhibits the highest computational complexity among all the schemes, especially for large system parameters. It is also observed that OTFS-QSM has the lowest computational complexity among all schemes for each case, while the OTFS, OTFS-SSK, OTFS-SM, and OTFS-MBM schemes have identical computational complexity.

\subsection{Error Performance Comparisons of OTFS-IM Systems}
In this section, the error performance results of traditional OTFS and various OTFS-IM schemes are presented and compared over high-mobility Rayleigh fading channels. All simulations use the Monte Carlo method and are performed in MATLAB. Unless otherwise specified, the ML detector is employed at the receiver side for all schemes, except for the OTFS-CIM technique, which utilizes the despreading-based ML detector. The signal-to-noise ratio (SNR) is defined as $\mathrm{SNR~(dB)} = 10 \log_{10} \left( \frac{E_b}{N_0} \right)$, where $E_b$ denotes the bit energy.

\begin{figure}[t!]
    \centering
    \includegraphics[width=0.9\linewidth]{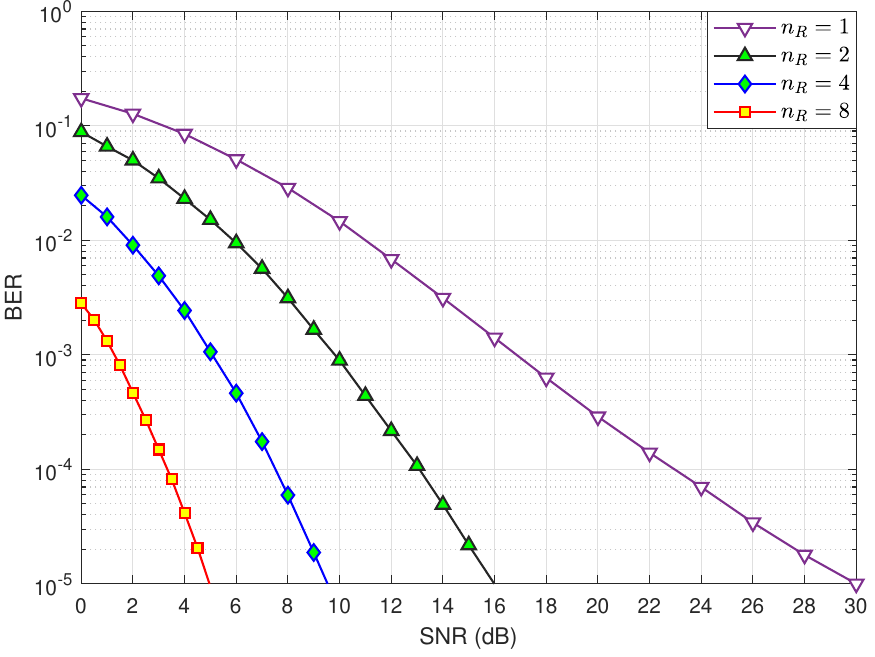}
    \caption{Error performance of the traditional OTFS system in SISO and SIMO schemes.}
    \label{OTFSNRs}
\end{figure}

Fig.~\ref{OTFSNRs} shows the error performance of the traditional OTFS system for different numbers of receive antennas ($n_R=2,4,8$) in SISO ($n_R = 1$) and SIMO configurations. The OTFS system employs the system parameters $N = 2$ and $M = 2$. As $n_R$ increases, the error performance improves due to the spatial diversity gain. The results show that multiple receive antennas significantly enhance OTFS performance by exploiting spatial diversity, leading to improved reliability under challenging channel conditions.

\begin{figure}[t!]
\centering{\includegraphics[width=0.45\textwidth]{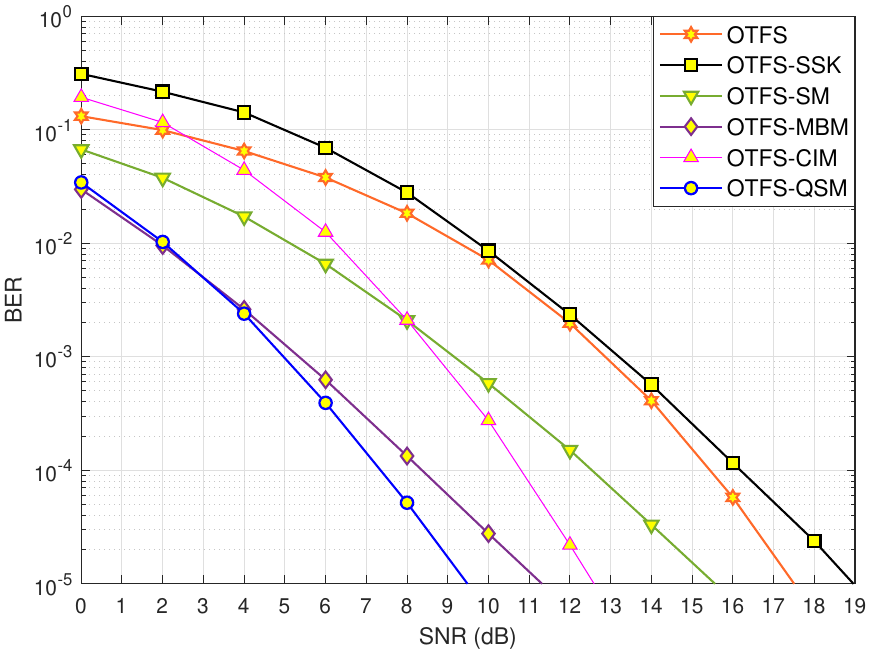}}	
	\caption{Error performance comparison of conventional OTFS and various OTFS-IM schemes for $24$ bpcu.}
	\label{OTFS_IMs1} 
\end{figure}

Fig. \ref{OTFS_IMs1} presents the error performance comparison of conventional OTFS and various OTFS-IM schemes under high-mobility Rayleigh channels for $24$ bpcu. It illustrates the comparative BER curves for conventional OTFS, OTFS-SSK, OTFS-SM, OTFS-MBM, OTFS-CIM, and OTFS-QSM systems. For all schemes, $n_R = 4$, $N = 2$, and $M = 2$, while the OTFS system uses $n_T = 1$, $M_Q = 64$, the OTFS-CIM system uses $n_T =1$, $M_Q = 4$, $n_C = 4$; the OTFS-SM and OTFS-QSM systems use $n_T = 4$ with $M_Q = 16$ and $M_Q = 4$, respectively, the OTFS-SSK system uses $n_T = 64$, and the OTFS-MBM system employs $n_{RF} = 4$, $M_Q = 4$, $n_T =1$. It can be observed that OTFS-QSM and OTFS-MBM achieve better error performance for 24 bits, as they utilize a larger modulation space through richer index domains. In contrast, the conventional OTFS and OTFS-SSK schemes demonstrate inferior error performance due to the absence of enhanced index resources. As a result, Fig. \ref{OTFS_IMs1} shows that integrating OTFS with advanced IM schemes enhances system reliability under high-mobility channel conditions.

\begin{figure}[t!]
\centering{\includegraphics[width=0.45\textwidth]{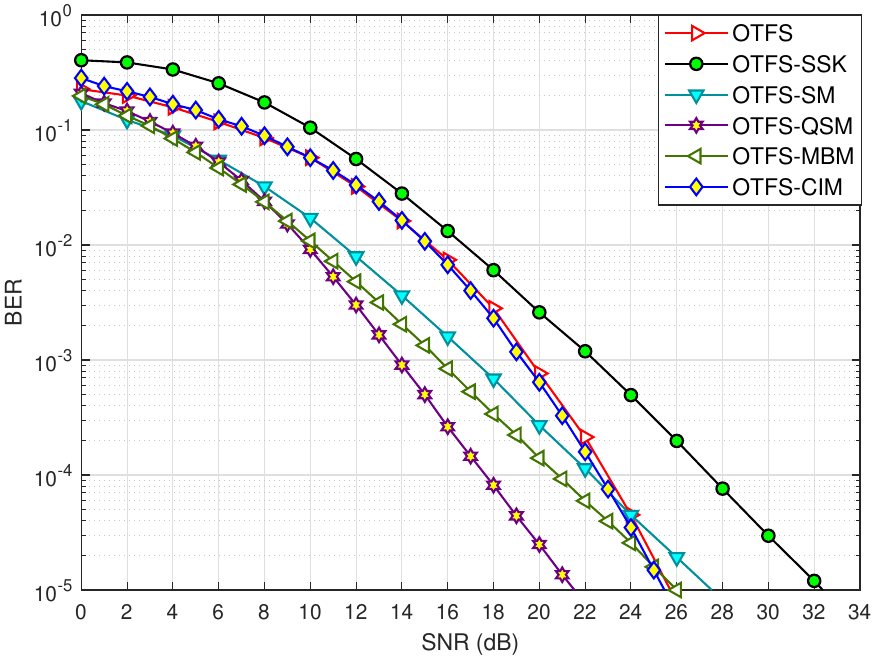}}	
	\caption{Error performance comparison of conventional OTFS and various OTFS-IM schemes for $28$ bpcu.}
	\label{OTFS_IMs2} 
\end{figure}

Similarly, Fig. \ref{OTFS_IMs2} compares the BER performance of conventional OTFS, OTFS-SSK, OTFS-SM, OTFS-MBM, OTFS-CIM, and OTFS-QSM systems for $28$ bpcu. All schemes employ $n_R = 2$, $N = 2$, and $M = 2$ receive antennas. Specifically, conventional OTFS uses $n_T = 1$ and $M_Q = 128$; OTFS-CIM employs $n_T = 1$, $M_Q = 8$, $L=8$, and $n_C = 4$; OTFS-SM and OTFS-QSM use $n_T = 4$, with $M_Q = 32$ for OTFS-SM and $M_Q = 8$ for OTFS-QSM; OTFS-SSK has $n_T = 128$; and OTFS-MBM utilizes $n_T = 1$, $n_{RF} = 4$, and $M_Q = 8$. As can be seen from Fig. \ref{OTFS_IMs2}, the OTFS-QSM provides the best performance, followed by OTFS-CIM, conventional OTFS, OTFS-MBM, OTFS-SM, and OTFS-SSK. Compared to Fig. \ref{OTFS_IMs1}, OTFS-IM systems do not exhibit the expected high performance sufficiently in Fig. \ref{OTFS_IMs2} due to the reduced number of bits carried in the indices. Also, in the OTFS-SSK system, it is observed that a very high number of transmit antennas ($n_T = 128$) are used, and therefore, the correlation between antennas increases greatly, thus reducing diversity and decreasing performance.

\begin{figure}[t!]
  \centering
  \includegraphics[width=0.9 \linewidth]{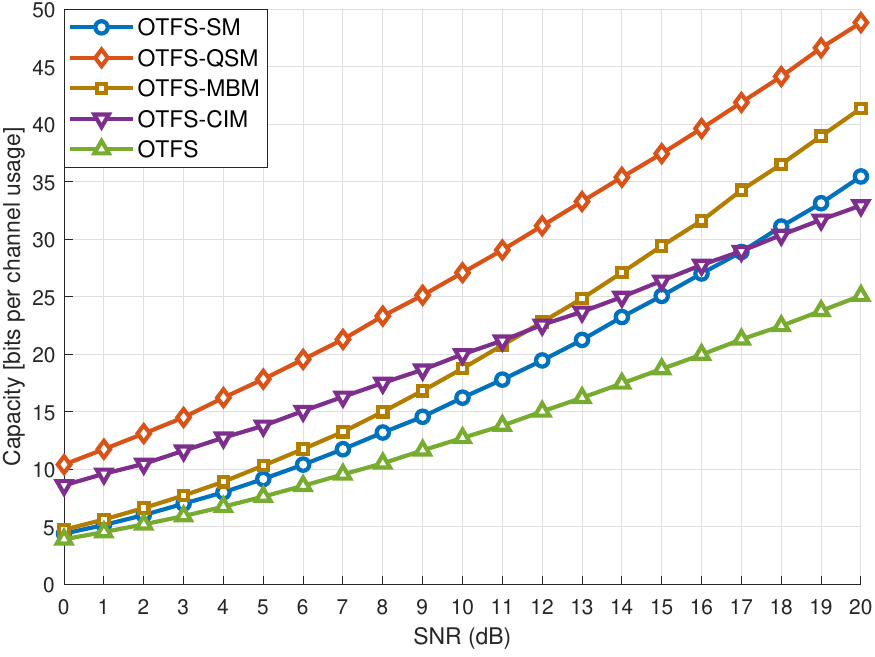}
  \caption{Capacity comparison of OTFS and OTFS-IM schemes.}
  \label{fig:otfs-capacity}
\end{figure}


\subsection{Capacity Comparisons of OTFS-IM Systems}
This section presents a derivation of the ergodic capacity of a MIMO system under the assumption that the receiver has perfect CSI and the channel is modeled as flat-fading. Furthermore, the capacities of OTFS-IM systems are compared for different SNR values. However, detailed capacity derivations of OTFS-IM systems are not provided due to concerns about exceeding the scope of this survey. 

The instantaneous channel capacity of the MIMO system is given as follows \cite{capacity1, capacity2}:
\begin{equation}\label{channel1}
\mathcal{C} =  \max_{\substack{\mathbf{Q}\succeq 0\\\operatorname{tr}(\mathbf{Q})\le P_T}} \log_2 \det \left( \mathbf{I}_{n_R} + \frac{1}{\sigma^2} \mathbf{H} \mathbf{Q} \mathbf{H}^H  \right) \text{bits/s/Hz}
\end{equation}
where $\mathbf{H} \in \mathbb{C}^{n_R \times n_T}$ is the MIMO channel matrix (e.g., Rayleigh fading channel), $\sigma^2$ is the noise power, $\mathbf{I}_{n_R}$ denotes the identity matrix of size $n_R$, $P_T$ is the average transmit power, and $\mathbf{Q} = \mathbb{E}[\mathbf{x} \mathbf{x}^H]$. If the transmitter does not have CSI, the covariance matrix $\mathbf{Q}  =\frac{P_T}{n_T} \mathbf{I}_{n_T}$. $\mathbf{Q}$ and SNR expression $\rho = \frac{P_T}{\sigma^2}$ are substituted into (\ref{channel1}), and the instantaneous channel capacity of the MIMO system  can be rewritten as follows \cite{capacity3, capacity4}:
\begin{equation}
\mathcal{C} = \log_2 \det \left( \mathbf{I}_{n_R} + \frac{P_T}{n_T \sigma^2} \mathbf{H} \mathbf{H}^H \right)
\end{equation}
The ergodic capacity is obtained by averaging the instantaneous capacity over all possible realizations of the random channel matrix. As a result, the ergodic capacity of the MIMO system is given as follows:
\begin{equation}
\mathcal{C} = \mathbb{E}_{\mathbf{H}} \left[ \log_2 \det \left( \mathbf{I}_{n_R} + \frac{\rho}{n_T} \mathbf{H} \mathbf{H}^H \right) \right]
\end{equation}

Fig.~\ref{fig:otfs-capacity} presents the capacity performance comparison of conventional OTFS and OTFS-SM, OTFS-QSM, OTFS-MBM, and OTFS-CIM schemes under high-mobility Rayleigh channels with system parameters $N = 2$, $M = 2$, $n_R = 2$, $n_T = 2$, $n_C = 2$, and $L = 4$. The OTFS-QSM and OTFS-MBM schemes achieve the highest capacity due to their ability to exploit richer index resources. The OTFS-CIM and OTFS-SM systems outperform conventional OTFS system due to their ability to convey extra information through index selection. The conventional OTFS scheme, which does not leverage any IM, exhibits the lowest capacity. Overall, the results demonstrate that integrating OTFS with advanced IM techniques can significantly enhance capacity under high-mobility conditions.

\begin{table*}[h!]
\centering
\caption{Energy-saving percentages of OTFS-IM systems compared to the conventional OTFS system under different system parameters (Cases 1–3).}
\label{tab:energy_saving}
\renewcommand{\arraystretch}{1.4}
\setlength{\tabcolsep}{30pt}
\begin{tabular}{|cccc|}
\rowcolor{headerblue}
\textbf{\textcolor{white}{Cases}} &
\textbf{\textcolor{white}{Case 1}} &
\textbf{\textcolor{white}{Case 2}} &
\textbf{\textcolor{white}{Case 3}} \\
\rowcolor{lightblue}
\multicolumn{1}{c}{Schemes}
& \cellcolor{lightblue} \makecell{\scriptsize($M_Q$=4, $n_T$=4, $N$=2)\\ \scriptsize($M$=2, $n_{RF}$=4, $N_C$=4)}
& \cellcolor{lightblue} \makecell{\scriptsize($M_Q$=8, $n_T$=8, $N$=4)\\ \scriptsize($M$=4, $n_{RF}$=8, $N_C$=8)}
& \cellcolor{lightblue} \makecell{\scriptsize($M_Q$=8, $n_T$=16, $N$=8)\\ \scriptsize($M$=8, $n_{RF}$=16, $N_C$=16)} \\
\midrule
\textbf{OTFS-SSK}
& \cellcolor{lightred}0.00\%
& \cellcolor{lightred}0.00\%
& \cellcolor{lightgreen}25.00\% \\
\textbf{OTFS-SM}
& \cellcolor{lightgreen}50.00\%
& \cellcolor{lightgreen}50.00\%
& \cellcolor{lightgreen}57.14\% \\
\textbf{OTFS-QSM}
& \cellcolor{lightgreen}66.67\%
& \cellcolor{lightgreen}66.67\%
& \cellcolor{lightgreen}72.73\% \\
\textbf{OTFS-MBM}
& \cellcolor{lightgreen}66.67\%
& \cellcolor{lightgreen}72.73\%
& \cellcolor{lightgreen}84.21\% \\
\textbf{OTFS-CIM}
& \cellcolor{lightgreen}66.67\%
& \cellcolor{lightgreen}66.67\%
& \cellcolor{lightgreen}72.73\% \\
\bottomrule
\end{tabular}
 \vspace{-1em}
\end{table*}

\begin{figure}[t]
\centering{\includegraphics[width=0.49\textwidth]{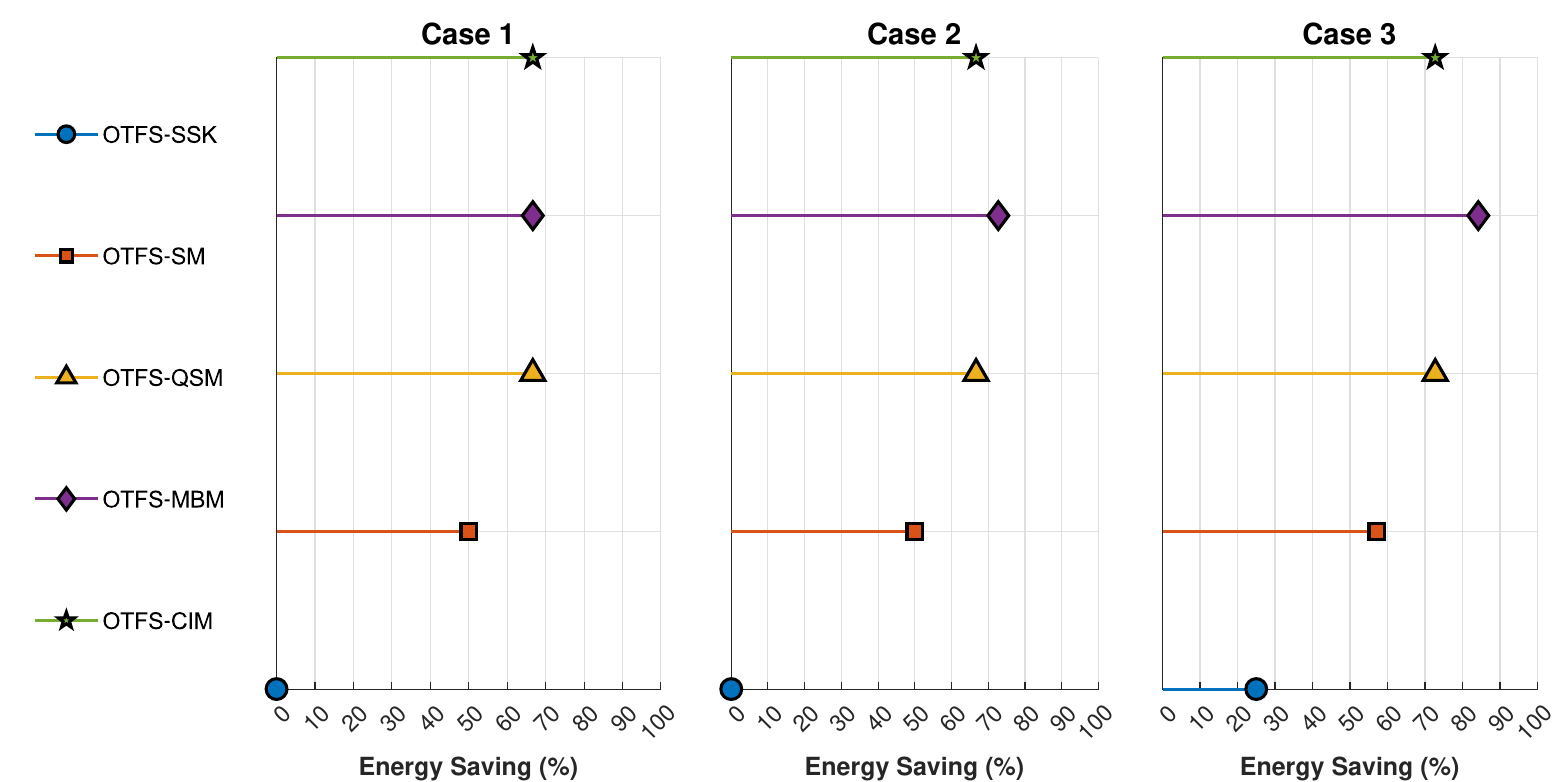}}	
	  \caption{Energy-saving percentages of OTFS-IM systems compared to the conventional OTFS system.}
	\label{OTFS_IM_energy} 
\end{figure}

\subsection{Energy Saving Comparisons of OTFS-IM Systems}
Energy saving is crucial for modern wireless systems to remain efficient and sustainable. OTFS-IM systems significantly reduce energy consumption compared to traditional OTFS systems by carrying most of the data in indices. The energy saving percentage $E_{\text{sav}}$ of the OTFS-IM systems is calculated as follows:
\begin{equation}\label{energy}
E_{\text{sav}}=\Big(1-\frac{\eta_{\text{OTFS}}}{\eta_\text{OTFS-IM}}\Big)E_b\%,
\end{equation}
where $\eta_\text{OTFS-IM}$ represents the spectral efficiency of the OTFS system, while $\eta_\text{OTFS-IM}$ represents the spectral efficiencies of the OTFS-IM systems. The energy saving percentages provided by OTFS-SSK, OTFS-SM, OTFS-QSM, OTFS-MBM, and OTFS-CIM systems compared to the traditional OTFS system are presented in Table~\ref{tab:energy_saving} and Fig.~\ref{OTFS_IM_energy} for three cases with the following system parameters: Case 1: ($M_Q=4$, $n_T=4$, $N=2$, $M=2$, $n_{RF}=4$, $N_C=4$);
Case 2: ($M_Q=8$, $n_T=8$, $N=4$, $M=4$, $n_{RF}=8$, $N_C=8$);
Case 3: ($M_Q=8$, $n_T=16$, $N=8$, $M=8$, $n_{RF}=16$, $N_C=16$). Table~\ref{tab:energy_saving} and Fig.~\ref{OTFS_IM_energy} show that the OTFS-MBM, OTFS-QSM, and OTFS-CIM systems achieve the highest energy saving percentages, especially in Case 3.  In Case 3, the OTFS-MBM system provides the maximum energy saving at $84.21\%$, while the OTFS-QSM and OTFS-CIM systems achieve $72.73\%$ energy saving. Overall, the findings clearly indicate that all OTFS-IM schemes provide high energy saving, and OTFS-IM techniques are significantly more energy-efficient than conventional OTFS.

\begin{table*}[t!]
\centering
\addtolength{\tabcolsep}{-1pt}
\caption{Spectral efficiency comparisons of OTFS-IM schemes.}
\label{SE_OTFS_IM_nu}
\begin{tabular}{|c|c|c|c|c|c|c|c|c|c|}
\hline
\textbf{Scheme} & \textbf{Formula} & \textbf{Cases} & $N$ & $M$ & $M_Q$ & $n_T$ & $n_{RF}$ & $n_C$ & \textbf{Spectral Efficiency} \\ \hline
\multirow{3}{*}{OTFS} 
  & \multirow{3}{*}{$\eta_{\text{OTFS}} = NM\,\log_2(M_Q)$} 
  & Case 1 & 2 & 2 & 4  & 1 & - & - & 8  \\ \cline{3-10}
  &  & Case 2 & 4 & 4 & 4  & 1 & - & - & 32 \\ \cline{3-10}
  &  & Case 3 & 4 & 4 & 8  & 1 & - & - & 48 \\ \hline

\multirow{3}{*}{OTFS-SSK} 
  & \multirow{3}{*}{$\eta_{\text{OTFS-SSK}} = NM\,\log_2(n_T)$} 
  & Case 1 & 2 & 2 & - & 2 & - & - & 4  \\ \cline{3-10}
  &  & Case 2 & 4 & 4 & - & 4 & - & - & 32 \\ \cline{3-10}
  &  & Case 3 & 4 & 4 & - & 8 & - & - & 48 \\ \hline

\multirow{3}{*}{OTFS-SM} 
  & \multirow{3}{*}{$\eta_{\text{OTFS-SM}} = NM\,\log_2(M_Q n_T)$} 
  & Case 1 & 2 & 2 & 4 & 2 & - & - & 12 \\ \cline{3-10}
  &  & Case 2 & 4 & 4 & 4 & 4 & - & - & 64 \\ \cline{3-10}
  &  & Case 3 & 4 & 4 & 8 & 8 & - & - & 96 \\ \hline

\multirow{3}{*}{OTFS-QSM} 
  & \multirow{3}{*}{$\eta_{\text{OTFS-QSM}} = NM\,\bigl[\log_2(M_Q) + 2\log_2(n_T)\bigr]$} 
  & Case 1 & 2 & 2 & 4 & 2 & - & - & 16 \\ \cline{3-10}
  &  & Case 2 & 4 & 4 & 4 & 4 & - & - & 96 \\ \cline{3-10}
  &  & Case 3 & 4 & 4 & 8 & 8 & - & - & 144 \\ \hline

\multirow{3}{*}{OTFS-MBM} 
  & \multirow{3}{*}{$\eta_{\text{OTFS-MBM}} = NM\,\bigl[\log_2(M_Q) + n_{RF}\bigr]$} 
  & Case 1 & 2 & 2 & 4 & 1 & 2 & - & 16 \\ \cline{3-10}
  &  & Case 2 & 4 & 4 & 4 & 1 & 4 & - & 96 \\ \cline{3-10}
  &  & Case 3 & 4 & 4 & 8 & 1 & 8 & - & 176 \\ \hline

\multirow{3}{*}{OTFS-CIM} 
  & \multirow{3}{*}{$\eta_{\text{OTFS-CIM}} = NM\,\bigl[\log_2(M_Q) + 2\log_2(n_C)\bigr]$} 
  & Case 1 & 2 & 2 & 4 & 1 & - & 2 & 16 \\ \cline{3-10}
  &  & Case 2 & 4 & 4 & 4 & 1 & - & 4 & 96 \\ \cline{3-10}
  &  & Case 3 & 4 & 4 & 8 & 1 & - & 8 & 144 \\ \hline  
\end{tabular}
 \vspace{-1em}
\end{table*}

\begin{figure}[t!]
\centering{\includegraphics[width=0.49\textwidth]{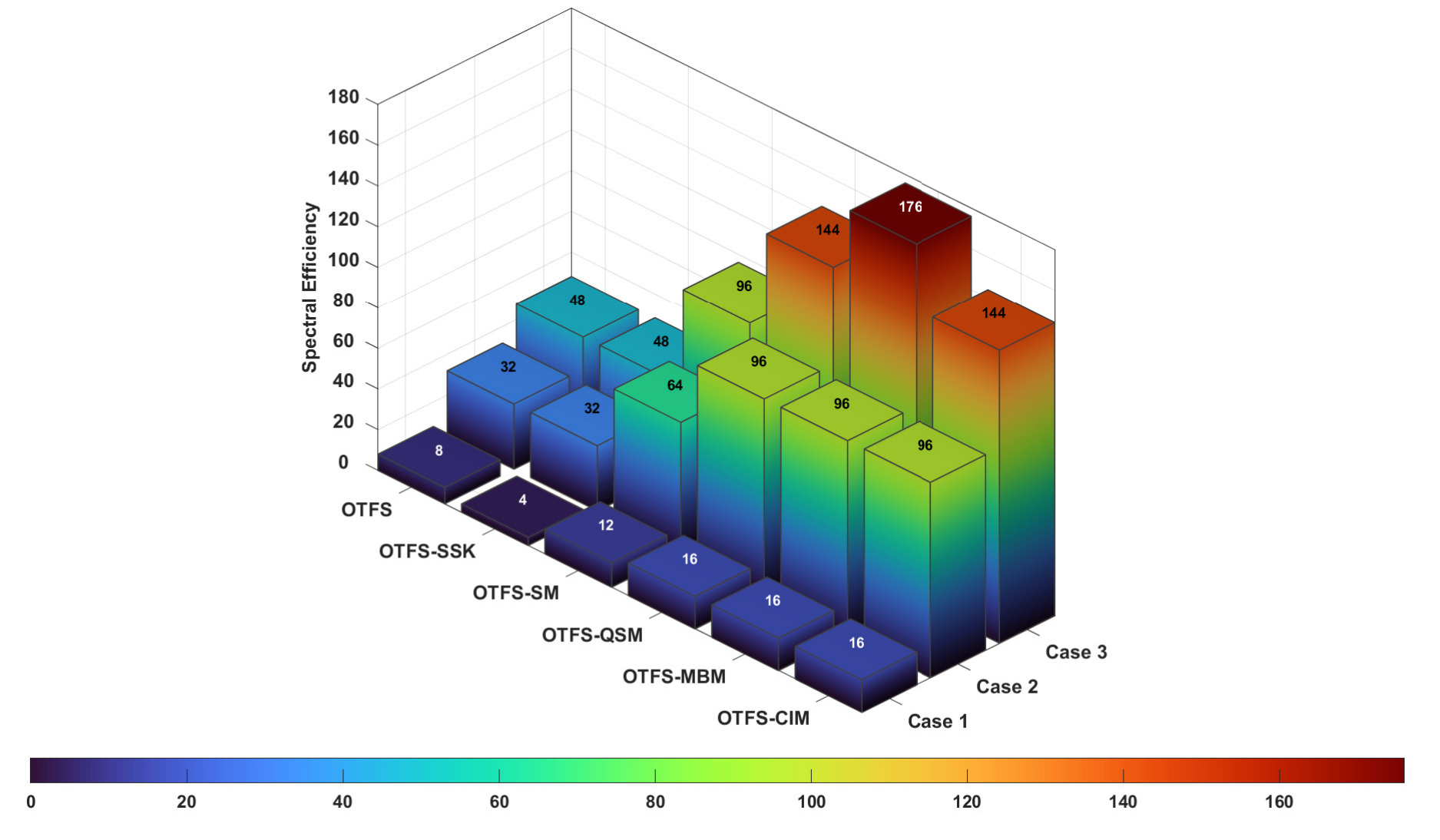}}	
	  \caption{Spectral efficiency comparison of traditional OTFS and OTFS-IM systems for different cases.}
	\label{OTFS_IM_spec} 
\end{figure}

\definecolor{headerblue}{RGB}{60,90,210}     
\definecolor{headertext}{RGB}{255,255,255}   
\definecolor{rowlightblue}{RGB}{232,240,254} 
\definecolor{rowwhite}{RGB}{255,255,255}     

\begin{table*}[t!]
\centering
\addtolength{\tabcolsep}{-2.5pt}
\caption{Spectral efficiency of OTFS–IM schemes.}
\label{tab:SE_OTFS_IM}
\begin{tabular}{|p{2.3cm}|p{2.2cm}|p{5.5cm}|p{6.7cm}|} 
\hline
\hline
\rowcolor{Gray}
 \textbf{\begin{tabular}[c]{@{}c@{}}\small Reference \\ \end{tabular}} & 
 \textbf{\begin{tabular}[c]{@{}c@{}} \small Scheme \\ \end{tabular}} & 
 \textbf{\begin{tabular}[c]{@{}c@{}} \small Spectral efficiency $\eta$ (bits/channel‐use) \\ \end{tabular}} & 
 \textbf{\begin{tabular}[c]{@{}c@{}} \small Explanation \\ \end{tabular}}  \\ \hline\hline
 \rowcolor{rowlightblue}
 \begin{tabular}[c]{@{}c@{}}  \cite{OTFS_IM1} \end{tabular} & 
 \begin{tabular}[c]{@{}c@{}} JDDIM-OTFS \end{tabular} & 
\begin{tabular}[c]{@{}c@{}} $\frac{1+\left\lfloor\log_{2}\!\binom{n}{k}\right\rfloor
      +k\left\lfloor\log_{2} n_{c}\right\rfloor
      +nk\log_{2}M_Q}{n^{2}}$ \end{tabular} &   \begin{tabular}[c]{@{}c@{}}          $k$: number of active  resource  blocks (DD bins) per \\ subframe; \\
$n_{c}$: number of different  constellations; \\
$n$: number of total  resource blocks  in a subframe;\\
$M_Q$: modulation order. \end{tabular}   \\ \hline 

\rowcolor{white}
\begin{tabular}[c]{@{}c@{}} \cite{OTFS_IM2}, \cite{OTFS_IM3}, \cite{OTFS_IM20} \end{tabular} & 
 \begin{tabular}[c]{@{}c@{}} DeIM-OTFS / \\ DoIM-OTFS \end{tabular} & 
 \begin{tabular}[c]{@{}c@{}} $
\frac{ \log_2 \binom{ M }{ k } + k \log_2(M_Q) N }{ M N }
$ \end{tabular} &   
 \begin{tabular}[c]{@{}c@{}} $M$ and $N$ represent the total available numbers of  \\ subcarriers and time slots.  \end{tabular}   \\ \hline 

\rowcolor{rowlightblue}
\begin{tabular}[c]{@{}c@{}} \cite{OTFS_IM4} \end{tabular} & 
 \begin{tabular}[c]{@{}c@{}} IM-FSC  \end{tabular} & 
 \begin{tabular}[c]{@{}c@{}} $\frac{ \left\lfloor \log_2 \binom{a}{k} \right\rfloor + k \log_2 M_Q }{ a }$ \end{tabular} &   
 \begin{tabular}[c]{@{}c@{}} $a = \frac{MN}{G}$, where $G$ is the number of subblocks. \end{tabular}   \\ \hline 
 
\rowcolor{white}
 \begin{tabular}[c]{@{}c@{}} \cite{OTFS_IM5}, \cite{OTFS_IM19}, \cite{OTFS_IM22}, \\ \cite{OTFS_IM24}, \cite{OTFS_IM25}, \cite{OTFS_IM31}, \\ \cite{OTFS_IM32}, \cite{OTFS_IM37}, \cite{OTFS_IM38}, \\ \cite{OTFS_IM41} \end{tabular} & 
 \begin{tabular}[c]{@{}c@{}} OTFS-IM    \end{tabular} & 
 \begin{tabular}[c]{@{}c@{}} $\frac{1}{n}\log_2 \binom{n}{k} + \frac{k}{n}\log_2 M_Q$ \end{tabular} &   
 \begin{tabular}[c]{@{}c@{}} -- \end{tabular}   \\ \hline 

 \rowcolor{rowlightblue}
 \begin{tabular}[c]{@{}c@{}} \cite{OTFS_IM6} \end{tabular} & 
 \begin{tabular}[c]{@{}c@{}} OTFS-IIM  \end{tabular} & 
 \begin{tabular}[c]{@{}c@{}} $\frac{1}{2n} \left\lfloor \log_2 \binom{n}{k} \right\rfloor + \frac{k}{n} \log_2 M_Q$  \end{tabular} &   
 \begin{tabular}[c]{@{}c@{}} -- \end{tabular}   \\ \hline 
 
\rowcolor{white}
 \begin{tabular}[c]{@{}c@{}} \cite{OTFS_IM7}, \cite{OTFS_IM10} \end{tabular} & 
 \begin{tabular}[c]{@{}c@{}} OTFS-DM-IM,\\ W-3D-OTFS-\\ DM-IM \end{tabular} & 
 \begin{tabular}[c]{@{}c@{}} $\frac{1}{n} \left\lfloor \log_2 \binom{n}{k} \right\rfloor + \frac{k}{n} \log_2 M_A + \frac{n-k}{n} \log_2 M_B$  \end{tabular} &   
 \begin{tabular}[c]{@{}c@{}} $M_A$ and $M_B$ denote the modulation order for the \\ active and inactive sub-grids, respectively. \end{tabular}   \\ \hline 

 \rowcolor{rowlightblue}
 \begin{tabular}[c]{@{}c@{}} \cite{OTFS_IM8}  \end{tabular} & 
 \begin{tabular}[c]{@{}c@{}} OTFS-IMMA    \end{tabular} & 
 \begin{tabular}[c]{@{}c@{}} $\left\lfloor \log_2 \binom{n}{k} \right\rfloor + k\log_2(M_Q)
$  \end{tabular} &   
 \begin{tabular}[c]{@{}c@{}} -- \end{tabular}   \\ \hline 

 \rowcolor{white}
 \begin{tabular}[c]{@{}c@{}} -- \end{tabular} & 
 \begin{tabular}[c]{@{}c@{}}  OTFS-SSK  \end{tabular} & 
 \begin{tabular}[c]{@{}c@{}} $NM\log_2(n_T)$ \end{tabular} &   
 \begin{tabular}[c]{@{}c@{}} $n_T$: the number of transmit antennas.  \end{tabular}   \\ \hline 

 \rowcolor{rowlightblue}
 \begin{tabular}[c]{@{}c@{}} \cite{SMOTFS3}, \cite{SMOTFS4},  \cite{SMOTFS7},\\ \cite{SMOTFS10}, \cite{SMOTFS12}  \cite{SMOTFS13}, \\ \cite{SMOTFS15}, \cite{SMOTFS16} \end{tabular} & 
 \begin{tabular}[c]{@{}c@{}} OTFS-SM  \end{tabular} & 
 \begin{tabular}[c]{@{}c@{}} $NM( \log_2 (n_T M_Q))$ \end{tabular} &   
 \begin{tabular}[c]{@{}c@{}} -- \end{tabular}   \\ \hline 

  \rowcolor{white}
 \begin{tabular}[c]{@{}c@{}} \cite{SMOTFS11} \end{tabular} & 
 \begin{tabular}[c]{@{}c@{}} Antenna selection- \\ based OTFS-SM    \end{tabular} & 
 \begin{tabular}[c]{@{}c@{}} $NM( \log_2 (n_S M_Q))$ \end{tabular} &   
 \begin{tabular}[c]{@{}c@{}} $n_S$ is the number of antennas selected from the $n_T$ \\ antennas. \end{tabular}   \\ \hline 

 \rowcolor{rowlightblue}
 \begin{tabular}[c]{@{}c@{}} \cite{OTFS_IM13} \end{tabular} & 
 \begin{tabular}[c]{@{}c@{}} TM-OTFS-IM   \end{tabular} & 
 \begin{tabular}[c]{@{}c@{}} $ \frac{a_1 \log_2(\mathcal{M}_1) + a_2 \log_2(\mathcal{M}_2) + \left\lfloor \log_2 \left( \binom{n}{a} \times \binom{a}{a_1} \right) \right\rfloor}{n}$ \end{tabular} &   
 \begin{tabular}[c]{@{}c@{}} $a_1$: the number of active subblocks  modulated by $\mathcal{M}_1$,\\
$a_2$: the number of active subblocks  modulated by $\mathcal{M}_2$,\\
$\mathcal{M}_1$: the modulation order of  the first constellation,\\
$\mathcal{M}_2$: the modulation order of  the second constellation,\\
$n$:the number of subblocks  in a frame,\\
$a$: the number of active subblocks. \end{tabular}   \\ \hline 

  \rowcolor{white}
 \begin{tabular}[c]{@{}c@{}} \cite{OTFS_IM14} \end{tabular} & 
 \begin{tabular}[c]{@{}c@{}} OTFS-CM-IM \\  (Type-I / Type-II \\ indexing)  \end{tabular} & 
 \begin{tabular}[c]{@{}c@{}} $\eta_{\mathrm{I}} = \frac{1}{NM} \Big[\, C \big( \log_{2}(\frac{n^{2}}{k}) + (n^{2} - k) \times $  \\  $ \log_{2} M_Q \big) +    n_\mathrm{RF} \Big]$ \\
$\eta_{\mathrm{II}} = \frac{1}{NM} \Big[\, C \big( 1 + \log_{2}(\frac{n}{k}) + k \log_{2} n_{c} +  $  \\  $ n k \log_{2} M_Q \big) + n_\mathrm{RF} \Big]$ \end{tabular} &   
 \begin{tabular}[c]{@{}c@{}} $n_c$: number of distinct constellations,\\
$C$: number of repeated sub-grids. \end{tabular}   \\ \hline 

 \rowcolor{rowlightblue}
 \begin{tabular}[c]{@{}c@{}} \cite{OTFS_IM15} \end{tabular} & 
 \begin{tabular}[c]{@{}c@{}} OTFS-DFIM  \end{tabular} & 
 \begin{tabular}[c]{@{}c@{}} $\eta_{\text{DFIM}} = \frac{2 \lfloor \log_2 \binom{b}{g} \rfloor + n_\epsilon + g \log_2 M_Q}{2b}$ \end{tabular} &   
 \begin{tabular}[c]{@{}c@{}} $b$: half of the subblock length, \\
$g$: number of active grid points in a subblock, \\
$n_\epsilon$: length of the indicator bits. \end{tabular}   \\ \hline 

   \rowcolor{white}
 \begin{tabular}[c]{@{}c@{}} \cite{OTFS_IM16} \end{tabular} & 
 \begin{tabular}[c]{@{}c@{}} E-OTFS-IM  \end{tabular} & 
 \begin{tabular}[c]{@{}c@{}} $\frac{1}{n} \left\lfloor \log_2 \big| \sum_{i=1}^{R} \binom{n}{k_i} |S_i|^{k_i}\big| \right\rfloor$ \end{tabular} &   
 \begin{tabular}[c]{@{}c@{}} $R$: the number of possible grid  activation patterns (GAPs)\\
$k_i$: the number of active grids  for the $i$-th GAP\\
$|S_i|$: constellation size for  the $i$-th GAP\\ \end{tabular}   \\ \hline 

 \rowcolor{rowlightblue}
 \begin{tabular}[c]{@{}c@{}} \cite{OTFS_IM17} \end{tabular} & 
 \begin{tabular}[c]{@{}c@{}}  OTFS-GDM-IM   \end{tabular} & 
 \begin{tabular}[c]{@{}c@{}}  $\left\lfloor \log_{2} \Bigg( \sum_{i=1}^{T} \Big( \binom{n}{k_i} M_{A}^{k_i} M_{B}^{n-k_i} \Big) \Bigg) \right\rfloor$ \end{tabular} &   
 \begin{tabular}[c]{@{}c@{}} $T$: the total number of  possible configurations. \end{tabular}   \\ \hline 

   \rowcolor{white}
 \begin{tabular}[c]{@{}c@{}} \cite{OTFS_IM18}, \cite{OTFS_IM40} \end{tabular} & 
 \begin{tabular}[c]{@{}c@{}} AEE-JDDIM-OTFS  \end{tabular} & 
 \begin{tabular}[c]{@{}c@{}}  $\frac{1 + \left\lfloor \log_2 \binom{n}{k} \right\rfloor + kn\,\frac{\log_2 M_Q}{n}}{n^2}$   \end{tabular} &   
 \begin{tabular}[c]{@{}c@{}} -- \end{tabular}   \\ \hline 

 \rowcolor{rowlightblue}
 \begin{tabular}[c]{@{}c@{}} \cite{OTFS_IM21}  \end{tabular} & 
 \begin{tabular}[c]{@{}c@{}} SIM-OTFS  \end{tabular} & 
 \begin{tabular}[c]{@{}c@{}} $\frac{1}{n} \left\lfloor \log_2 \binom{n}{k} \right\rfloor + \frac{k}{n} \log_2 M_Q$  \end{tabular} &   
 \begin{tabular}[c]{@{}c@{}} -- \end{tabular}   \\ \hline 

   \rowcolor{white}
 \begin{tabular}[c]{@{}c@{}} \cite{OTFS_IM22} \end{tabular} & 
 \begin{tabular}[c]{@{}c@{}} OTFS-IM-NOMA   \end{tabular} & 
 \begin{tabular}[c]{@{}c@{}} $\frac{1}{n} \Big( \lfloor \log_2 \binom{n}{k} \rfloor + k \log_2 M_Q \Big)$  \end{tabular} &   
 \begin{tabular}[c]{@{}c@{}} -- \end{tabular}   \\ \hline 

 \rowcolor{rowlightblue}
 \begin{tabular}[c]{@{}c@{}} \cite{OTFS_IM23} \end{tabular} & 
 \begin{tabular}[c]{@{}c@{}} OTFS-IM \\ (DaF relaying) \end{tabular} & 
 \begin{tabular}[c]{@{}c@{}} $\frac{1}{2n} \Big( \lfloor \log_2 \binom{n}{k} \rfloor + k \log_2 M_Q \Big)$  \end{tabular} &   
 \begin{tabular}[c]{@{}c@{}} -- \end{tabular}   \\ \hline 

    \rowcolor{white}
 \begin{tabular}[c]{@{}c@{}} \cite{OTFS_IM26}, \cite{OTFS_IM35} \end{tabular} & 
 \begin{tabular}[c]{@{}c@{}} SIM-OTFS  \end{tabular} & 
 \begin{tabular}[c]{@{}c@{}} $\log_2 (n_T M N M_Q)$ \end{tabular} &   
 \begin{tabular}[c]{@{}c@{}} -- \end{tabular}   \\ \hline 

 \rowcolor{rowlightblue}
 \begin{tabular}[c]{@{}c@{}} \cite{OTFS_IM27} \end{tabular} & 
 \begin{tabular}[c]{@{}c@{}} VBLAST-OTFS-IM \end{tabular} & 
 \begin{tabular}[c]{@{}c@{}} $\eta_{\text{VBLAST}} = n_T \Bigg( \frac{ \lfloor \log_2 \binom{n}{k} \rfloor + k \log_2 M_Q }{n} \Bigg)$ \end{tabular} &   
 \begin{tabular}[c]{@{}c@{}} -- \end{tabular}   \\ \hline 

    \rowcolor{white}
 \begin{tabular}[c]{@{}c@{}} \cite{OTFS_IM28}  \end{tabular} & 
 \begin{tabular}[c]{@{}c@{}} GIM-MIMO-OTFS  \end{tabular} & 
 \begin{tabular}[c]{@{}c@{}} $\frac{
\Big\lfloor \log_2 \binom{n_T}{n_a} \Big\rfloor 
+ n_a G \Big( \Big\lfloor \log_2 \binom{n}{k} \Big\rfloor + k \log_2 M_Q \Big)
}{MN}$ \end{tabular} &   
 \begin{tabular}[c]{@{}c@{}} $G$: the number of subblocks. \end{tabular}   \\ \hline 

 \rowcolor{rowlightblue}
 \begin{tabular}[c]{@{}c@{}} \cite{OTFS_IM29} \end{tabular} & 
 \begin{tabular}[c]{@{}c@{}} GSDDIM-OTFS   \end{tabular} & 
 \begin{tabular}[c]{@{}c@{}} $ \frac{ \big( \lfloor \log_2 \binom{n n_T}{k} \rfloor + k \log_2 M_Q \big) G }{MN}$ \end{tabular} &   
 \begin{tabular}[c]{@{}c@{}} -- \end{tabular}   \\ \hline 

    \rowcolor{white}
 \begin{tabular}[c]{@{}c@{}} \cite{OTFS_IM30} \end{tabular} & 
 \begin{tabular}[c]{@{}c@{}} GSIM-OTFS  \end{tabular} & 
 \begin{tabular}[c]{@{}c@{}} $\lfloor \log_2 \binom{n_T}{n_a} \rfloor +   \log_2(N) + $ \\ $\log_2(M) + \log_2(M_Q) \Big\rfloor$ \end{tabular} &   
 \begin{tabular}[c]{@{}c@{}} -- \end{tabular}   \\ \hline 

 \rowcolor{rowlightblue}
 \begin{tabular}[c]{@{}c@{}} \cite{OTFS_IM31} \end{tabular} & 
 \begin{tabular}[c]{@{}c@{}} IM-OTFS  \end{tabular} & 
 \begin{tabular}[c]{@{}c@{}} $\lfloor \log_2 \binom{n}{k} \rfloor g + k \log_2 M_Q \, g \quad$ \end{tabular} &   
 \begin{tabular}[c]{@{}c@{}} $g$: the number of groups.  \end{tabular}     \\ \hline  \hline 
\end{tabular}
\end{table*}

\begin{table*}[t!]
\ContinuedFloat
\centering
\addtolength{\tabcolsep}{-2.5pt}
\caption{Spectral efficiency of OTFS–IM schemes (Continued).}
\begin{tabular}{|p{2.3cm}|p{2.2cm}|p{5.5cm}|p{6.7cm}|} 
\hline
\hline

\rowcolor{Gray}
 \textbf{\begin{tabular}[c]{@{}c@{}}\small Reference \\ \end{tabular}} & 
 \textbf{\begin{tabular}[c]{@{}c@{}} \small Scheme \\ \end{tabular}} & 
 \textbf{\begin{tabular}[c]{@{}c@{}} \small Spectral efficiency $\eta$ (bits/channel‐use) \\ \end{tabular}} & 
 \textbf{\begin{tabular}[c]{@{}c@{}} \small Explanation \\ \end{tabular}}  \\ \hline\hline

\rowcolor{rowlightblue}
 \begin{tabular}[c]{@{}c@{}} \cite{OTFS_IM33}  \end{tabular} & 
 \begin{tabular}[c]{@{}c@{}} MM-OTFS-IM  \end{tabular} & 
 \begin{tabular}[c]{@{}c@{}} $\frac{
\Big\lfloor \log_{2} \big( (NM)! \big) \Big\rfloor + NM\,\big( \log_{2}(M_Q) \big)
}{NM}$ \end{tabular} &   
 \begin{tabular}[c]{@{}c@{}} -- \end{tabular}   \\ \hline 

 \rowcolor{white}
 \begin{tabular}[c]{@{}c@{}} \cite{OTFS_IM34} \end{tabular} & 
 \begin{tabular}[c]{@{}c@{}} OTFS-I/Q-IM  \end{tabular} & 
 \begin{tabular}[c]{@{}c@{}} $ \frac{2}{n} \Big( \lfloor \log_2 \binom{n}{k} \rfloor + k \log_2 M_Q \Big)$  \end{tabular} &   
 \begin{tabular}[c]{@{}c@{}} -- \end{tabular}  \\ \hline 
 
 \rowcolor{rowlightblue}
 \begin{tabular}[c]{@{}c@{}} \cite{OTFS_IM36}  \end{tabular} & 
 \begin{tabular}[c]{@{}c@{}} OTFS-IM \\ ($IQ$ imb. comp.)  \end{tabular} & 
 \begin{tabular}[c]{@{}c@{}} $\frac{G \big( \lfloor \log_2 \binom{n}{k} \rfloor + k \log_2 M_Q \big) }{NM}$ \end{tabular} &   
 \begin{tabular}[c]{@{}c@{}} -- \end{tabular}   \\ \hline

\rowcolor{white}
 \begin{tabular}[c]{@{}c@{}} \cite{OTFS_IM39} \end{tabular} & 
 \begin{tabular}[c]{@{}c@{}} OTFS-IM4   \end{tabular} & 
 \begin{tabular}[c]{@{}c@{}} $ \frac{1}{n} \left( \left\lfloor \log_2 \left( \frac{n}{k} \right) \right\rfloor + \frac{k}{2} \log_2(M_c) \right) $ \end{tabular} &   
 \begin{tabular}[c]{@{}c@{}} $M_c$ denotes the size of the four-dimensional \\ spherical code. \end{tabular}   \\ \hline 

  \rowcolor{rowlightblue}
 \begin{tabular}[c]{@{}c@{}} \cite{OTFS_IM42} \end{tabular} & 
 \begin{tabular}[c]{@{}c@{}} ODDM-HMIM    \end{tabular} & 
 \begin{tabular}[c]{@{}c@{}} $\log_2 Q_1 + \frac{\log_2 Q_2}{N_b}$ \end{tabular} &   
 \begin{tabular}[c]{@{}c@{}} $Q_1$: QAM order, $Q_2$ is the mode order. \\ $N_b$: the block length. \end{tabular}   \\ \hline 

\rowcolor{white}
 \begin{tabular}[c]{@{}c@{}} \cite{SMOTFS1}, \cite{SMOTFS2}  \end{tabular} & 
 \begin{tabular}[c]{@{}c@{}}  GSM-OTFS    \end{tabular} & 
 \begin{tabular}[c]{@{}c@{}} $\left\lfloor \log_2 \binom{n_T}{n_a} \right\rfloor + n_a \log_2 M_Q$ \end{tabular} &   
 \begin{tabular}[c]{@{}c@{}} $n_a$: the number of active transmit antennas. \end{tabular}   \\ \hline 

  \rowcolor{rowlightblue}
 \begin{tabular}[c]{@{}c@{}} \cite{SMOTFS5}  \end{tabular} & 
 \begin{tabular}[c]{@{}c@{}} ESM-OTFS  \end{tabular} & 
 \begin{tabular}[c]{@{}c@{}} $NM( \log_2(\mathbb{C}) )$ \end{tabular} &   
 \begin{tabular}[c]{@{}c@{}} $\mathbb{C}$ is the size of the ESM constellation.  \end{tabular}   \\ \hline 

\rowcolor{white}
 \begin{tabular}[c]{@{}c@{}} \cite{SMOTFS9}  \end{tabular} & 
 \begin{tabular}[c]{@{}c@{}} MMIM-OTFS-SM  \end{tabular} & 
 \begin{tabular}[c]{@{}c@{}} $\frac{
    \log_{2}(n_{T}) + G\left( \left\lfloor \log_{2}(L!) \right\rfloor + L \log_{2}(M_Q) \right)}{MN}$ \end{tabular} &   
 \begin{tabular}[c]{@{}c@{}}  $L = \frac{MN}{G}$ \end{tabular}   \\ \hline 

  \rowcolor{rowlightblue}
 \begin{tabular}[c]{@{}c@{}} \cite{SMOTFS8}  \end{tabular} & 
 \begin{tabular}[c]{@{}c@{}} SM-STBC-OTFS     \end{tabular} & 
 \begin{tabular}[c]{@{}c@{}} $\left\lfloor \log_2 \binom{n_T}{2} \right\rfloor + 2 \log_2 M_Q$ \end{tabular} &   
 \begin{tabular}[c]{@{}c@{}} -- \end{tabular}   \\ \hline 

\rowcolor{white}
 \begin{tabular}[c]{@{}c@{}} \cite{OTFS_IM9} \end{tabular} & 
 \begin{tabular}[c]{@{}c@{}} OTFS-SBIM \end{tabular} & 
 \begin{tabular}[c]{@{}c@{}} $ \frac{k \log_2 M_Q + \left\lfloor \log_2 \binom{n}{k} \right\rfloor + 1}{n} $  \end{tabular} &   
 \begin{tabular}[c]{@{}c@{}} -- \end{tabular}   \\ \hline 

  \rowcolor{rowlightblue}
 \begin{tabular}[c]{@{}c@{}} \cite{OTFS_IM11} \end{tabular} & 
 \begin{tabular}[c]{@{}c@{}} OTFS-MBM   \end{tabular} & 
 \begin{tabular}[c]{@{}c@{}} $NM \left(log_2(M_Q) + n_\mathrm{RF}\right)$ \end{tabular} &   
 \begin{tabular}[c]{@{}c@{}} $n_\mathrm{RF}$: the number of RF mirrors. \end{tabular}   \\ \hline 

\rowcolor{white}
 \begin{tabular}[c]{@{}c@{}} \cite{OTFS_IM12}  \end{tabular} & 
 \begin{tabular}[c]{@{}c@{}} OTFS-CIM  \end{tabular} & 
 \begin{tabular}[c]{@{}c@{}} $NM \left( \log_2(M_Q) + 2 \log_2(N_C) \right)$  \end{tabular} &   
 \begin{tabular}[c]{@{}c@{}} $N_C$: the number of spreading codes.  \end{tabular}   \\ \hline 

   \rowcolor{rowlightblue}
 \begin{tabular}[c]{@{}c@{}} \cite{OTFSQSM1}  \end{tabular} & 
 \begin{tabular}[c]{@{}c@{}} QSM-OTFS  \end{tabular} & 
 \begin{tabular}[c]{@{}c@{}} $\log_2(M_Q) + 2 \left\lfloor \log_2(n_T) \right\rfloor$ \end{tabular} &   
 \begin{tabular}[c]{@{}c@{}} -- \end{tabular}   \\ \hline 

\rowcolor{white}
 \begin{tabular}[c]{@{}c@{}} \cite{OTFSQSM2}  \end{tabular} & 
 \begin{tabular}[c]{@{}c@{}} DAQSM-OTFS   \end{tabular} & 
 \begin{tabular}[c]{@{}c@{}} $\frac{ P\log_{2}M_{Q} \;+\;2\log_{2}G \;+\;2\Bigl\lfloor \log_{2}\binom{Q}{P} \Bigr\rfloor }{2}$  \end{tabular} &   
 \begin{tabular}[c]{@{}c@{}} $Q$ is the number of  dispersion matrices. \\ $P$ is the number of  constellation symbols. \end{tabular}   \\ \hline \hline 
\end{tabular}
\end{table*}

\subsection{Spectral Efficiency Comparisons of OTFS-IM Systems}
In this subsection, the spectral efficiency of conventional OTFS and OTFS-IM schemes is compared under different system configurations, as detailed in Table \ref{SE_OTFS_IM_nu}. Three cases are considered for each scheme: Case 1 uses $N = 2$ delay bins, $M = 2$ Doppler bins, and modulation order $M_Q = 4$; Case 2 uses $N = 4$, $M = 4$, $M_Q = 4$; and Case 3 uses $N = 4$, $M = 4$, $M_Q = 8$. For OTFS, the number of transmit antennas is fixed at $n_T = 1$. In the OTFS-SSK, OTFS-SM, and OTFS-QSM schemes, the number of transmit antennas increases with each case, specifically $n_T = 2, 4, 8$ for Cases 1, 2, and 3, respectively. In OTFS-MBM, the number of RF mirrors is set to $n_{RF} = 2, 4, 8$, while $n_T = 1$. For OTFS-CIM, the number of spreading codes is $n_C = 2, 4, 8$ in Cases 1, 2, and 3, with $n_T = 1$. Fig. \ref{OTFS_IM_spec} shows that OTFS-IM schemes deliver higher spectral efficiency than conventional OTFS. In addition, depending on the chosen system parameters, OTFS-MBM exhibits the fastest increase in spectral efficiency across the considered cases. Because the spectral efficiency increase of OTFS-MBM is linear rather than logarithmic, unlike the other schemes. In addition, although OTFS-SM and OTFS-QSM systems use the same system parameters, the OTFS-QSM system achieves significantly higher spectral efficiency. Also, Table \ref{tab:SE_OTFS_IM} presents a comprehensive list of the spectral efficiency expressions of all OTFS-IM schemes in the literature.

\begin{figure}[t!]
  \centering
  \includegraphics[width=0.9\linewidth]{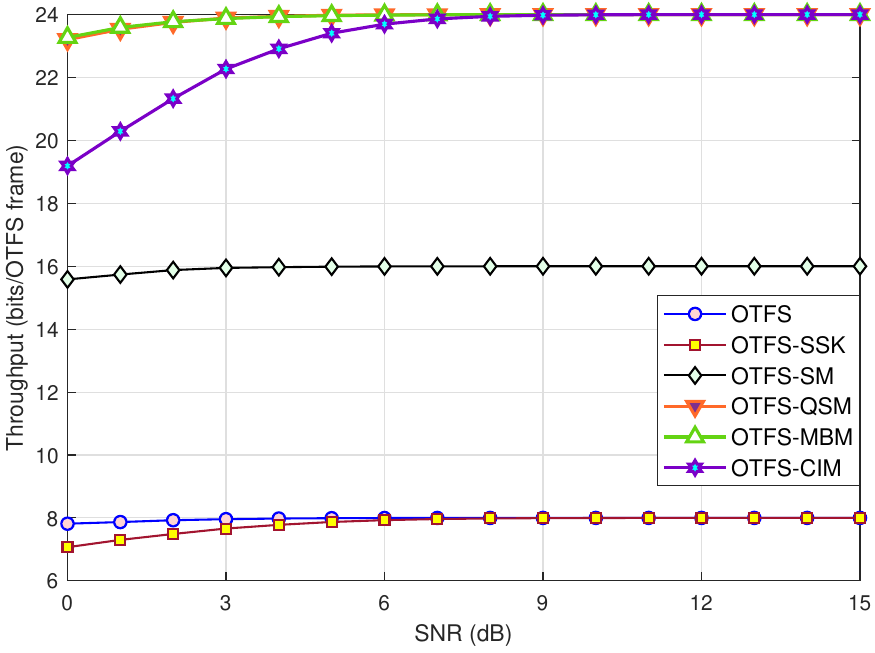}
  \caption{Throughput comparison of OTFS and OTFS-IM schemes.}
  \label{fig:otfs-throughput}
\end{figure}

\subsection{Throughput Comparisons of OTFS-IM Systems}
In OTFS-based IM systems, information bits are encoded not only in traditional constellation symbols, but also within indices of various active system parameters, thus increasing the total number of bits transmitted per symbol. The throughput of a wireless communication system can be expressed mathematically as follows:
\begin{equation}\label{throughputeq}
\mathcal{T}   =  \frac{\big(1-\text{BER}\big)}{T_x}\eta,
\end{equation} 
where BER and $\eta$ are the bit error rate and spectral efficiency of the considered system, respectively, and $T_x$ is the symbol transmission duration. Fig. \ref{fig:otfs-throughput} illustrates the throughput performance of various OTFS and OTFS-IM systems. All systems use $n_R = 4$ and $M_Q = 4$. The OTFS-SSK, OTFS-SM, and OTFS-QSM systems employ $n_T = 4$; OTFS-MBM system uses $n_{RF} = 4$; and OTFS-CIM system is configured with $N_C = 4$ and $L = 8$. In  Fig. \ref{fig:otfs-throughput}, it is observed that OTFS-IM systems achieve higher throughput than conventional OTFS systems, thanks to IM. Among the OTFS-IM schemes, OTFS-MBM, OTFS-QSM, and OTFS-CIM achieve the highest throughput, followed by OTFS-SM and OTFS-SSK.

\section{Challenges, Benefits, and Future Directions for OTFS-IM Systems}
OTFS-IM systems offer many advantages for next-generation wireless communication. However, they also have several challenges that need to be addressed. The key challenges, benefits, and potential future directions of OTFS-IM systems are discussed as follows:

\begin{enumerate}
    \item As shown in Fig. \ref{OTFS_IM_comps}, OTFS-IM systems have an acceptable computational complexity compared to conventional OTFS systems for the same spectral efficiency. However, in OTFS-IM systems, increasing the data rate requires increasing the number of system parameters, which leads to higher hardware complexity compared to conventional OTFS systems. Therefore, it is important to design low-complexity detectors for OTFS-IM systems. In \cite{OTFS_IM3}, a low-complexity CMP algorithm is designed for the proposed DoIM-OTFS scheme. A low-complexity detector based on integrated MMSE-ML detection is proposed for the OTFS-IM system in \cite{OTFS_IM5}. To reduce the complexity of the receiver of the OTFS-IIM system, a low-complexity energy-based detector is proposed in \cite{OTFS_IM6, OTFS_IM21}. At the receiver in the IM-MIMO-OTFS system, a two-stage index detection is explored, enabling separate detection of spatial indices and DD indices with low complexity in \cite{OTFS_IM28}. In \cite{OTFS_IM32}, a low-complexity SS-EP-based detector is developed for OTFS-IM systems. A low-complexity detector is designed for the OTFS-SM system in \cite{SMOTFS7, SMOTFS16}. In OTFS-IM systems, the joint detection of transmitted symbols and indices, together with the addition of the $NM$ DD domain size, leads to higher receiver complexity.  Thus, although low-complexity detectors have been proposed for OTFS-IM systems in the literature, further development of advanced low-complexity detectors is necessary for 6G and beyond wireless networks.

    \item There are currently no studies in the literature that have integrated the OTFS system into a MISO architecture. The OTFS technique can be integrated into existing MISO systems, such as RIS-based spatial MBM (SMBM) in \cite{MBM3}, and RIS-based SSK in \cite{RISCHAL} schemes, for improved spectral efficiency.

    \item The spectral efficiency of conventional OTFS modulation is significantly increased through the integration of IM techniques. However, to further increase spectral efficiency, the OTFS technique can be combined with high spectral efficiency IM schemes such as SMBM in \cite{SMBMilk, MBM3}, double SM in \cite{DSM}, quadrature SMBM in \cite{QSMMBM}, CIM-based SMBM (CIM-SMBM) in \cite{MBM6}, double SMBM in \cite{doubleSMBM}, CIM-based SM (CIM-SM) in \cite{CIM7}, and CIM-based QSM (CIM-QSM) in \cite{CIM2}.

    \item Even though OTFS-IM systems can achieve high data rates, their maximum data rate is ultimately limited by the finite number of indexable communication resources.

    \item The integration of IM schemes with OTFS improves not only spectral efficiency but also error performance, throughput, and energy efficiency, positioning it as a strong candidate for meeting the challenging reliability and performance requirements of 6G systems.

    \item As also described in \cite{OTFSSURVEY14} and \cite{OTFSchalle1}, the OTFS technique spreads information over a longer block, providing robustness against channel fluctuations. However, this also increases latency. Unlike the OFDM technique, with the same number of subcarriers and subcarrier spacing, each OTFS signal lasts $N$ times longer than a traditional OFDM symbol.

    \item The fractional Doppler spreads the effective channel in the DD domain, making accurate channel estimation difficult, especially when the signal duration is limited to reduce latency \cite{OTFSchalle2, OTFSchalle3, OTFSchalle4, OTFSchalle5, OTFSchalle6, OTFSchalle7}.

    \item  In OTFS systems, modulation and demodulation rely on successive domain transformations, specifically the ISFFT and Heisenberg transform at the transmitter, and the Wigner transform and SFFT at the receiver, which compared to OFDM (especially for large block sizes), result in considerable complexity and hardware-related challenges, making the efficient design and implementation of transceiver architectures and hardware mandatory.

    \item In OTFS-SM systems, spectral efficiency increases logarithmically with the number of transmit antennas; however, the use of a large number of antennas leads to higher spatial correlation, which deteriorates the error performance and imposes significant hardware complexity. Furthermore, achieving very high spectral efficiency requires a large number of antennas in practice, which makes physical implementation impossible. To overcome these limitations, MBM-based IM techniques provide a linear increase in spectral efficiency by increasing only the number of RF mirrors, independent of the number of multiple RF chains and antennas, and can be integrated with OTFS. In \cite{OTFS_IM11}, OTFS and MBM techniques are combined in a SIMO architecture. However, MBM-based OTFS schemes have not been extensively explored in the literature.

    \item Rectangular pulses are widely used in OTFS systems, but they cause out-of-band emissions that lead to adjacent channel interference and deteriorate error performance \cite{OTFSSURVEY5, OTFSPRAC, OTFSPRAC2}.

    \item The combination of IM with OTFS-RIS systems has only been investigated in \cite{OTFS_RIS_IM_1}. OTFS-RIS can be integrated with high-performance IM schemes such as SSK, SM, QSM, MBM, and CIM to achieve higher spectral and energy efficiency as well as improved error performance.

    \item Although the OTFS-based systems are better than the OFDM-based systems in terms of PAPR \cite{OTFSSURVEY8, OTFS_IM34, OTFS_IM39}, the OTFS-SBIM system proposed in \cite{OTFS_IM9} achieves lower BER and PAPR compared to the traditional OTFS-IM system. However, there is still a need to further reduce the PAPR in OTFS systems.

    \item Due to its structural similarity to OFDM, OTFS is inherently susceptible to synchronization errors such as carrier frequency offset and sampling time offset; however, carrier frequency offset does not pose an issue for OTFS, as it manifests merely as a Doppler shift in the DD domain \cite{OTFSENK1, OTFSENK2, OTFSENK3, OTFSENK4, OTFSENK5}.

    \item Combining OTFS and CIM techniques provides benefits for physical layer security. CIM transmits additional data bits by encoding in the indices of selected spreading codes, which not only increases spectral and energy efficiency but also improves signal randomness and secrecy by varying the spreading code indices for each transmission. OTFS further enhances this by spreading the signal in the DD domain, resulting in a highly random and unpredictable channel environment, making interception and decoding by eavesdroppers significantly more difficult. However, studies on combining OTFS and CIM techniques within a wireless communication system are quite limited \cite{OTFS_IM12}.

    \item OTFS-IM systems are highly sensitive to hardware impairments such as phase noise, $IQ$ imbalance, and carrier frequency offset. Robust synchronization and compensation algorithms specifically designed for OTFS-IM are essential for reliable system operation. \cite{OTFS_IM36} proposes an IM–aided $IQ$ imbalance compensation scheme for OTFS systems to address hardware impairments in next-generation wireless communications. 

    \item Machine learning and deep learning based detectors improve the reliability and efficiency of symbol and index detection in OTFS-IM systems. In \cite{OTFS_IM40}, deep learning is employed to optimize the mapping and demapping operations in the proposed AEE-JDDIM-OTFS system, which is an OTFS-IM scheme. However, other OTFS-IM schemes in the literature have not yet been integrated with machine learning and deep learning techniques.
\end{enumerate}

\section{Conclusion}
A comprehensive survey of OTFS-based wireless communication systems has been conducted, with particular emphasis placed on OTFS-IM schemes. The fundamental principles, transceiver model, and error performance of conventional OTFS modulation in SISO and MISO architectures have been outlined, and all OTFS-IM schemes reported in the literature have been systematically classified according to their system architectures, employed detectors, and performance metrics, including capacity, PAPR, diversity, complexity, imperfect CSI, spectral efficiency, and outage probability. The operating principles and system models of OTFS-SM, OTFS-QSM, OTFS-MBM, OTFS-CIM, and OTFS-SSK variants have been described, and comparative evaluations in terms of computational complexity, error performance, capacity, energy saving, spectral efficiency, and throughput have been provided. Furthermore, the principal challenges, potential benefits, and open research directions for OTFS-IM systems have been addressed, addressing aspects such as complexity reduction, spectral efficiency enhancement, latency minimization, accurate channel estimation, hardware constraints, synchronization, PAPR mitigation, physical layer security, and integration with advanced wireless communication technologies. The robustness of OTFS-IM schemes against challenging channel conditions, combined with their ability to efficiently utilize available resources, positions them as a promising approach for enabling faster, more reliable, and more energy-efficient communication. In conclusion, OTFS-IM schemes are expected to play a key role in enabling 6G and future generation wireless communication systems with higher capacity, enhanced reliability, and improved efficiency.

\bibliographystyle{IEEEtran}
\bibliography{IEEEabrv,Referanslar}


\vfill

\end{document}